\documentclass[12pt,a4paper]{amsart}
\usepackage[T1]{fontenc}
\usepackage{mathdots}
\usepackage[a4paper,left=2.25cm,right=2.25cm,top=3cm,bottom=3cm]{geometry}
\usepackage{enumerate}
\usepackage{tikz}\usetikzlibrary{decorations.markings,patterns}
\usepackage{mathtools}
\usepackage{accents}
\usepackage{breqn}
\usepackage{xcolor}
\newcommand{\bra}[1]{\left\langle #1\right|}
\newcommand{\ket}[1]{\left|#1\right\rangle}
\newcommand{\kets}[1]{|#1\rangle}
\newcommand{\braket}[2]{\left< #1 \vphantom{#2}\right|
\! \left. #2 \vphantom{#1} \right>}%

\newcommand{\nn}{\nonumber}
\newcommand{\eps}{\epsilon}

\newcommand{\figref}[1]{Figure~\ref{fig:#1}}
\newcommand{\secref}[1]{Section~\ref{sec:#1}}
\newcommand{\propref}[1]{Proposition~\ref{prop:#1}}
\newcommand{\lmref}[1]{Lemma~\ref{lm:#1}}
\newcommand{\thmref}[1]{Theorem~\ref{thm:#1}}
\newcommand{\corref}[1]{Corollary~\ref{cor:#1}}
\newcommand{\appref}[1]{Appendix~\ref{app:#1}}
\newcommand{\conjref}[1]{Conjecture~\ref{conj:#1}}
\newcommand{\tabref}[1]{Table~\ref{tab:#1}}
\newcommand\scs{\scriptstyle}
\def\Ker{\operatorname{Ker}}
\def\Im{\operatorname{Im}}

\def\rkm{\operatorname{rm}}
\def\rk{\operatorname{rk}}
\def\mdeg{\operatorname{mdeg}}
\newcommand{\stack}[2]{\begin{subarray}{c} #1 \\ #2 \end{subarray}}
\newcommand\cl{cl}
\def\Pf{\operatorname{Pf}}
\DeclarePairedDelimiter\ceil{\lceil}{\rceil}
\DeclarePairedDelimiter\floor{\lfloor}{\rfloor}
\newcommand\Z{{\mathbb Z}}
\newcommand\ZN{{\Z\bmod N}}
\newcommand\ttilde[1]{\tilde{\vphantom{\bar #1}\smash{\tilde{#1}}}}
\newcommand\divdif{\partial'}
\newcommand\C{\mathbb C}
\newcommand\M{\mathcal M}
\newcommand\g{\mathfrak g}

\newcommand\gl{\mathfrak{gl}}
\renewcommand\sp{\mathfrak{sp}}
\newcommand\La{\Lambda}
\newcommand\U{\Upsilon}
\allowdisplaybreaks[1]
\theoremstyle{plain}
\newtheorem{thm}{Theorem}[section]
\newtheorem{lm}[thm]{Lemma}
\newtheorem{prop}[thm]{Proposition}
\newtheorem{cor}[thm]{Corollary}
\theoremstyle{definition}
\newtheorem{defn}{Definition}[section]
\newtheorem{conj}{Conjecture}[section]
\newtheorem{exmp}{Example}[section]
\theoremstyle{remark}
\newtheorem*{remark}{Remark}
\newtheorem*{thm*}{Theorem}

\author{Anita Ponsaing}
\author{Paul Zinn-Justin}
\thanks{The authors are supported by ERC grant 278124 ``LIC''.
The computer experimentation that led to the results of this paper have been performed
with the help of \cite{M2}.}
\address{Laboratoire de Physique Th\'eorique et Hautes \'Energies, CNRS UMR 7589 and Universit\'e Pierre et Marie Curie (Paris 6), 4 place Jussieu, 75252 Paris cedex 05, France}
\title{Type $\hat{\mathrm C}$ Brauer loop schemes and loop model with boundaries}
\begin{document}
\begin{abstract}
In this paper we study the Brauer loop model on a strip and the associated quantum Knizhnik--Zamolodchikov (qKZ) equation. We show that the minimal degree solution of the Brauer qKZ equation with one of four different possible boundary conditions, gives the multidegrees of the irreducible components of generalizations of the Brauer loop scheme of \cite[Knutson--Zinn-Justin '07]{artic33} with
one of four kinds of symplectic-type symmetry. This is accomplished by studying these irreducible components, which are indexed by link patterns, and describing the geometric action of Brauer
generators on them. We also provide recurrence relations for the multidegrees and compute the sum rule (multidegree of the whole schemes).
\end{abstract}

\maketitle
\tableofcontents

\section{Introduction}
\subsection{Background}
Recently there has been renewed interest in the connection
between quantum integrable systems and algebraic geometry, see e.g.~\cite{GK-K,GRTV,MO-qg,Motegi-Groth}.
A particularly explicit realization of this correspondence can be found in the case of the Brauer
loop model \cite{MNR-crossloop,dGN-brauer,artic32} and its geometric counterpart,
the Brauer loop scheme
\cite{artic33,artic39}. As part of the dictionary between these two subjects (cf.~\cite{MO-qg}),
the type of symmetry of the geometric object determines the boundary conditions of the integrable model.
In the setup of \cite{artic33,artic39}, the Brauer loop scheme has symmetry governed by
an algebra of type $\hat{\mathrm A}$, so that the corresponding $R$-matrix
satisfies the ordinary Yang--Baxter equation and the Brauer loop model has periodic boundary conditions.
It is expected that other types will lead to other boundary conditions
(see \cite{artic35} for some experiments in that general direction).
Among them, type $\hat{\mathrm C}$ is particularly natural:
according to its Dynkin diagram (see \secref{dynkin}),
it should correspond to models defined on a strip, with a bulk defined in terms of
ordinary $R$-matrices satisfying the Yang--Baxter equation, and two
boundaries, each with a $K$-matrix satisfying the reflection
(or boundary Yang--Baxter) equation.

The purpose of this paper
is to validate this hypothesis by, on the one hand, introducing and studying
the Brauer loop model on a strip with various boundary conditions, and on the other,
describing their geometric counterparts, leading to four distinct
type $\hat{\mathrm C}$ Brauer loop schemes.

\subsection{Results}
More specifically, we study the type $\hat{\mathrm C}$ quantum Knizhnik--Zamolodchikov (qKZ) equation -- by this we mean the generalization due to Cherednik \cite{Che-qKZ} of
the original qKZ equation \cite{FR-qKZ} to all types -- associated
to the Brauer $R$-matrix and to either a trivial or nontrivial $K$-matrix at both boundaries, with a further refinement
consisting in identifying or not the two boundaries.
Taking into account the obvious left/right symmetry, these boundary conditions lead to four possibilities,
denoted by the superscripts $\mathrm i$ (identified), $\mathrm c$ (closed), $\mathrm o$ (open), $\mathrm m$ (mixed).
We are interested in a polynomial solution for a specific value of the shift parameter
of the qKZ equation.\footnote{We expect polynomiality to be only possible for discrete values of this parameter,
and the one we choose to give the lowest possible degree among these polynomial solutions.}
Note that the Brauer qKZ equation is very nontrivial because, in contradistinction with the more usual case of the
Hecke algebra,
the $R$-matrix has three terms, so that a polynomial solution is not obviously related to the action of an algebra
(say, the double affine Hecke algebra \cite{Che}) on polynomials,
i.e., it does not immediately reduce to a standard representation theory problem.

We study in \secref{qKZ} {\em polynomial}\/ solutions of the qKZ equation.
We discuss various properties they possess, including recurrence relations.
We determine in particular a lower bound on
their degree, and that if there exists a solution which saturates this bound,
then it is unique up to multiplication by a constant;
however, contrary to type $\hat{\mathrm A}$, we cannot show at this stage the existence of
such a solution in types $\hat{\mathrm C}$.

Before turning to the geometry, let us mention some motivation and physical applications. In \secref{loopmodel}, we recall that setting the loop weight to $1$ results in the shift parameter of the qKZ equation being zero, and show that the polynomial solution mentioned above is an eigenvector of the inhomogeneous transfer matrix of the Brauer loop model with the same boundary conditions. In the physical range of parameters where the Boltzmann weights are positive, it is in fact the {\em ground state}\/ of the transfer matrix, making
it particularly interesting to calculate. Equivalently, since the transfer matrix is stochastic, the entries of the ground state can be interpreted
as (unnormalized) probabilities of the connectivity of boundary points on a half-infinite strip. (The normalization
is in fact computed in the present work). This paves the way to the calculation of more physically interesting quantities
such as correlation functions.
The model on a strip is particularly interesting because it should help with the computation of the boundary-to-boundary current,
similarly to the work \cite{dGNP} on the noncrossing loop model.

Next we come to the geometric construction. In \secref{brauer}, we define four (conical, affine) schemes, which we call type $\hat{\mathrm C}$ Brauer loop schemes, corresponding to the four cases $\mathrm i,\mathrm c,\mathrm o,\mathrm m$ mentioned above. We provide different descriptions, either as infinite periodic matrices or as flat limits of certain orbits generalizing nilpotent orbits. We also define the group action that these schemes are naturally equipped with, and in particular the torus action. We then describe their irreducible components, following a similar study in \cite{artic33} in type $\hat{\mathrm{A}}$, in terms of link patterns, giving a first hint of the connection to the loop model, since these link patterns form the natural basis of the space on which the Brauer algebra acts. The construction makes use of several type C analogues of the classification of $B$-orbits of upper triangular matrices which square to zero \cite{Meln}. As a byproduct, we point out the connection to a symplectic analogue of the commuting variety, which was one of the motivations of \cite{artic33}.

Finally, \secref{brauertoloop} provides the exact connection between the Brauer loop model and the Brauer loop scheme in various types. This can be summarized by the following meta-theorem:
\begin{thm*}
The multidegrees of the irreducible components of the Brauer loop scheme form a polynomial solution of the qKZ equation.
\end{thm*}
(Multidegrees are a convenient reformulation of equivariant cohomology in our setting;\footnote{Since our solution of the Yang--Baxter equation is {\em rational}, we obtain on the geometric side ordinary cohomology, as opposed to $K$-theory or elliptic cohomology.}
equivalently, they can be thought of as equivariant volumes up to overall normalization.)
This statement will be made more precise later (see~\thmref{main}). The proof involves the detailed analysis of the action of certain $SL_2$ subgroups on the Brauer loop schemes.

An interesting feature is that in type $\hat{\mathrm C}$, it is not possible to solve the qKZ equation explicitly in order to exhibit a polynomial solution (contrary to the case of type $\hat{\mathrm A}$, see \cite{artic32}, where one can at least in principle compute it inductively). Therefore, the geometry provides an explicit solution of the qKZ problem, which by definition has all the desired properties (polynomiality in all variables, minimal degree).

The analysis of \secref{brauertoloop} is rather technical, and
unfortunately, we have not been able to do without a case-by-case analysis depending on
the boundary condition, a fully general approach
(similar to the analysis of \cite{MO-qg}) being outside the scope of this
paper. We have therefore decided to give the full proofs only in types $\mathrm i,\mathrm c$.
In addition,
we have mentioned in parallel the case of type $\hat{\mathrm A}$ (denoted $\mathrm p$ for periodic),
not only to summarize the main results of \cite{artic33,artic39} and for comparison purposes,
but also because some of the proofs and results we give are new even in type $\hat{\mathrm A}$.

The conclusion wraps up the proof of the main theorem and discuss {\em sum rules}. 
On the geometric side, these correspond to the multidegrees of the full Brauer loop schemes. (In fact, using a combination of flat deformation and
equivariant localization, some formulae, albeit not particularly explicit, are already provided in \secref{brauer}). On the physical side,
they are the normalization constants for the probabilities of the connectivity of boundary points on the half-infinite strip.
Using recurrence relations, we provide alternative formulae for them as Pfaffians or determinants.

In the appendices we give some small size solutions of the qKZ system as well as some technical results that are needed in the proofs.

\subsection{Dynkin diagrams}\label{sec:dynkin}
Since the models and geometry we consider are based on the affine
Dynkin diagrams of type $\hat{\mathrm A}$
and $\hat{\mathrm C}$, we briefly describe our conventions concerning these.

First introduce the following notation: given an integer $L\geq 2$, define the group $\Gamma$ acting on $\Z$ by generators
\begin{itemize}
\item $i\mapsto i+L$ in type $\hat{\mathrm A}$;
\item $i\mapsto i+2L$, $i\mapsto 2L-i+1$ in type $\hat{\mathrm C}$.
\end{itemize}
Equivalence classes in $\Z/\Gamma$ are canonically identified with edges in the Dynkin diagram of $\hat A_{L-1}$, $\hat C_L$, respectively. We denote by $\cl(i)$ the class of $i\in\Z$ in $\Z/\Gamma$. The standard choice of representatives is $1,\ldots,L$ (they also correspond to the choice of variables $z_1,\ldots,z_L$ in the weights, see below), and when there is no risk of confusion we identify such representatives $i$ and $\cl(i)$. For future purposes, also define the action of $\Gamma=\Gamma_e\cup\Gamma_o$ (where $\Gamma_e$ are translations and $\Gamma_o$ are reflections when acting on $\Z$) on $\Z\times\Z$ by $(i,j)\mapsto (\gamma(i),\gamma(j))$ if $\gamma\in \Gamma_e$, $(i,j)\mapsto (\gamma(j),\gamma(i))$ if $\gamma\in \Gamma_o$, and $\cl(i,j)$ to be the class of $(i,j)$ under this action.

\begin{figure}[ht]
\tikzset{dynkin/.style={circle,fill,inner sep=0.5mm}}
\tikzset{arrow/.style={postaction={decorate,thick,decoration={markings,mark = at position #1 with {\arrow{>}}}}},arrow/.default=0.5}
\begin{tikzpicture}(baseline=(current  bounding  box.center),scale=0.75)
\draw (0:1) node[dynkin] {}
\foreach \angle/\lab in {45/2,90,135,180/\vdots,225,270,315/L,360/1} {
-- (\angle:1)
node[dynkin,label={\angle:$\ifx\angle\lab\else\scs\lab\fi$}] {}
};
\begin{scope}[shift={(5,0)}]
\draw[double,arrow] (0,0) node[dynkin,label=below:$\scs 0$] {} -- (1,0) node[dynkin,label=below:$\scs 1$] {};
\draw (1,0) \foreach \x in {2,...,5} { -- (\x,0) node[dynkin] {} };
\draw[double,arrow] (6,0) node[dynkin,label=below:$\scs L$] {} -- (5,0) node[dynkin,,label=below:$\scs L-1$] {};
\end{scope}
\end{tikzpicture}
\caption{Dynkin diagrams for the affine $\hat{\mathrm A}_{L-1}$ (left) and $\hat{\mathrm C}_{L}$ (right) root systems.}
\label{fig:dynkin}
\end{figure}
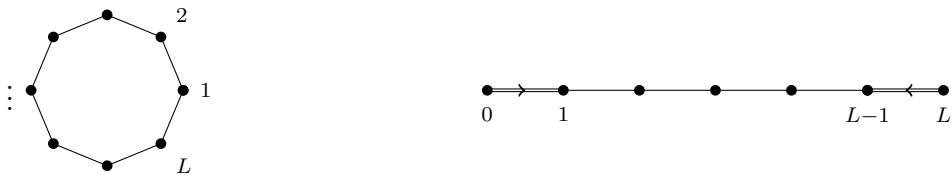

We also choose the somewhat clumsy (but standard) convention to index a node by the edge to its left, except for the leftmost node of type $\hat{\mathrm C}$ which is labelled $0$, see \figref{dynkin}.

Similarly, the root lattices are defined as follows. We start with a countable set of generators of the form $s$ and $z_i$, $i\in\Z$. Then we take the quotient of the abelian group they generate with the relations (for all $i\in\Z$):
\begin{itemize}
\item $z_{i+L}=z_i+s$ in type $\hat{\mathrm A}_{L-1}$;
\item $z_{i+2L}=z_i+s$, $z_{2L-i+1}=-z_i$ in type $\hat{\mathrm C}$.
\end{itemize}
The result is isomorphic to $\Z^{L+1}$, a possible choice of generators being $s,z_1,\ldots,z_L$.

In all types, the simple roots are then defined by $\alpha_i=z_i-z_{i+1}$, $i$ index of the Dynkin diagram. More explicitly, and if we use the variables $s,z_1,\ldots,z_L$, we have:
\begin{itemize}
\item In type $\hat{\mathrm A}_{L-1}$, $\alpha_i=z_i-z_{i+1}$ for $1\le i\le L-1$, and $\alpha_L=z_L-z_1-s$;
\item In type $\hat{\mathrm C}_{L}$, $\alpha_i=z_i-z_{i+1}$ for $1\leq i\leq L-1$,
$\alpha_0=-2z_1-s$ and $\alpha_L=2z_L$.
\end{itemize}

We also need the commutative ring generated by $s$ and $z_i$, $i\in\Z$, i.e., the quotient of $\Z[s,z_i,i\in\Z]$ with the same relations above. Reflections w.r.t.\ simple roots act on it in the following way (with the same choice of variables):
\begin{itemize}
\item In type $\hat{\mathrm A}_{L-1}$,
\[
\tau_i f(\ldots,z_i,z_{i+1},\ldots)=f(\ldots,z_{i+1},z_i,\ldots)\quad 1\le i\le L-1,
\quad
\tau_L f(z_1,\ldots,z_L)=f(z_L-s,\ldots,z_1+s);
\]
\item In type $\hat{\mathrm C}_{L}$
\begin{gather*}
\tau_if(\ldots,z_i,z_{i+1},\ldots)=f(\ldots,z_{i+1},z_i,\ldots)\quad 1\le i\le L-1,\\
 \tau_0 f(z_1,\ldots)=f(-z_1-s,\ldots),\quad \tau_Lf(\ldots,z_L)=f(\ldots,-z_L).
\end{gather*}
\end{itemize}
Finally, we define the divided difference operators, acting on functions of $s,z_1,\dots,z_L$:
\begin{equation}
\label{eq:divdiffops}
\partial_i:=\frac{1}{\alpha_i}(1-\tau_i).
\end{equation}

\section{The quantum Knizhnik--Zamolodchikov equation}
\label{sec:qKZ}
The qKZ equation\footnote{also called ``difference Knizhnik--Zamolochikov equation'' because it is naturally expressed in terms of the {\em additive}\/ spectral parameter. This is due to the fact that our solution of the Yang--Baxter equation is {\em rational}.} that we will be studying is based on the Brauer algebra. We will look at five different boundary conditions, which we will refer to as periodic ($\mathrm{p}$), closed ($\mathrm{c}$), identified ($\mathrm{i}$), open ($\mathrm{o}$), and mixed ($\mathrm{m}$). The first of these corresponds to the type $\hat{\mathrm A}$ root system while the other four correspond to the type $\hat{\mathrm C}$ root system. We state some previous results for periodic \cite{dGN-brauer,artic32} and closed \cite{PDF-open}, but for the other boundary conditions the results are original.

Our aim, as explained in the introduction, is to give a geometric meaning to the qKZ solutions of type $\hat{\mathrm C}$, as was done for the periodic case in \cite{dGN-brauer,artic32,artic33,artic39}, however we will include the periodic case in all our statements in order to make references and comparisons to it.

\subsection{The Brauer algebra}
We list here the aspects of the Brauer algebra \cite{Brauer,Wenzl-Brauer} that are common to all boundary conditions. The Brauer algebra is built from the Temperley--Lieb generators $\{e_i\ |\ i=1,\dots, L-1\}$ and the crossing generators $\{f_i\ |\ i=1,\dots, L-1\}$, graphically depicted as
\[ e_i:=\raisebox{9pt}{
\begin{tikzpicture}[baseline=(current  bounding  box.center),scale=0.75]
\draw[smooth] (1,0) to[out=90,in=180] (1.5,0.5) to[out=0,in=90] (2,0);
\draw[smooth] (1,1.5) to[out=270,in=180] (1.5,1) to[out=0,in=270] (2,1.5);
\draw[dotted] (0,1.5) -- (0.5,1.5);
\draw (0.5,1.5)--(2.5,1.5);
\draw[dotted] (2.5,1.5) -- (3,1.5);
\node[above] at (1,1.5) {${\scriptstyle i}$};
\node[above] at (2,1.5) {${\scriptstyle i+1}$};
\end{tikzpicture}}
\qquad f_i:=\raisebox{9pt}{
\begin{tikzpicture}[baseline=(current  bounding  box.center),scale=0.75]
\draw (1,0) -- (2,1.5);
\draw (1,1.5) -- (2,0);
\draw[dotted] (0,1.5) -- (0.5,1.5);
\draw (0.5,1.5)--(2.5,1.5);
\draw[dotted] (2.5,1.5) -- (3,1.5);
\node[above] at (1,1.5) {${\scriptstyle i}$};
\node[above] at (2,1.5) {${\scriptstyle i+1}$};
\end{tikzpicture}}
,\]
which satisfy the rule-of-thumb ``strings are pulled tight, closed loops give a weight of $\beta$'', explicitly (see e.g.~\cite{Schuler-BWM})
\begin{align}
\label{eq:TLrelns} e_i^2&=\beta\ e_i, & f_i e_i&=e_i,\\
\nn f_i^2&=1, & e_i f_i&=e_i,\\
\nn e_i e_{i\pm 1} e_i&=e_i, & f_i e_{i\pm 1} e_i&=f_{i\pm 1} e_i,\\
\nn f_i f_{i+1} f_i&=f_{i+1} f_i f_{i+1},\qquad\qquad & e_i e_{i\pm 1} f_i&=e_i f_{i\pm 1}.
\end{align}
One can show that all relations that can be derived from the graphical depiction are a consequence of
\eqref{eq:TLrelns}, so that the Brauer is defined by generators $e_i$, $f_i$, $i=1,\ldots,L-1$, and
relations \eqref{eq:TLrelns}.

Using the parametrization $\beta=\frac{A-\eps}{A-\eps/2}$, and the definition 
\[ r(z):=(A+z)(2A-z-\eps), \]
we define for $i=1,\dots,L-1$ the $R$-matrices
\begin{equation}
\label{eq:checkR}
\check R_i(z):=\frac{(2A-\eps)(A-z)}{r(z)}+\frac{(2A-\eps)z}{r(z)}e_i+\frac{(A-z)z}{r(z)}f_i,
\end{equation}
with the graphical depiction
\[
\check R_i(z_i-z_{i+1})=\begin{tikzpicture}[scale=0.75,baseline=(current bounding box.center)]
\draw[thick,->] (0,0) -- (1.5,2);
\draw[thick,->] (1.5,0) -- (0,2);
\node[below] at (0,0) {$z_i$};
\node[below] at (1.5,0) {$z_{i+1}$};
\end{tikzpicture}.
 \]
By the relations \eqref{eq:TLrelns} the $R$-matrices satisfy unitarity
\begin{equation}\label{eq:unit}
 \check R_i(z)\check R_i(-z)=1,
\end{equation}
and the Yang--Baxter equation (YBE)
\begin{equation}\label{eq:ybe}
\check R_i(z)\check R_{i+1}(z+w)\check R_i(w)=\check R_{i+1}(w)\check R_i(z+w)\check R_{i+1}(z).
\end{equation}

The $R$-matrix also has the important property
\begin{equation}
\label{eq:Rpoint}
\check R_i(A)=\frac1{\beta}e_i.
\end{equation}

An important remark is that
equations \eqref{eq:unit} and \eqref{eq:ybe} 
are nothing but the relations of the symmetric group
for the operator $\tau_i \check R_i(z_i-z_{i+1})$. 
In the two paragraphs that follow, we shall extend the Brauer algebra
(i.e., make a choice of boundary conditions for the loop model) in order
to obtain relations of the {\em affine}\/ Weyl groups $\hat{\mathrm A}$
or $\hat{\mathrm C}$.

\subsubsection{Type \texorpdfstring{$\hat{\mathrm A}$}{A}}
The periodic Brauer algebra has two additional generators, $e_L$ and $f_L$, which act between sites $L$ and $1$ and satisfy all the relations \eqref{eq:TLrelns} under the identification $L+1\equiv 1$. The graphical depiction is the natural analogue
of the one for the ordinary Brauer algebra, where diagrams are drawn on a (periodic) strip.

One may wish to add the following relations involving idempotent elements $I_1$ and $I_2$:
\begin{align*}
I_1I_2I_1=\beta^2 I_1&\qquad I_2I_1I_2=\beta^2 I_2,\\
I_1:=e_1e_3\dots e_{L-1},&\qquad I_2:=e_2e_4\dots e_L,\qquad\qquad L\text{ even},\\
I_1:=e_1e_3\dots e_L,&\qquad I_2:=e_2e_4\dots e_{L-1},\qquad\qquad L\text{ odd}.
\end{align*}
(in particular, they will be satisfied in the representation below).

Defining the $R$-matrix using the same formula
\eqref{eq:checkR} for $i=1,\ldots,L$,
equations \eqref{eq:unit} and \eqref{eq:ybe} are satisfied with indices mod $L$, i.e., we obtain the relations of the affine Weyl group of type $\hat {\mathrm A}_{L-1}$.

\begin{figure}[ht]
\begin{tikzpicture}[baseline=(current  bounding  box.center),scale=0.75]
\draw[thick] (0,0) circle(1);
\draw[smooth] (22.5:1) to[out=202.5,in=337.5] (157.5:1);
\draw[smooth] (112.5:1) to[out=292.5,in=22.5] (202.5:1);
\draw[smooth] (67.5:1) to[out=247.5,in=112.5] (292.5:1);
\draw[smooth] (247.5:1) to[out=67.5,in=157.5] (337.5:1);
\node at (202.5:1.3) {${\scriptstyle 1}$};
\node at (247.5:1.3) {${\scriptstyle 2}$};
\node at (292.5:1.3) {${\scriptstyle 3}$};
\node at (337.5:1.3) {${\scriptstyle 4}$};
\node at (22.5:1.3) {${\scriptstyle 5}$};
\node at (67.5:1.3) {${\scriptstyle 6}$};
\node at (112.5:1.3) {${\scriptstyle 7}$};
\node at (157.5:1.3) {${\scriptstyle 8}$};
\end{tikzpicture}
\caption{An example periodic link pattern for $L=8$.}
\label{fig:Alp}
\end{figure}
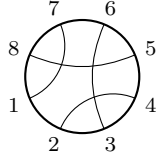

The periodic Brauer algebra has a representation on the vector space with canonical basis
indexed by {\em link patterns}. In the periodic case, the latter are chord diagrams that connect the $L$ points around a circle in pairs, see \figref{Alp}. If $L$ is odd one site is left unpaired, referred to either as a fixed point or as a connection to infinity. We refer to the set of periodic link patterns of size $L$ as $\mathrm{LP}_L^{\mathrm p}$, and it has $(2\ceil{L/2}-1)!!$ elements.

The Brauer generators act on the link patterns in the natural graphical way,
by pasting the strip around the disk; for example,
\[ e_2\quad
\begin{tikzpicture}[baseline=(current  bounding  box.center),scale=0.75]
\draw[thick] (0,0) circle(1);
\draw[smooth] (22.5:1) to[out=202.5,in=337.5] (157.5:1);
\draw[smooth] (112.5:1) to[out=292.5,in=22.5] (202.5:1);
\draw[smooth] (67.5:1) to[out=247.5,in=112.5] (292.5:1);
\draw[smooth] (247.5:1) to[out=67.5,in=157.5] (337.5:1);
\node at (247.5:1.3) {${\scriptstyle 2}$};
\node at (292.5:1.3) {${\scriptstyle 3}$};
\end{tikzpicture}
\;=\;
\begin{tikzpicture}[baseline=(current  bounding  box.center),scale=0.75]
\draw[thick] (0,0) circle(1);
\draw[smooth] (22.5:1) to[out=202.5,in=337.5] (157.5:1);
\draw[smooth] (112.5:1) to[out=292.5,in=22.5] (202.5:1);
\draw[smooth] (67.5:1) to[out=247.5,in=157.5] (337.5:1);
\draw[smooth] (247.5:1) to[out=67.5,in=112.5] (292.5:1);
\node at (247.5:1.3) {${\scriptstyle 2}$};
\node at (292.5:1.3) {${\scriptstyle 3}$};
\end{tikzpicture}
\;,\qquad\qquad f_2\quad
\begin{tikzpicture}[baseline=(current  bounding  box.center),scale=0.75]
\draw[thick] (0,0) circle(1);
\draw[smooth] (22.5:1) to[out=202.5,in=337.5] (157.5:1);
\draw[smooth] (112.5:1) to[out=292.5,in=22.5] (202.5:1);
\draw[smooth] (67.5:1) to[out=247.5,in=112.5] (292.5:1);
\draw[smooth] (247.5:1) to[out=67.5,in=157.5] (337.5:1);
\node at (247.5:1.3) {${\scriptstyle 2}$};
\node at (292.5:1.3) {${\scriptstyle 3}$};
\end{tikzpicture}
\;=\;
\begin{tikzpicture}[baseline=(current  bounding  box.center),scale=0.75]
\draw[thick] (0,0) circle(1);
\draw[smooth] (22.5:1) to[out=202.5,in=337.5] (157.5:1);
\draw[smooth] (112.5:1) to[out=292.5,in=22.5] (202.5:1);
\draw[smooth] (67.5:1) to[out=247.5,in=67.5] (247.5:1);
\draw[smooth] (292.5:1) to[out=112.5,in=157.5] (337.5:1);
\node at (247.5:1.3) {${\scriptstyle 2}$};
\node at (292.5:1.3) {${\scriptstyle 3}$};
\end{tikzpicture}
\;.
\]

\subsubsection{Type \texorpdfstring{$\hat{\mathrm C}$}{C}}
Let us now add two more generators to the ordinary Brauer algebra,
$e_0$ and $e_L$, with relations
\begin{align}
\label{eq:TLbdrelns} e_0^2&=e_0, & e_L^2&=e_L,\\
\nn e_1e_0 e_1&=e_1, & e_{L-1}e_Le_{L-1}&=e_{L-1},\\
\nn e_0 f_1 e_0&=e_0 e_1e_0, & e_L f_{L-1} e_L&=e_Le_{L-1}e_L.
\end{align}
These do not appear in every version of the type $\hat{\mathrm C}$ boundary conditions. In the closed case, we shall use neither, i.e., stick to the ordinary Brauer algebra; in the mixed case,
we shall need the one-boundary Brauer algebra, i.e., add $e_L$;
and in the identified and or open cases, we shall use the two-boundary 
Brauer algebra including both $e_0$ and $e_L$.

For identified and open boundaries we also have the idempotent relations
\begin{align*}
I_1I_2I_1=I_1&\qquad I_2I_1I_2=I_2,\\
I_1:=e_0 e_2\dots e_L,&\qquad I_2:=e_1e_3\dots e_{L-1},\qquad\qquad L\text{ even},\\
I_1:=e_0 e_2\dots e_{L-1},&\qquad I_2:=e_1e_3\dots e_L,\qquad\qquad L\text{ odd}.
\end{align*}

\begin{figure}[ht]
\begin{tikzpicture}[baseline=0,scale=0.75]
\draw (0,1.5)--(2.5,1.5);
\draw[smooth] (0.5,1.5) to[out=270,in=180] (1,1) to[out=0,in=270] (1.5,1.5);
\draw[smooth] (1,1.5) to[out=270,in=180] (1.5,1) to[out=0,in=270] (2,1.5);
\node[above] at (0.5,1.5) {${\scriptstyle 1}$};
\node[above] at (1,1.5) {${\scriptstyle 2}$};
\node[above] at (1.5,1.5) {${\scriptstyle 3}$};
\node[above] at (2,1.5) {${\scriptstyle 4}$};
\end{tikzpicture}\hspace{1cm}
\begin{tikzpicture}[baseline=0,scale=0.75]
\draw (0,1.5)--(2.5,1.5);
\draw[smooth] (0.5,1.5) to[out=270,in=180] (1,1) to[out=0,in=270] (1.5,1.5);
\draw (1,1.5) -- (1,0.5);\node[circle,fill=black,inner sep=0pt,minimum size=0.1cm] at (1,0.5) {};
\draw (2,1.5) -- (2,0.5);\node[circle,fill=black,inner sep=0pt,minimum size=0.1cm] at (2,0.5) {};
\end{tikzpicture}\hspace{1cm}
\begin{tikzpicture}[baseline=0,scale=0.75]
\draw (0,1.5)--(2.5,1.5);
\draw[thick] (2.5,1.5)--(2.5,0);
\draw[thick] (0,1.5)--(0,0);
\draw[smooth] (0.5,1.5) to[out=270,in=180] (1,1) to[out=0,in=270] (1.5,1.5);
\draw[smooth] (1,1.5) to[out=270,in=180] (2.5,0.5);
\draw[smooth] (2,1.5) to[out=270,in=0] (0,0.5);
\end{tikzpicture}\hspace{1cm}
\begin{tikzpicture}[baseline=0,scale=0.75]
\draw (0,1.5)--(2.5,1.5);
\draw[thick] (2.5,1.5)--(2.5,0);
\draw[smooth] (0.5,1.5) to[out=270,in=180] (1,1) to[out=0,in=270] (1.5,1.5);
\draw[smooth] (1,1.5) to[out=270,in=180] (2.5,0.5);
\draw[smooth] (2,1.5) to[out=270,in=180] (2.5,1);
\end{tikzpicture}
\caption{Example closed, identified, open and mixed link patterns for $L=4$.}
\label{fig:Clp}
\end{figure}
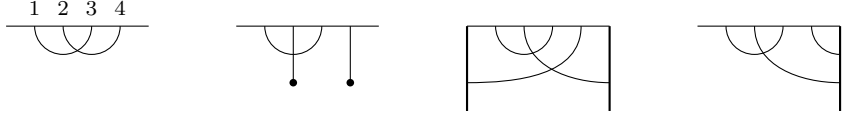

The type $\hat{\mathrm C}$ link patterns are a string of sites numbered from $1$ to $L$, connected to each other in pairs or (if allowed by the boundary conditions) to a boundary, or (in the odd size closed case) left unpaired. See \figref{Clp} for examples.
We can represent the link patterns by tuples of numbers, where the $i$th represents the site that site $i$ is connected to, with $l$ representing the left boundary, $r$ representing the right, $b$ representing the generic boundary, and $\bullet$ representing the unpaired site (or connection to infinity) in the odd closed case. For example, the link pattern in \figref{Alp} is $\{(7,4,6,2,8,3,1,5)\}$, and the link patterns in \figref{Clp} are $\{(3,4,1,2),\ (3,b,1,b),$ $(3,r,1,l),\ (3,r,1,r)\}$. The link pattern sets and respective sizes are
\begin{align*}
\mathrm{LP}_L^{\mathrm i}:&\qquad\sum_{j=0}^{\left\lfloor L/2 \right\rfloor} \binom{L}{2j} (2j-1)!! & \mathrm{LP}_L^{\mathrm c}:&\qquad(2\lceil L/2\rceil-1)!!\\
\mathrm{LP}_L^{\mathrm o}:&\qquad\sum_{j=0}^{\left\lfloor L/2 \right\rfloor} 2^{L-2j}\binom{L}{2j} (2j-1)!! & \mathrm{LP}_L^{\mathrm m}:&\qquad\sum_{j=0}^{\left\lfloor L/2 \right\rfloor} \binom{L}{2j} (2j-1)!!\ .
\end{align*}
We will sometimes refer to the number of chords in a link pattern: this refers to the number of links connecting sites to sites or sites to boundaries. It does not count the unpaired site in the odd closed or periodic cases. For example, the numbers of chords in the link patterns in \figref{Clp} are $2$, $3$, $3$, and $3$ respectively.

The standard graphical depiction for $e_0$ and $e_L$ (used in the open and mixed cases) is
\[ e_0=\raisebox{9pt}{
\begin{tikzpicture}[baseline=(current  bounding  box.center),scale=0.75]
\draw[smooth] (0.5,0) to[out=90,in=0] (0,0.5);
\draw[smooth] (0,1) to[out=0,in=270] (0.5,1.5);
\draw[dotted] (1,1.5) -- (1.5,1.5);
\draw (0,1.5)--(1,1.5);
\draw[thick] (0,1.5)--(0,0);
\node[above] at (0.5,1.5) {${\scriptstyle 1}$};
\end{tikzpicture}}
\;, \qquad e_L=\raisebox{9pt}{
\begin{tikzpicture}[baseline=(current  bounding  box.center),scale=0.75]
\draw[smooth] (1.5,1) to[out=180,in=270] (1,1.5);
\draw[smooth] (1,0) to[out=90,in=180] (1.5,0.5);
\draw[dotted] (0,1.5) -- (0.5,1.5);
\draw (0.5,1.5)--(1.5,1.5);
\draw[thick] (1.5,1.5)--(1.5,0);
\node[above] at (1,1.5) {${\scriptstyle L}$};
\end{tikzpicture}}
\;, \]
however for identified boundaries the graphical depiction is
\[ e_0=\raisebox{9pt}{
\begin{tikzpicture}[baseline=(current  bounding  box.center),scale=0.75]
\draw (0.5,0) -- (0.5,0.5);\node[circle,fill=black,inner sep=0pt,minimum size=0.1cm] at (0.5,0.5) {};
\draw (0.5,1) -- (0.5,1.5);\node[circle,fill=black,inner sep=0pt,minimum size=0.1cm] at (0.5,1) {};
\draw[dotted] (1,1.5) -- (1.5,1.5);
\draw (0,1.5)--(1,1.5);
\draw[thick] (0,1.5)--(0,0);
\node[above] at (0.5,1.5) {${\scriptstyle 1}$};
\end{tikzpicture}}
\;, \qquad e_L=\raisebox{9pt}{
\begin{tikzpicture}[baseline=(current  bounding  box.center),scale=0.75]
\draw (1,1) -- (1,1.5);
\draw (1,0) -- (1,0.5);
\node[circle,fill=black,inner sep=0pt,minimum size=0.1cm] at (1,1) {};
\node[circle,fill=black,inner sep=0pt,minimum size=0.1cm] at (1,0.5) {};
\draw[dotted] (0,1.5) -- (0.5,1.5);
\draw (0.5,1.5)--(1.5,1.5);
\draw[thick] (1.5,1.5)--(1.5,0);
\node[above] at (1,1.5) {${\scriptstyle L}$};
\end{tikzpicture}}
\;, \]
where the dots signify a connection to a single generic boundary. 
In other words, identified and open Brauer algebras are the same, but we
use different representations and therefore corresponding graphical depictions.

The boundary generators act on link pattern in a similar way to the bulk generators. The generator is placed on top of the link pattern, and a new link pattern is formed according to the resulting connections between the sites. For example, in the open case
\[ e_0\quad
\begin{tikzpicture}[baseline=(current bounding box).center,scale=0.75]
\node[above] at (0.5,1.5) {${\scriptstyle 1}$};
\draw (0,1.5)--(2.5,1.5);
\draw[thick] (2.5,1.5)--(2.5,-0.5);
\draw[thick] (0,1.5)--(0,-0.5);
\draw[smooth] (0.5,1.5) to[out=270,in=180] (1,1) to[out=0,in=270] (1.5,1.5);
\draw[smooth] (1,1.5) to[out=270,in=180] (2.5,0.5);
\draw[smooth] (2,1.5) to[out=270,in=0] (0,0);
\end{tikzpicture}
\;=\;
\begin{tikzpicture}[baseline=(current bounding box).center,scale=0.75]
\node[above] at (0.5,1.5) {${\scriptstyle 1}$};
\draw (0,1.5)--(2.5,1.5);
\draw[thick] (2.5,1.5)--(2.5,-0.5);
\draw[thick] (0,1.5)--(0,-0.5);
\draw[smooth] (0.5,1.5) to[out=270,in=0] (0,1);
\draw[smooth] (1.5,1.5) to[out=270,in=0] (0,0.5);
\draw[smooth] (1,1.5) to[out=270,in=180] (2.5,0.5);
\draw[smooth] (2,1.5) to[out=270,in=0] (0,0);
\end{tikzpicture}
\;, \]
and in the identified case
\[e_L\quad
\begin{tikzpicture}[baseline=25,scale=0.75]
\node[above] at (2,1.5) {${\scriptstyle L}$};
\draw (0,1.5)--(2.5,1.5);
\draw[smooth] (0.5,1.5) to[out=270,in=180] (1,1) to[out=0,in=270] (1.5,1.5);
\draw[smooth] (1,1.5) to[out=270,in=180] (1.5,1) to[out=0,in=270] (2,1.5);
\end{tikzpicture}
\;=\;
\begin{tikzpicture}[baseline=25,scale=0.75]
\node[above] at (2,1.5) {${\scriptstyle L}$};
\draw (0,1.5)--(2.5,1.5);
\draw[smooth] (0.5,1.5) to[out=270,in=180] (1,1) to[out=0,in=270] (1.5,1.5);
\draw (1,1.5) -- (1,0.5);\node[circle,fill=black,inner sep=0pt,minimum size=0.1cm] at (1,0.5) {};
\draw (2,1.5) -- (2,0.5);\node[circle,fill=black,inner sep=0pt,minimum size=0.1cm] at (2,0.5) {};
\end{tikzpicture}
\;.
\]

We now define the $K$-matrices, graphically denoted by
\begin{equation*}
\check K_0(w)=:
\begin{tikzpicture}[baseline=(current  bounding  box.center),>=stealth,scale=0.75]
\usetikzlibrary{arrows}
\draw (0.8,1) -- (0,2) -- (0,0) -- cycle;
\draw[thick,<-<] (0.8,1.5) to[out=180,in=90] (0,1) to[out=270,in=180] (0.8,0.5);
\node[right] at (0.8,1.5) {${\scriptstyle w-s/2}$};
\node[right] at (0.8,0.5) {${\scriptstyle -w-s/2}$};
\node[circle,fill=black,draw=black,inner sep=0pt,minimum size=0.15cm] at (0,1) {};
\end{tikzpicture}\;,\qquad
\check K_{L}(w)=:
\begin{tikzpicture}[baseline=(current  bounding  box.center),>=stealth,scale=0.75]
\usetikzlibrary{arrows}
\draw (0,1) -- (0.8,2) -- (0.8,0) -- cycle;
\node[circle,fill=black,draw=black,inner sep=0pt,minimum size=0.15cm] at (0.8,1) {};
\draw[thick,<-<] (0,1.5) to[out=0,in=90] (0.8,1) to[out=270,in=0] (0,0.5);
\node[left] at (0,1.5) {${\scriptstyle -w}$};
\node[left] at (0,0.5) {${\scriptstyle w}$};
\end{tikzpicture}\;.
\end{equation*}
Let
\[ k(w):=(A+2w). \]
For a given boundary condition, if $e_0$ or $e_L$ exists we define the $K$-matrix to be
\begin{equation}
\label{eq:checkK}
\check K_{0,L}(w):=\frac{(A-2w)}{k(w)}+\frac{4w}{k(w)}\ e_{0,L}.
\end{equation}
If the boundary generator doesn't exist, we define the $K$-matrix to be the identity. With these two possible definitions, and by \eqref{eq:TLbdrelns}, the $K$-matrices satisfy unitarity
\[ \check K_0(w)\check K_0(-w)=\check K_L(w)\check K_L(-w)=1, \]
and the boundary Yang--Baxter equation (reflection equation)
\begin{align*}
\check K_0(z)\check R_1(z+w)\check K_0(w)\check R_1(w-z)&=\check R_1(w-z)\check K_0(w)\check R_1(z+w)\check K_0(z),\\
\check K_L(z)\check R_{L-1}(z+w)\check K_L(w)\check R_{L-1}(w-z)&=\check R_{L-1}(w-z)\check K_L(w)\check R_{L-1}(z+w)\check K_L(z).
\end{align*}
Again, these relations are simply those satisfied by boundary generators
of the affine Weyl group $\hat{\mathrm C}_L$.

\subsection{The quantum Knizhnik--Zamolodchikov equation}

\subsubsection{Type \texorpdfstring{$\hat{\mathrm A}$}{A}}
For each $i$ we define the scattering matrix
\begin{align*}
 S_i&:=\check R_{i-1}(z_i-z_{i-1}-s) \dots \check R_1(z_i-z_1-s) \sigma^{-1} \check R_{L-1}(z_i-z_L) \dots \check R_i(z_i-z_{i+1})\\
&=\;
\begin{tikzpicture}[baseline=(current  bounding  box.center),>=stealth,scale=0.75]
\usetikzlibrary{arrows}
\draw (4,0) to[out=90,in=180] (4.5,1) -- (7,1) to[out=0,in=270] (7.5,1.5);
\draw[dashed] (7.5,1.5) to[out=90,in=0] (7,2) -- (1,2) to[out=180,in=270] (0.5,2.5);
\draw[->] (0.5,2.5) to[out=90,in=180] (1,3) -- (3.5,3) to[out=0,in=270] (4,4);
\draw[->] (1,0) -- (1,4);
\draw[->] (3,0) -- (3,4);
\draw[->] (5,0) -- (5,4);
\draw[->] (7,0) -- (7,4);
\node at (2,0.3) {\dots};
\node at (2,3.6) {\dots};
\node at (6,0.3) {\dots};
\node at (6,3.6) {\dots};
\node[below] at (1,0) {$z_1$};
\node[below] at (3,0) {$z_{i-1}$};
\node[below] at (5,0) {$z_{i+1}$};
\node[below] at (7,0) {$z_L$};
\node[below] at (4,0) {$z_i$};
\node[above] at (4,4) {$z_i-s$};
\end{tikzpicture}
\;,
\end{align*}
where $\sigma$ is an operator that rotates a link pattern by one clockwise step,\footnote{Adding $\sigma$, rather than simply the $\tau_i \check R_i$, corresponds to considering the {\em extended}\/ affine Weyl group of type $\hat{\mathrm A}_{L-1}$.} and $s$ is a new (nonzero) parameter. The quantum Knizhnik--Zamolodchikov (qKZ) equation is then \cite{FR-qKZ}
\begin{equation}
\label{eq:qKZp}
S_i\ket{\Psi(\dots,z_i,\dots)}=\ket{\Psi(\dots,z_i-s,\dots)},
\end{equation}
where $\ket\Psi$ is a vector belonging to the space spanned by $\mathrm{LP}_L^{\mathrm p}$,
\[ \ket{\Psi(z_1,\dots,z_L)}=\sum_{\pi\in\mathrm{LP}_L^{\mathrm p}} \psi_{\pi}(z_1,\dots,z_L) \ket\pi. \]
We will sometimes include a subscript to indicate the size $L$ of the system if necessary. Here we are interested in a stronger version, called the qKZ system, which is the following system:
\begin{align}
\label{eq:qKZbulk}\check R_i(z_i-z_{i+1})\ket{\Psi(\dots,z_i,z_{i+1},\dots)}&=\ket{\Psi(\dots,z_{i+1},z_i,\dots)}, \qquad 1\leq i\leq L-1,\\
\label{eq:qKZrotate}\sigma\ket{\Psi(z_1,\dots,z_L)}&=\ket{\Psi(z_2,\dots,z_L,z_1+s)}.
\end{align}
It is easy to show that \eqref{eq:qKZbulk}--\eqref{eq:qKZrotate} implies \eqref{eq:qKZp}, though the converse is in general not true. We refer to \eqref{eq:qKZbulk} as exchange relations, and \eqref{eq:qKZrotate} as the rotation equation.

\subsubsection{Type \texorpdfstring{$\hat{\mathrm C}$}{C}}
Cherednik considered generalizations of the qKZ equation to other types \cite{Che-qKZ}. For type $\hat{\mathrm C}$ there is no longer a need for $\sigma$, but instead we use the boundary operators. The scattering matrix is defined for each $i$ as
\begin{align*}
S_i&:=\check R_{i-1}(z_i-z_{i-1}-s)\dots \check R_{1}(z_i-z_1-s)\check K_0(z_i-s/2)\check R_{1}(z_1+z_i)\dots \check R_{L-1}(z_L+z_i)\\
&\qquad\times\check K_L(z_i)\check R_{L-1}(z_i-z_L) \dots \check R_i(z_i-z_{i+1})\\
&=\;
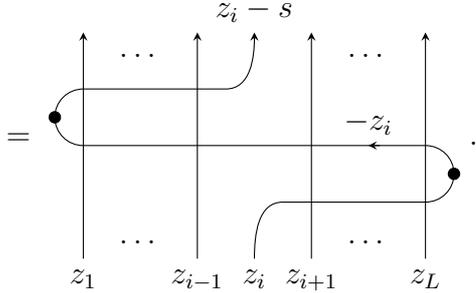
\begin{tikzpicture}[baseline=(current  bounding  box.center),>=stealth,scale=0.75]
\usetikzlibrary{arrows}
\draw[->] (4,0) to[out=90,in=180] (4.5,1) -- (7,1) to[out=0,in=270] (7.5,1.5) to[out=90,in=0] (7,2) -- (6,2);
\draw[->] (6.2,2) -- (1,2) to[out=180,in=270] (0.5,2.5) to[out=90,in=180] (1,3) -- (3.5,3) to[out=0,in=270] (4,4);
\draw[->] (1,0) -- (1,4);
\draw[->] (3,0) -- (3,4);
\draw[->] (5,0) -- (5,4);
\draw[->] (7,0) -- (7,4);
\node at (2,0.3) {\dots};
\node at (2,3.6) {\dots};
\node at (6,0.3) {\dots};
\node at (6,3.6) {\dots};
\node[circle,fill=black,draw=black,inner sep=0pt,minimum size=0.15cm] at (0.5,2.5) {};
\node[circle,fill=black,draw=black,inner sep=0pt,minimum size=0.15cm] at (7.5,1.5) {};
\node[below] at (1,0) {$z_1$};
\node[below] at (3,0) {$z_{i-1}$};
\node[below] at (5,0) {$z_{i+1}$};
\node[below] at (7,0) {$z_L$};
\node[below] at (4,0) {$z_i$};
\node[above] at (6,2) {$-z_i$};
\node[above] at (4,4) {$z_i-s$};
\end{tikzpicture}
\;.
\end{align*}

The qKZ equation is then as before
\begin{equation}
S_i\ket{\Psi(\dots,z_i,\dots)}=\ket{\Psi(\dots,z_i-s,\dots)},
\end{equation}
where
\[ \ket{\Psi(z_1,\dots,z_L)}=\sum_{\pi\in\mathrm{LP}_L^{\mathrm a}} \psi_{\pi}(z_1,\dots,z_L) \ket\pi, \qquad \mathrm a\in\{\mathrm i,\mathrm c,\mathrm o,\mathrm{m}\}, \]
and the qKZ system is
\begin{align}
\label{eq:qKZCR}\check R_i(z_i-z_{i+1})\ket{\Psi(\dots,z_i,z_{i+1},\dots)}&=\ket{\Psi(\dots,z_{i+1},z_i,\dots)}, \qquad 1\leq i\leq L-1,\\
\label{eq:qKZCK0}\check K_0(-z_1-s/2)\ket{\Psi(z_1,\dots,z_L)}&=\ket{\Psi(-z_1-s,\dots,z_L)},\\
\label{eq:qKZCKL}\check K_L(z_L)\ket{\Psi(z_1,\dots,z_L)}&=\ket{\Psi(z_1,\dots,-z_L)},
\end{align}
which implies the qKZ equation. We will refer to \eqref{eq:qKZCK0} and \eqref{eq:qKZCKL} as boundary exchange relations.

\subsection{Solution}
\label{sec:qKZsoln}
In what follows we shall be interested in solutions of the systems \eqref{eq:qKZbulk}--\eqref{eq:qKZrotate} and \eqref{eq:qKZCR}--\eqref{eq:qKZCKL} which are {\em polynomials}\/ in their arguments $z_1,\ldots,z_L$ as well as $A$ and $\eps$. It is easily seen that these polynomials will be homogeneous. The shift parameter is taken to be $s^{\mathrm{p,i,c}}=\eps$ or $s^{\mathrm{o,m}}=\eps/2$, because as we shall show in \secref{brauertoloop}, polynomial solutions exist at these values.
We will also show that the minimal degree solution is unique up to a constant.

The qKZ system gives us a set of relationships between the components of $\ket\Psi$, as well as giving special linear factors and symmetries that must appear in some components based on the associated link pattern. We will explore these in detail in the next section. We will also show that any solution of the qKZ system must satisfy an infinite number of recurrence relations.

\begin{remark}
In the periodic and closed cases all the components can be written as divided difference operators acting on just one component, but in the other cases this is not possible. We will not use this approach, for more details see \cite{PDF-open,artic32}.
\end{remark}

\subsubsection{Factors and symmetries}
\label{sec:qKZfacsyms}
Here we will list all the symmetries and factors dictated by the qKZ system, both for the bulk and the boundaries. Note that for the periodic model only the bulk rules apply, and the rotation equation \eqref{eq:qKZrotate} is an extra restriction.

First we define the modified divided difference operator for $0<i<L$, with $\alpha_i=z_i-z_{i+1}$ as in \secref{dynkin},
\[
\partial'_i := (A+\alpha_i) \partial_i \frac{1}{A+\alpha_i}.
\]
The rules obtained from the qKZ system \eqref{eq:qKZCR}--\eqref{eq:qKZCKL} are listed below for a component $\psi_\pi$ corresponding to a link pattern $\pi$. Again, the rules for the boundaries only apply if there is a non-trivial $K$-matrix, since if the $K$-matrix is trivial the boundary exchange relation merely implies a symmetry.\\
\noindent For $i=0$:
\begin{enumerate}[i.]
\item If $\pi(1)\neq l,b$, then there is no link pattern $\rho$ for which $e_0\ket{\rho}=\ket{\pi}$. So we have
\[ \partial_0\ \frac{\psi_\pi}{k(-z_1-s/2)}=0, \]
implying that $\psi_\pi=k(-z_1-s/2)S^0_\pi$, where $S^0_\pi$ is a polynomial in $z_1,\ldots,z_L$ that is invariant under $(z_1+s/2)\leftrightarrow(-z_1-s/2)$.\\[-3mm]
\item Otherwise there is a small link from site $1$ to the left boundary (or the generic boundary if identified), and the boundary exchange relation gives us the relationship
\begin{equation}
\label{eq:qKZl}
k(-z_1-s/2)(-\partial_0)\psi_\pi=2\sum_{\rho\neq\pi:\ e_0\ket\rho=\ket\pi}\psi_\rho.
\end{equation}
\end{enumerate}
For $i=L$:
\begin{enumerate}[i.]
\item If $\pi(L)\neq r,b$, then there is no $\rho$ for which $e_L\ket{\rho}=\ket{\pi}$. So we have
\[ \partial_L\ \frac{\psi_\pi}{k(z_L)}=0, \]
implying that $\psi_\pi=k(z_L)S^L_\pi$, where $S^L_\pi$ is a polynomial in $z_1,\ldots,z_L$ that is even in $z_L$.\\[-3mm]
\item Otherwise there is a small link from site $L$ to the right boundary (or the generic boundary if identified), and the boundary exchange relation gives us the relationship
\begin{equation}
\label{eq:qKZr}
k(z_L)(-\partial_L)\psi_\pi=2\sum_{\rho\neq\pi:\ e_L\ket\rho=\ket\pi}\psi_\rho.
\end{equation}
\end{enumerate}
For general $i$, $0<i<L$:
\begin{enumerate}[i.]
\item If $\pi(i)\neq i+1$, then there is no $\rho$ for which $e_i\ket{\rho}=\ket{\pi}$. So we have
\begin{align}
\label{eq:qKZf} (2A-z_i+z_{i+1}-\eps)(-\partial'_i)\psi_\pi&=\psi_\pi+\psi_{f_i\pi},\\
\nn (2A-z_i+z_{i+1}-\eps)(-\partial'_i)\psi_{f_i\pi}&=\psi_\pi+\psi_{f_i\pi}.
\end{align}
Specializing the coefficient of $\ket\pi$ in the $i$th exchange relation to $z_{i}=A+z_{i+1}$ gives
\[ \tau_i\psi_\pi\big|_{z_i=A+z_{i+1}}=0, \]
implying that $\psi_\pi$ (and equivalently, $\psi_{f_i\pi}$) contains a factor of $(A+z_i-z_{i+1})$.

Additionally, if $(\pi(i),\pi(i+1))=(l,l)$, $(r,r)$ or $(b,b)$, then $f_i\ket{\pi}=\ket{\pi}$, and \eqref{eq:qKZf} becomes
\[ \partial_i\ \frac{\psi_\pi}{r(z_i-z_{i+1})}=0, \]
implying that $\psi_\pi=r(z_i-z_{i+1})S^i_\pi$, where $S^i_\pi$ is a polynomial in $z_1,\ldots,z_L$ that is symmetric in $z_i$ and $z_{i+1}$.\\[-3mm]
\item Otherwise there is a small link from site $i$ to site $i+1$, and the $i$th exchange relation gives us the relationship
\begin{equation}
\label{eq:qKZi}
r(z_i-z_{i+1})(-\partial_i)\psi_\pi=(2A-\eps)\sum_{\rho\neq\pi:\ e_i\ket\rho=\ket\pi}\psi_\rho.
\end{equation}
\end{enumerate}

\subsubsection{Maximally factorized components}
The rules above give many linear factors for certain components, some of which are a result of the symmetry conditions. In every case except closed, there are some components for which there are a quadratic (in $L$) number of these small factors. We refer to these as the maximally factorized components. In the even periodic case, there is one such component, from which all the others can be determined by means of the qKZ system (see however our remark at the start of \secref{qKZsoln}). In the open, identified and odd periodic cases there is more than one maximally factorized component.

We use $\psi_\Omega$ to refer to the maximally factorized components, and they are labelled by the following link patterns:
\begin{align*}
\Omega^{\mathrm{p}}&:=(L/2+1,\dots,L,1,\dots,L/2),\qquad L\text{ even}\\
\Omega^{\mathrm{p}}_k&:=\sigma^k\ ((L+1)/2,\dots,L-1,1,\dots,(L-1)/2,\bullet),\qquad L\text{ odd}\\
\Omega_1^{\mathrm{i}}&:=(b,\dots,b), \qquad\qquad \Omega_2^{\mathrm{i}}:=(L,b,\dots,b,1),\\
\Omega_k^{\mathrm{o}}&:=(\underbrace{r,\dots,r}_{k},\underbrace{l,\dots,l}_{L-k}),\qquad k=0,\dots,L,\\
\Omega^{\mathrm{m}}&:=(r,\dots,r).
\end{align*}
The explicit formulas for the components are (the periodic case comes from \cite{artic32}; only $\psi^L_{\Omega^{\mathrm p}_0}$ is given for odd size, the others can be obtained by application of \eqref{eq:qKZrotate}):
\begin{align}
\label{eq:psis} &\psi^L_{\Omega^{\mathrm{p}}}=\prod_{\stack{1\leq i<j\leq L}{j-i<L/2}}(A+z_i-z_j)\prod_{\stack{1\leq i<j\leq L}{j-i>L/2}}(A-z_i+z_j-s) S^{\emptyset}_{\Omega^{\mathrm{p}}}(z_1,\dots,z_L),\\[1mm]
\nn &\psi^L_{\Omega_0^{\mathrm{p}}}=\prod_{\stack{1\leq i<j\leq L}{j-i<(L-1)/2 \text{ or } i>(L-1)/2}}(A+z_i-z_j)\prod_{\stack{1\leq i<j\leq L}{j-i>(L-1)/2}}(A-z_i+z_j-s) S^{\emptyset}_{\Omega_0^{\mathrm{p}}}(z_1,\dots,z_L),\\[1mm]
\nn &\psi^L_{\Omega_1^{\mathrm{i}}}=2^L\prod_{1\leq i<j\leq L}r(z_i-z_j)S^{\{1,\dots,L-1\}}_{\Omega_1^{\mathrm{i}}}(z_1,\dots,z_L),\\[1mm]
\nn &\psi^L_{\Omega_2^{\mathrm{i}}}=2^{L-1}\ k(-z_1-s/2)k(z_L)\prod_{2\leq i<j\leq L-1}r(z_i-z_j)\prod_{i=2}^{L-1}\Big( (A+z_1-z_i)(A+z_i-z_L)\\
\nn &\times(A-z_1-z_i-s)(A+z_i+z_L)\Big) S^{\{0,2,\dots,L-2,L\}}_{\Omega_2^{\mathrm{i}}}(z_1,\dots,z_L),\\[1mm]
\nn &\psi^L_{\Omega_k^{\mathrm{o}}}=2^L\prod_{i=1}^k k(-z_i-s/2)\prod_{i=k+1}^L k(z_i) \prod_{1\leq i<j\leq k} r(z_i-z_j)r(-z_i-z_j-s)\\
\nn &\times \prod_{k+1\leq i<j\leq L}r(z_i-z_j)r(z_i+z_j)\prod_{i=1}^k\prod_{j=k+1}^L \Big( (A+z_i-z_j)(A-z_i-z_j-s)\\
\nn &\times (A+z_i+z_j)(A-z_i+z_j-s)\Big) S^{\{0,\dots,k-1,k+1\dots,L\}}_{\Omega_k^{\mathrm{o}}}(z_1,\dots,z_L),\\[1mm]
\nn &\psi^L_{\Omega^{\mathrm{m}}}=2^L\prod_{1\leq i<j\leq L}r(z_i-z_j)r(-z_i-z_j-s)S^{\{0,\dots,L-1\}}_{\Omega^{\mathrm{m}}}(z_1,\dots,z_L),
\end{align}
where the $S$ functions are polynomials whose superscripts denote their symmetries as defined in the previous section.

\subsubsection{Recurrence relations}
Here we will describe recurrence relations that are satisfied by solutions to the qKZ system \eqref{eq:qKZbulk}--\eqref{eq:qKZrotate} or \eqref{eq:qKZCR}--\eqref{eq:qKZCKL}. The first proposition describes a `bulk' recurrence relation, which involves setting one of the variables to be dependent on another, and the second considers a `boundary' recurrence relation, which is only valid in cases with a nontrivial boundary, and involves setting one variable to a constant. Recall that we have set $s^{\mathrm{p,i,c}}=\eps$ and $s^{\mathrm{o,m}}=\eps/2$.

The proofs of the two propositions depend on the following two lemmas.
\begin{lm}
\label{lm:symfac}
If, for a polynomial vector $\kets{\Phi^{(1)}}$, some integer $0<i<L$, and a polynomial $f$ which is coprime to $\tau_i f$ and does not contain the factor $r(z_{i+1}-z_i)$, we have
\[ \check R_i(z_i-z_{i+1})\kets{\Phi^{(1)}} = \frac{f}{(\tau_i f)}\tau_i \kets{\Phi^{(1)}}, \]
then $\kets{\Phi^{(1)}}=f\kets{\Phi^{(2)}}$ where $\kets{\Phi^{(2)}}$ is a polynomial vector that satisfies the $i$th exchange relation.

If in addition $\kets{\Phi^{(1)}}$ satisfies the $k$th exchange relation for some $k\neq i$, then
\[ \check R_k(z_k-z_{k+1})\kets{\Phi^{(2)}} = \frac{(\tau_k f)}{f}\tau_k \kets{\Phi^{(2)}}. \]
Thus if $f$ does not contain the factor $r(z_k-z_{k+1})$, $\kets{\Phi^{(1)}}$ also contains any factor of $\tau_k f$ that is not in $f$.
\end{lm}
\begin{remark}
An equivalent statement can be made for the $K$-matrices.
\end{remark}
\begin{proof}
Straightforward, using polynomiality of the vectors and the fact that the only denominator in the equations which is not explicit is that of the $R$-matrix.
\end{proof}
\begin{lm}
\label{lm:RKids}
We define the following operators on link patterns: Let $\varphi_i$ acting on a link pattern insert a small loop from site $i$ to $i+1$ while increasing the size of the link pattern by $2$, and let $\tilde\varphi_0$ (resp.~$\tilde\varphi_L$) insert a small loop from the first (resp.~last) site to the left (resp.~right) boundary, which increases the size of the link pattern by $1$.

We have the following identities:
\begin{multline}
\label{eq:RRRRR}
\check R_{j-1}(z_j-z_{j+2})\check R_j(A+z_j-z_{j+2})\check R_{j+1}(z_{j-1}-z_{j+2})\check R_j(z_{j-1}-(A+z_j))\\
\times\check R_{j-1}(z_{j-1}-z_j)\varphi_j=\frac{r(A+z_j-z_{j-1})r(z_{j+2}-z_j)}{r(z_{j-1}-z_j)r(A+z_j-z_{j+2})}\ \varphi_j\circ \check R_{j-1}(z_{j-1}-z_{j+2});
\end{multline}
\begin{equation}
\label{eq:YBphi}
\check R_j(z_{j-1}-(A+z_j))\check R_{j-1}(z_{j-1}-z_j)\varphi_j=\frac{r(A+z_j-z_{j-1})}{r(z_{j-1}-z_j)}\varphi_{j-1};
\end{equation}
\begin{multline}
\label{eq:K0varphi}
\check K_0(-A-z_1-s/2)\check R_1(-A-2z_1-s)\check K_0(-z_1-s/2)\ \varphi_1\\
=\frac{k(A+z_1+s/2) k(-z_1-s/2-\eps/2)}{k(-z_1-s/2) k(A+z_1+s/2-\eps/2)}\ \varphi_1;
\end{multline}
\begin{equation}
\label{eq:KLvarphi}
\check K_L(z_{L-1})\check R_{L-1}(A+2z_{L-1})\check K_L(A+z_{L-1})\ \varphi_{L-1}
=\frac{k(-z_{L-1}) k(A+z_{L-1}-\eps/2)}{k(A+z_{L-1}) k(-z_{L-1}-\eps/2)}\ \varphi_{L-1};
\end{equation}
\begin{multline}
\label{eq:lYBphi}
\check R_1(-A/2-z_2-s/2) \check K_0(-z_2-s/2) \check R_1(A/2-z_2-s/2)\ \tilde\varphi_0\\
= \frac{ k(A+z_2+s/2) k(A-z_2-s/2-\eps) }{ k(A-z_2-s/2) k(A+z_2+s/2-\eps) }\ \tilde\varphi_0 \circ \check K_0(-z_2-s/2);
\end{multline}
\begin{multline}
\label{eq:rYBphi}
\check R_{L-1}(-A/2+z_{L-1}) \check K_L(z_{L-1}) \check R_1(A/2+z_{L-1})\ \tilde\varphi_L \\
= \frac{ k(A-z_{L-1}) k(A+z_{L-1}-\eps) }{ k(A+z_{L-1}) k(A-z_{L-1}-\eps) }\ \tilde\varphi_L \circ \check K_L(z_{L-1});
\end{multline}
where \eqref{eq:K0varphi}--\eqref{eq:rYBphi} are only true in the boundary cases where the relevant $K$-matrix is nontrivial.
\end{lm}
\begin{proof}
These are easily proved by the definitions, in the same way as the Yang--Baxter and boundary Yang--Baxter equations.
\end{proof}
We will also repeatedly use the fact $r(A+z)=r(-z-\eps)$.

\begin{prop}
\label{prop:recur}
Given a polynomial solution $\ket{\Psi_L}$ of the qKZ system for size $L$, we can construct a polynomial solution $\ket{\Psi_{L-2}}$ of the qKZ system for size $L-2$ by taking out any two neighbouring sites, by
\begin{multline}
\label{eq:psirecur}\ket{\Psi_L(z_1,\dots,z_j,A+z_j,\dots,z_L)}\\
=p_j(z_j|z_1,\dots,\hat z_j,\hat z_{j+1},\dots,z_L)\ \varphi_j\ket{\Psi_{L-2}(z_j|z_1,\dots,\hat z_j,\hat z_{j+1},\dots,z_L)},
\end{multline}
where the notation $\hat z_i$ means that $z_i$ is missing from the list of arguments. The proportionality factors for different boundary conditions are
\begin{align}
\label{eq:propfacs} &p_j^{\mathrm{p}}(z_j|\dots,\hat z_j,\hat z_{j+1},\dots)=2\ \prod_{i<j}r(z_i-z_j) \prod_{i>j+1}r(A+z_j-z_i),\\[1mm]
\nn &p_j^{\mathrm{i}}(z_j|\dots,\hat z_j,\hat z_{j+1},\dots)=2\ k(-z_j-\eps/2)k(A+z_j)\prod_{i<j} r(z_i-z_j)\prod_{i>j+1} r(A+z_j-z_i)\\
\nn &\times \prod_{i\neq j,j+1}r(A+z_i+z_j),\\[1mm]
\nn &p_j^{\mathrm{c}}(z_j|\dots,\hat z_j,\hat z_{j+1},\dots)=2\ \prod_{i<j}r(z_i-z_j) \prod_{i>j+1}r(A+z_j-z_i)\prod_{i\neq j,j+1}r(A+z_i+z_j),\\[1mm]
\nn &p_j^{\mathrm{o}}(z_j|\dots,\hat z_j,\hat z_{j+1},\dots)=2\ (A-\eps/2)(2A-\eps/2) k(-z_j-\eps/2) k(A+z_j) k(-z_j-\eps/4) \\
\nn &\times k(A+z_j-\eps/4)\prod_{i<j} r(z_i-z_j) \prod_{i>j+1} r(A+z_j-z_i) \\
\nn &\times\prod_{i\neq j,j+1}r(A+z_i+z_j)r(z_i-z_j-\eps/2) r(-z_j-z_i-\eps/2),\\[1mm]
\nn &p_j^{\mathrm{m}}(z_j|\dots,\hat z_j,\hat z_{j+1},\dots)=2\ (A-\eps/2)(2A-\eps/2)k(-z_j-\eps/2) k(A+z_j)\prod_{i<j} r(z_i-z_j) \\
\nn &\times\prod_{i>j+1} r(A+z_j-z_i)\prod_{i\neq j,j+1}r(A+z_i+z_j)r(z_i-z_j-\eps/2) r(-z_j-z_i-\eps/2).
\end{align}
(The constant factors are included for technical reasons that will be explained later.)
\end{prop}

\begin{proof}[Proof of \propref{recur}]
First we note that the $j$th exchange relation implies
\begin{equation}
\label{eq:varphipsi}
 \ket{\Psi_L(z_{j+1}=A+z_j)}=\varphi_j\kets{\Psi^{(1)}_{L-2,j}(z_j|z_1,\dots,z_{j-1}|z_{j+2},\dots,z_L)},
\end{equation}
for some vector in the space of link patterns of size $L-2$. We note that the exchange relations for $i\neq j-1,j,j+1$ are still valid for this new vector $\kets{\Psi^{(1)}_{L-2,j}}$.

We will drop the subscript $L-2$ here. Applying both sides of \eqref{eq:RRRRR} to $\kets{\Psi^{(1)}_j}$, we have via \eqref{eq:varphipsi}
\[ \varphi_j\tilde\tau_{j-1}\kets{\Psi^{(1)}_j}=\frac{r(A+z_j-z_{j-1})r(z_{j+2}-z_j)}{r(z_{j-1}-z_j)r(A+z_j-z_{j+2})} \check R_{j-1}(z_{j-1}-z_{j+2})\varphi_j\kets{\Psi^{(1)}_j}, \]
where $\tilde\tau_{j-1}$ swaps $z_{j-1}$ and $z_{j+2}$. We define $\varphi_j^\dag$ as an upside-down loop between sites $j$ and $j+1$, so that $\varphi_j\varphi_j^\dag=e_j$ and $\varphi_j^\dag\varphi_j=\beta$. We can then multiply by $\varphi_j^\dag$ and use \lmref{symfac} to get
\begin{equation}
\label{eq:psi1}
\kets{\Psi^{(1)}_j}=\prod_{i<j} r(z_i-z_j) \prod_{i>j+1} r(A+z_j-z_i)\ \kets{\Psi^{(2)}_j},
\end{equation}
with $\kets{\Psi^{(2)}_j}$ satisfying the exchange relations
\begin{align*}
\check R_{i}(z_i-z_{i+1})\kets{\Psi^{(2)}_j}&=\tau_i \kets{\Psi^{(2)}_j} \qquad i\neq j-1,j,j+1,\\
\check R_{j-1}(z_{j-1}-z_{j+2})\kets{\Psi^{(2)}_j}&=\tilde\tau_{j-1} \kets{\Psi^{(2)}_j}.
\end{align*}

By applying $\check R_j(z_{j-1}-(z_j+A))\check R_{j-1}(z_{j-1}-z_j)$ to $\ket{\Psi_L(z_{j+1}=A+z_j)}$, and using \eqref{eq:YBphi} and \eqref{eq:varphipsi}, we have
\[ \kets{\Psi^{(2)}_j(z_j|z_1,\dots,z_{j-1}|z_{j+2},\dots,z_L)}=\kets{\Psi^{(2)}_{j-1}(z_j|z_1,\dots,z_{j-2}|z_{j-1},z_{j+2},\dots,z_L)}. \]
In other words, $\kets{\Psi^{(2)}_j}$ and $\kets{\Psi^{(2)}_k}$ are related by an obvious rearrangement of arguments. Thus we can drop the subscript $j$ and simplify the argument notation: $\kets{\Psi^{(2)}(z_j|z_1,\dots,z_{j-1},z_{j+2},\dots,z_L)}$.

Finally we consider the different boundary conditions separately:
\begin{itemize}
\item Periodic:
The rotation equation for $\ket{\Psi_L}$ at $z_{j+1}=A+z_j$ leads to the rotation equation for $\kets{\Psi^{(2)}}$, so $\kets{\Psi^{(2)}}$ is a solution to the qKZ system of size $L-2$. Thus $\ket{\Psi_{L-2}}=\kets{\Psi^{(2)}}$ and the proportionality factor \eqref{eq:psi1} is the same as in \eqref{eq:propfacs}.

\item Closed and Identified:
From \eqref{eq:qKZCK0} we have
\begin{align}
\label{eq:KsPsi2} \check K_0(-z_1-s/2) r(z_1-z_j) &\kets{\Psi^{(2)}(z_j|\dots)}\\
\nn &=r(-z_1-z_j-s)\ \tau_0\kets{\Psi^{(2)}(z_j|\dots)},\qquad j> 1,\\
\nn \check K_L(z_L) r(A+z_j-z_L) &\kets{\Psi^{(2)}(z_j|\dots)}\\
\nn &=r(A+z_j+z_L)\ \tau_L\kets{\Psi^{(2)}(z_j|\dots)},\qquad j< L-1.
\end{align}
This implies via the $K$-matrix version of \lmref{symfac} that
\[ \kets{\Psi^{(2)}(z_j|\dots)}=\prod_{i\neq j,j+1}r(A+z_i+z_j)\kets{\Psi^{(3)}(z_j|\dots)},\qquad \forall j. \]

Applying \eqref{eq:K0varphi} to $\kets{\Psi^{(1)}}$ leads to
\[ \kets{\Psi^{(3)}(-A-z_1-\eps|z_3,\dots)}=\frac{k(A+z_1+\eps/2) k(-z_1-\eps)}{k(-z_1-\eps/2) k(A+z_1)} \kets{\Psi^{(3)}(z_1|z_3,\dots)}, \]
implying
\[ \kets{\Psi^{(3)}(z_j|\dots)}=k(-z_j-\eps/2) k(A+z_j)\kets{\Psi^{(4)}(z_j|\dots)}. \]
A similar argument can be made for $j=L-1$ using \eqref{eq:KLvarphi} but this results in the same factors.

Now $\ket{\Psi_{L-2}}=\kets{\Psi^{(3)}}$ for the closed case and $\ket{\Psi_{L-2}}=\kets{\Psi^{(4)}}$ for the identified case, and the proportionality factors are as given in \eqref{eq:propfacs}.
\item Mixed and Open:
We have again the equations \eqref{eq:KsPsi2}, but this time $s=\eps/2$ so they imply separate factors. Using \lmref{symfac} and its $K$-matrix version, we thus have
\[ \kets{\Psi^{(2)}(z_j|\dots)}=\prod_{i\neq j,j+1}r(A+z_i+z_j)r(-z_i-z_j-\eps/2)r(z_i-z_j-\eps/2)\kets{\Psi^{(3)}(z_j|\dots)},\qquad \forall j. \]

Again we can consider \eqref{eq:K0varphi} and \eqref{eq:KLvarphi}, and this time the two produce different factors, implying that for the mixed case
\[ \kets{\Psi^{(3)}(z_j|\dots)}=k(-z_j-\eps/2) k(A+z_j)\kets{\Psi^{(4)}(z_j|\dots)}, \]
and for the open case,
\[ \kets{\Psi^{(3)}(z_j|\dots)}=k(-z_j-\eps/2) k(A+z_j)k(-z_j-\eps/4) k(A+z_j-\eps/4)\kets{\Psi^{(4)}(z_j|\dots)}. \]
Now $\ket{\Psi_{L-2}}=\kets{\Psi^{(4)}}$ and the proportionality factors are as given in \eqref{eq:propfacs}.\qedhere
\end{itemize}
\end{proof}
Note that the recurrence relation at $j$ implies that if a link pattern does not have a small loop from $j$ to $j+1$, then the corresponding component in the solution of the qKZ system disappears when $z_{j+1}=A+z_j$, which is consistent with the factors found in \secref{qKZfacsyms}. A similar statement will be true of the following proposition.
\begin{prop}
\label{prop:bdrecur}
Given a polynomial solution $\ket{\Psi_L}$ of the type $\hat{\mathrm C}$ qKZ system for size $L$, we can construct a polynomial solution $\ket{\Psi_{L-1}}$ of the qKZ system for size $L-1$ by taking out the first or last site (iff the $K$-matrix at the chosen boundary is nontrivial), by
\begin{align}
\label{eq:lrecur} \ket{\Psi_L((A-s)/2,z_2,\dots,z_L)}=p_0(z_2,\dots,z_L)\ \tilde\varphi_0\ket{\Psi_{L-1}(z_2,\dots,z_L)},\\
\label{eq:rrecur} \ket{\Psi_L(z_1,\dots,z_{L-1},-A/2)}=p_L(z_1,\dots,z_{L-1})\ \tilde\varphi_L\ket{\Psi_{L-1}(z_1,\dots,z_{L-1})}.
\end{align}
The proportionality factors for different boundary conditions are
{\small
\begin{align}
&\label{eq:bdpropfacs} p_0^{\mathrm{i}}(z_2,\dots,z_L)=2\ \prod_{j=2}^L \frac{k(A-z_j-\eps/2)k(A+z_j-\eps/2)}{4},\\
&\nn p_0^{\mathrm{o}}(z_2,\dots,z_L)=2\ (2A-s)\prod_{j=2}^L \frac{k(A-z_j-\eps/4) k(A+z_j-3\eps/4) k(A+z_j-\eps/4) k(A-z_j-3\eps/4)}{16},\\
&\nn p_L^{\mathrm{i}}(z_1,\dots,z_{L-1})=2\ \prod_{j=1}^{L-1} \frac{k(A+z_j) k(A-z_j-\eps)}{4},\\
&\nn p_L^{\mathrm{o}}(z_1,\dots,z_{L-1})=2\ (2A-s)\prod_{j=1}^{L-1} \frac{k(A+z_j) k(A-z_j-\eps)k(A-z_j-\eps/2) k(A+z_j-\eps/2)}{16},\\
&\nn p_L^{\mathrm{m}}(z_1,\dots,z_{L-1})=2\ \prod_{j=1}^{L-1}\frac{k(A+z_j) k(A-z_j-\eps)k(A-z_j-\eps/2) k(A+z_j-\eps/2)}{16}.
\end{align}
}
(The constant factors are again included for technical reasons.)
\end{prop}
\begin{proof}
The proof is very similar to the bulk case, so we will skip some details. Let us first consider the left boundary. The left boundary exchange relation implies
\[ \ket{\Psi_L(z_1=(A-s)/2)}=\tilde \varphi_0\kets{\Psi^{(1)}(\hat z_1)}, \]
for some vector in the space of link patterns of size $L-1$.
Equation \eqref{eq:lYBphi} leads by the $K$-matrix version of \lmref{symfac} to, for the identified case,
\[ \kets{\Psi^{(1)}}=\prod_{j>1}k(A-z_j-\eps/2) k(A+z_j-\eps/2)\kets{\Psi^{(2)}}, \]
and for the open case,
\[ \kets{\Psi^{(1)}}=\prod_{j>1}k(A-z_j-\eps/4) k(A+z_j-3\eps/4) k(A+z_j-\eps/4) k(A-z_j-3\eps/4)\kets{\Psi^{(2)}}. \]

Similarly, \eqref{eq:rYBphi} leads to for the identified case
\[ \kets{\Psi^{(1)}}=\prod_{j<L}k(A+z_j) k(A-z_j-\eps)\kets{\Psi^{(2)}}, \]
and for the open and mixed cases
\[ \kets{\Psi^{(1)}}=\prod_{j<L}k(A+z_j) k(A-z_j-\eps)k(A-z_j-\eps/2) k(A+z_j-\eps/2)\kets{\Psi^{(2)}}. \]
In every case, $\ket{\Psi_{L-1}}=\kets{\Psi^{(2)}}$.
\end{proof}
%
%

\subsubsection{Uniqueness of minimal degree solutions}
We can show that any solution to the qKZ system satisfies an infinite number of recurrence relations. Choosing $z_L$ as the specialization variable, via the qKZ system we can show that $\ket{\Psi_L}$ has two-site recurrences at the following points, where $1\leq j\leq L-1$ and $t$ is a non-negative integer:
\begin{itemize}
\item For type $\hat{\mathrm A}$, $z_L = A+z_j-ts$ and $z_L=-A+z_j+(t+1)s$;
\item For type $\hat{\mathrm C}$, $z_L = \pm A\pm z_j\mp ts$ and $z_L=\pm A\mp z_j\mp (t+1)s$;
\end{itemize}
and the type $\hat{\mathrm C}$ cases have one site recurrences at the points
\begin{itemize}
\item For $\mathrm i,\mathrm o$, $z_L = (\pm A\mp (2t+1)s)/2$;
\item For $\mathrm i,\mathrm o,\mathrm m$, $z_L = -A/2+ts$ and $z_L = A/2-(t+1)s$.
\end{itemize}

Two polynomial solutions of the qKZ system are proportional iff their proportionality factor is a constant, a fact which is a direct consequence of the exchange relations. However it is conceivable that this constant could include one of the extra variables from a larger solution, for example $\ket{\Psi_1^{\mathrm{o}}(z_1)}$ could have an overall factor depending on $z_2$. We will concentrate on solutions whose overall factor is a constant with respect to the variables of a system of any size. We call these minimal solutions.

The infinite number of recurrence relations and the requirement that the solution is polynomial indicates that the minimal solution is unique.

In each boundary case, we can calculate the minimal solution of the qKZ system of smallest meaningful size to provide a grounding for the recurrence relations. In $\mathrm{p}$ and $\mathrm{c}$, because there is only a two-site recurrence, we take both $L=1$ and $L=2$. For $\mathrm{i}$ and $\mathrm{m}$, we only need $L=1$. In all of these cases the vector only has one element, which must therefore both be polynomial and satisfy $\psi(z_1)=\psi(z_1+\eps)$, an impossible requirement unless $\psi$ is in fact a constant. Thus for all these cases the smallest solution is a constant.

In $\mathrm{o}$, the smallest meaningful system is $L=1$, but $\ket{\Psi_1^{\mathrm{o}}(z_1)}$ already has two components. However it is possible to show that the solution is linear, by assuming a polynomial form of arbitrary degree and trying to solve the system. The solution is (up to a constant)
\begin{align*}
\psi^{\mathrm{o}}_\ell(z_1)&=2(A+2z_1),\\
\psi^{\mathrm{o}}_r(z_1)&=2(A-s-2z_1).
\end{align*}

The recurrence relations tell us the total degree (in the $z_i$ as well as in $A$ and $\eps$, making the solutions homogeneous) of a solution of any qKZ system of size $L$ given a solution of size $L-2$ or $L-1$. With the small size solutions given above, we thus know the degree of solutions to systems of all sizes:
\[
\begin{tabular}{|c|c|c|c|c|c|}
\hline
Total degree   & $\mathrm{p}$                                & $\mathrm{i}$ & $\mathrm{c}$                                & $\mathrm{o}$ & $\mathrm{m}$ \\ \hline
$p_j$          & $2(L-2)$                                    & $2(2L-3)$    & $4(L-2)$                                    & $2(4L-5)$    & $4(2L-3)$ \\
$p_{0,L}$      & -                                           & $2(L-1)$     & -                                           & $4L-3$       & $4(L-1)$ \\
$\ket{\Psi_L}$ & $2\floor{\frac{L}{2}}\floor{\frac{L-1}{2}}$ & $L(L-1)$     & $4\floor{\frac{L}{2}}\floor{\frac{L-1}{2}}$ & $L(2L-1)$    & $2L(L-1)$ \\ \hline
\end{tabular}
\]
Note that these degrees are exhausted by the factors listed in \eqref{eq:psis}, thus all the unknown symmetric functions in those expressions must be constant.

%

\section{The Brauer loop model}
\label{sec:loopmodel}

For this section we will set $\beta=1$, which implies that $\eps=s=0$. The Brauer loop model is a statistical mechanical model of crossing loops, based on the Brauer algebra. At this special point, the transfer matrix becomes stochastic and the ground state eigenvalue is $1$. In this section we will show that the ground state eigenvector is a solution to the qKZ system at $\eps=s=0$, and that the would-be unique minimal qKZ solution for general $\eps$ becomes the ground state eigenvector when $\eps\to 0$.

\subsection{Definition and transfer matrix}
The Brauer loop model is defined on a vertically semi-infinite square lattice, on a cylinder in type $\hat{\mathrm A}$ or on a strip in type $\hat{\mathrm C}$. Loops are drawn on the faces in three possible ways, and the model is integrable via the Yang--Baxter equation when the probabilities of these configurations are given by the (unchecked) $R$-matrix
\begin{align*}
R(w-z)&:= \frac{2A(A-w+z)}{r(w-z)}\;
\begin{tikzpicture}[baseline=(current  bounding  box.center),scale=0.75]
\draw (0,0) -- (0,1) -- (1,1) -- (1,0) -- cycle;
\draw[thick,smooth] (0.5,0) to[out=90,in=180] (1,0.5);
\draw[thick,smooth] (0,0.5) to[out=0,in=270] (0.5,1);
\end{tikzpicture}
\;+\frac{2A(w-z)}{r(w-z)}\;
\begin{tikzpicture}[baseline=(current  bounding  box.center),scale=0.75]
\draw (0,0) -- (0,1) -- (1,1) -- (1,0) -- cycle;
\draw[thick,smooth] (0.5,0) to[out=90,in=0] (0,0.5);
\draw[thick,smooth] (1,0.5) to[out=180,in=270] (0.5,1);
\end{tikzpicture}
\;+\frac{(A-w+z)(w-z)}{r(w-z)}\;
\begin{tikzpicture}[baseline=(current  bounding  box.center),scale=0.75]
\draw (0,0) -- (0,1) -- (1,1) -- (1,0) -- cycle;
\draw[thick] (0.5,0) -- (0.5,1);
\draw[thick] (0,0.5) -- (1,0.5);
\end{tikzpicture}
\\
&=:\;
\begin{tikzpicture}[baseline=(current  bounding  box.center),scale=0.75,>=stealth]
\usetikzlibrary{arrows}
\draw (0,0) -- (0,1) -- (1,1) -- (1,0) -- cycle;
\draw[thick,->] (-0.5,0.5) -- (1.5,0.5);
\draw[thick,->] (0.5,-0.5) -- (0.5,1.5);
\node[left] at (-0.5,0.5) {$w$};
\node[below] at (0.5,-0.5) {$z$};
\end{tikzpicture}
\;,
\end{align*}
with $r(z)=(A+z)(2A-z)$.

We describe the configuration probabilities of the smallest repeating element of the lattice by the transfer matrix $T(w|z_1,\dots,z_L)$, which acts on the vector space spanned by link patterns $\mathrm{LP}_N$. We will define the transfer matrix explicitly in the following sections, separately for types $\hat{\mathrm A}$ and $\hat{\mathrm C}$. The transfer matrix is stochastic, so the ground state eigenvalue is 1; one can check that in the physical range of parameters where Boltzmann weights are positive, the Perron--Frobenius theorem applies so that the associated eigenvector is unique up to normalization (so that this remains true for generic values of the parameters). We can also use the Yang--Baxter equation (and if necessary the boundary YBE) to show that two transfer matrices with different values of $w$ commute, so the eigenvector does not depend on $w$. Additionally, the entries of the transfer matrix are homogeneous rational functions of $w, z_i$ and $A$, thus the ground state can be normalized so that its entries are homogeneous polynomials in the $z_i$ and $A$ without a common factor. The ground state eigenvector equation is therefore
\[ T(w|z_1,\dots,z_L)\ket{\Psi(z_1,\dots,z_L)}=\ket{\Psi(z_1,\dots,z_L)}, \]
and $\ket{\Psi}$ is written in the basis of link patterns as
\[ \ket{\Psi(z_1,\dots,z_L)}=\sum_{\pi\in\mathrm{LP}_N}\psi_\pi(z_1,\dots,z_L)\ket\pi, \]
where $\psi_\pi(z_1,\dots,z_L)$ are coprime polynomials.

From the YBE, the transfer matrix satisfies the interlacing relation
\begin{equation}
\label{eq:bulkinterlace}
\check R_i(z_i-z_{i+1})T(w|\dots,z_i,z_{i+1},\dots)=T(w|\dots,z_{i+1},z_i,\dots)\check R_i(z_i-z_{i+1}).
\end{equation}
It also has the recurrence relation
\begin{equation}
\label{eq:transrecur}
T_L(w|z_1,\dots,z_i,A+z_i,\dots,z_L)\varphi_i=\varphi_i\circ T_{L-2}(w|z_1,\dots,\hat z_i,\hat z_{i+1},\dots,z_L),
\end{equation}
where $\varphi_i$ is as defined in \propref{recur}.

Each of the above statements is true for every boundary condition, but the proofs differ slightly. For further details see for example \cite{PDF-open,artic32}.

\subsubsection{Type \texorpdfstring{$\hat{\mathrm A}$}{A}}
The periodic Brauer loop model is drawn on a semi-infinite cylinder, of which the smallest repeating element is one row. The transfer matrix is therefore defined as
\[ T(w|z_1,\dots,z_L):=\mathrm{tr}_w\left(R(w-z_L)\dots R(w-z_1)\right), \]
which can be depicted graphically as
\def\a{3 }
\def\b{1 }
\def\th{50 }
\[T(w|z_1,\dots,z_L)=
\begin{tikzpicture}[baseline=(current  bounding  box.center),>=stealth,scale=0.75]
\usetikzlibrary{arrows}
\draw (-\a,0)--(-\a,-1);
\draw (\a,0)--(\a,-1);
\draw (-\a,0)--(-\a,-1);
\draw ({\a*cos(270)},{\b*sin(270)})--({\a*cos(270)},{\b*sin(270)-1});
\draw ({\a*cos(270+20)},{\b*sin(270+20)})--({\a*cos(270+20)},{\b*sin(270+20)-1});
\draw ({\a*cos(270-20)},{\b*sin(270-20)})--({\a*cos(270-20)},{\b*sin(270-20)-1});
\draw[thick,->] ({\a*cos(270-10)},{\b*sin(270-10)-1.5})--({\a*cos(270-10)},{\b*sin(270-10)+0.5});
\node[below] at ({\a*cos(270-10)},{\b*sin(270-10)-1.5}) {${\scriptstyle z_L}$};
\draw[thick,->] ({\a*cos(270+10)},{\b*sin(270+10)-1.5})--({\a*cos(270+10)},{\b*sin(270+10)+0.5});
\node[below] at ({\a*cos(270+10)},{\b*sin(270+10)-1.5}) {${\scriptstyle z_1}$};
\begin{scope}
\draw[dashed] ({-\th}:\a and \b) arc ({-\th}:{180+\th}:\a and \b);
\draw ({180+\th}:\a and \b) arc ({180+\th}:{360-\th}:\a and \b);
\end{scope}
\begin{scope}[shift={(0,-0.5)}]
\draw[thick] ({15}:\a and \b) arc ({15}:{180-15}:\a and \b);
\draw[thick] ({-\th-10}:\a and \b) arc ({-\th-10}:{0}:\a and \b);
\node[above] at ({-\th}:\a and \b) {${\scriptstyle w}$};
\draw[thick,->] ({180}:\a and \b) arc ({180}:{360-\th}:\a and \b);
\end{scope}
\begin{scope}[shift={(0,-1)}]
\draw[dashed] ({180-30}:\a and \b) arc ({180-30}:{30}:\a and \b);
\draw[dashed] ({360}:\a and \b) arc ({360}:{360-\th}:\a and \b);
\draw[dashed] ({180}:\a and \b) arc ({180}:{180+\th}:\a and \b);
\draw ({180+\th}:\a and \b) arc ({180+\th}:{360-\th}:\a and \b);
\end{scope}
\end{tikzpicture}\;.
\]
In addition to the properties already listed, this transfer matrix satisfies the rotation property
\begin{equation}
\label{eq:Tsigma} T(w|z_2,\dots,z_L,z_1)\ \sigma =\sigma\ T(w|z_1,\dots,z_L).
\end{equation}

\subsubsection{Type \texorpdfstring{$\hat{\mathrm C}$}{C}}
\def\Kla{
\begin{tikzpicture}[baseline=(current  bounding  box.center),scale=0.75]
\draw (0.8,1) -- (0,2) -- (0,0) -- cycle;
\draw[thick,smooth] (0.5,0.5) to[out=140,in=220] (0.5,1.5);
\end{tikzpicture}
}
\def\Klb{
\begin{tikzpicture}[baseline=(current  bounding  box.center),scale=0.75]
\draw (0.8,1) -- (0,2) -- (0,0) -- cycle;
\draw[thick] (0.5,0.5) -- (0,0.5);
\draw[thick] (0.5,1.5) -- (0,1.5);
\end{tikzpicture}
}
\def\Kidlb{
\begin{tikzpicture}[baseline=(current  bounding  box.center),scale=0.75]
\draw (0.8,1) -- (0,2) -- (0,0) -- cycle;
\draw[thick,smooth] (0.8,0.5) to[out=180,in=290] (0.3,0.8);\node[circle,fill=black,inner sep=0pt,minimum size=0.1cm] at (0.3,0.8) {};
\draw[thick,smooth] (0.3,1.2) to[out=70,in=180] (0.8,1.5);\node[circle,fill=black,inner sep=0pt,minimum size=0.1cm] at (0.3,1.2) {};
\end{tikzpicture}
}
\def\Kra{
\begin{tikzpicture}[baseline=(current  bounding  box.center),scale=0.75]
\draw (0,1) -- (0.8,2) -- (0.8,0) -- cycle;
\draw[thick,smooth] (0.3,0.5) to[out=40,in=320] (0.3,1.5);
\end{tikzpicture}
}
\def\Krb{
\begin{tikzpicture}[baseline=(current  bounding  box.center),scale=0.75]
\draw (0,1) -- (0.8,2) -- (0.8,0) -- cycle;
\draw[thick] (0.3,0.5) -- (0.8,0.5);
\draw[thick] (0.3,1.5) -- (0.8,1.5);
\end{tikzpicture}
}
\def\Kidrb{
\begin{tikzpicture}[baseline=(current  bounding  box.center),scale=0.75]
\draw (0,1) -- (0.8,2) -- (0.8,0) -- cycle;
\draw[thick,smooth] (0,0.5) to[out=0,in=250] (0.5,0.8);\node[circle,fill=black,inner sep=0pt,minimum size=0.1cm] at (0.5,0.8) {};
\draw[thick,smooth] (0.5,1.2) to[out=110,in=0] (0,1.5);\node[circle,fill=black,inner sep=0pt,minimum size=0.1cm] at (0.5,1.2) {};
\end{tikzpicture}
}
The Brauer loop model for type $\hat{\mathrm C}$ is defined on a semi-infinite strip. At the boundaries the loop configurations are described by the (unchecked) $K$-matrices, which are depicted as
\begin{equation}
\label{eq:K}
K_0(w)=:
\begin{tikzpicture}[baseline=(current  bounding  box.center),>=stealth,scale=0.75]
\usetikzlibrary{arrows}
\draw (0.8,1) -- (0,2) -- (0,0) -- cycle;
\draw[thick,>->] (0.8,1.5) to[out=180,in=90] (0,1) to[out=270,in=180] (0.8,0.5);
\node[right] at (0.8,1.5) {${\scriptstyle -w}$};
\node[right] at (0.8,0.5) {${\scriptstyle w}$};
\node[circle,fill=white,draw=black,inner sep=0pt,minimum size=0.15cm] at (0,1) {};
\end{tikzpicture}
\;,\qquad K_L(w)=:
\begin{tikzpicture}[baseline=(current  bounding  box.center),>=stealth,scale=0.75]
\usetikzlibrary{arrows}
\draw (0,1) -- (0.8,2) -- (0.8,0) -- cycle;
\node[circle,fill=black,draw=black,inner sep=0pt,minimum size=0.15cm] at (0.8,1) {};
\draw[thick,<-<] (0,1.5) to[out=0,in=90] (0.8,1) to[out=270,in=0] (0,0.5);
\node[left] at (0,1.5) {${\scriptstyle -w}$};
\node[left] at (0,0.5) {${\scriptstyle w}$};
\end{tikzpicture}
\;,
\end{equation}
and defined as follows for the different boundary conditions, with $k(w)=A+2w$:

Identified:
\begin{align*}
K_0^{\mathrm i}(w)&:=\frac{-A+2w}{k(A-w)}\;\Kla\;+\frac{4(A-w)}{k(A-w)}\;\Kidlb\;,\\[2mm]
K_L^{\mathrm i}(w)&:=\frac{A-2w}{k(w)}\;\Kra\;+\frac{4w}{k(w)}\;\Kidrb\;;
\end{align*}

Closed:
\[ K_0^{\mathrm c}(w)=K_L^{\mathrm c}(w):=1; \]

Open:
\begin{align*}
K_0^{\mathrm o}(w)&:=\frac{-A+2w}{k(A-w)}\;\Kla\;+\frac{4(A-w)}{k(A-w)}\;\Klb\;,\\[2mm]
K_L^{\mathrm o}(w)&:=\frac{A-2w}{k(w)}\;\Kra\;+\frac{4w}{k(w)}\;\Krb\;;
\end{align*}

Mixed:
\begin{align*}
K_0^{\mathrm m}(w)&:=1,\\
K_L^{\mathrm m}(w)&:=\frac{A-2w}{k(w)}\;\Kra\;+\frac{4w}{k(w)}\;\Krb\;.
\end{align*}

The transfer matrix for these models describes two rows of the lattice (the smallest repeating element),
\[ T(w|z_1,\dots,z_L):=\mathrm{tr}_w\left(K_0(w)R(z_1+w)\cdots R(z_L+w)K_L(w)R(w-z_L)\dots R(w-z_1)\right), \]
graphically depicted as
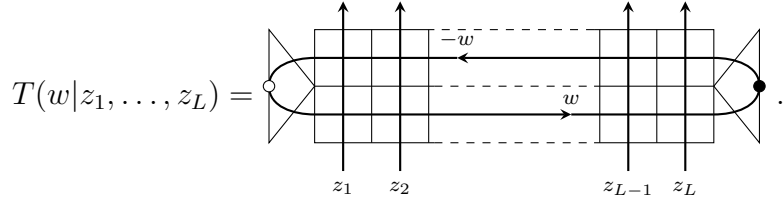
\[T(w|z_1,\dots,z_L)=
\begin{tikzpicture}[baseline=(current  bounding  box.center),>=stealth,scale=0.75]
\usetikzlibrary{arrows}
\draw (3.1,2) -- (1,2) -- (1,0) -- (3.1,0);
\draw[dashed](3.1,0) -- (5.9,0);
\draw (5.9,0) -- (8,0) -- (8,2) -- (5.9,2);
\draw[dashed] (5.9,2) -- (3.1,2);
\draw (1,1) -- (0.2,2) -- (0.2,0) -- cycle;
\draw (8,1) -- (8.8,2) -- (8.8,0) -- cycle;
\draw (1,1) -- (3.1,1);
\draw[dashed] (3.1,1) -- (5.9,1);
\draw (5.9,1) -- (8,1);
\draw (2,0) -- (2,2);
\draw (3,0) -- (3,2);
\draw (6,0) -- (6,2);
\draw (7,0) -- (7,2);
\draw[thick,->] (1.5,-0.5) -- (1.5,2.5); \node[below] at (1.5,-0.5) {${\scriptstyle z_1}$};
\draw[thick,->] (2.5,-0.5) -- (2.5,2.5); \node[below] at (2.5,-0.5) {${\scriptstyle z_2}$};
\draw[thick,->] (6.5,-0.5) -- (6.5,2.5); \node[below] at (6.5,-0.5) {${\scriptstyle z_{L-1}}$};
\draw[thick,->] (7.5,-0.5) -- (7.5,2.5); \node[below] at (7.5,-0.5) {${\scriptstyle z_L}$};
\draw[thick,->] (8,1.5) -- (3.5,1.5);
\node[above] at (3.5,1.5) {${\scriptstyle -w}$};
\draw[thick] (3.5,1.5) -- (1,1.5);
\draw[thick] (1,1.5) to[out=180,in=90] (0.2,1) to[out=270,in=180] (1,0.5);
\draw[thick,->] (1,0.5) -- (5.5,0.5);
\node[above] at (5.5,0.5) {${\scriptstyle w}$};
\draw[thick] (5.5,0.5) -- (8,0.5);
\draw[thick] (8,0.5) to[out=0,in=270] (8.8,1) to[out=90,in=0] (8,1.5);
\node[circle,fill=white,draw=black,inner sep=0pt,minimum size=0.15cm] at (0.2,1) {};
\node[circle,fill=black,draw=black,inner sep=0pt,minimum size=0.15cm] at (8.8,1) {};
\end{tikzpicture}\ .
\]
In addition to the interlacing condition \eqref{eq:bulkinterlace}, the boundary YBE implies the boundary interlacing conditions
\begin{align}
\label{eq:bdinterlace} \check K_0(-z_1)T(w|z_1,\dots)&=T(w|{-z_1},\dots)\check K_0(-z_1),\\
\nn \check K_L(z_L)T(w|\dots,z_L)&=T(w|\dots,-z_L)\check K_L(z_L).
\end{align}
In the cases where $K_{0,L}=1$, these turn into symmetries. There are also boundary recurrence relations in addition to the bulk recurrence relation \eqref{eq:transrecur}, which are only valid if the associated $K$-matrix is non-trivial:
\begin{align*}
T_L(w|A/2,z_2,\dots,z_L)\tilde\varphi_0&=\tilde\varphi_0\circ T_{L-1}(w|z_2,\dots,z_L),\\
T_L(w|z_1,\dots,z_{L-1},-A/2)\tilde\varphi_L&=\tilde\varphi_L\circ T_{L-1}(w|z_1,\dots,z_{L-1}),
\end{align*}
where $\tilde\varphi_0$ and $\tilde\varphi_L$ are defined as in \propref{bdrecur}. 

\subsection{Relationship between the loop model and the qKZ system}
For the purposes of this section, we will use $\ket{\Phi}$ to denote a solution of the qKZ system with $s=0$, and $\ket{\Psi}$ to denote the ground state of the Brauer loop model with loop weight $\beta=1$.

Some of the statements made about $\ket{\Phi}$ in \secref{qKZ} are not valid at $s=0$. In particular, the number of recurrence relations satisfied by the solution is no longer infinite -- there are $2(L-1)$ recurrence relations for the periodic case, $4(L-1)+2$ for identified, open and mixed, and $4(L-1)$ for closed. Thus uniqueness of the solution is not guaranteed; indeed, any solution could be multiplied by a polynomial that has the symmetry of the appropriate Weyl group to make a new solution. Finally, we note that the prefactor in \eqref{eq:RRRRR} is equal to $1$, so the calculation of the proportionality factor is not valid.

However it is still true that any solution of the qKZ system at $s=0$ has a recurrence to a smaller size solution. Thus there is a family of solutions to the qKZ system of different sizes, each of which can be obtained from a larger solution by recurrence. In this section we will show that any member of this family is a ground state of the Brauer loop model transfer matrix, and vice-versa, and that there is a unique recursive family of solutions with coprime entries.

Further, any recursive family of polynomial solutions to the qKZ system for general $s$ of the kind considered in \secref{qKZ}, when taken at $s=0$, will give the family of ground state eigenvectors of the Brauer transfer matrix, up to a symmetric factor.

\begin{prop}
\label{prop:evecqkz}
The ground state of the Brauer transfer matrix with coprime entries is a solution to the qKZ system at $s=0$.
\end{prop}
\begin{proof}[Proof for type $\hat{\mathrm A}$]
This proposition is also stated in \cite{artic32}. We apply the interlacing condition \eqref{eq:bulkinterlace} at any $i$ to the eigenvector $\ket\Psi$,
\[ \check R_i(z_i-z_{i+1})\ket{\Psi}=T(w|\dots,z_{i+1},z_i,\dots)\check R_i(z_i-z_{i+1})\ket{\Psi}. \]
Since the eigenvector is unique, we can deduce that
\[ \check R_i(z_i-z_{i+1})\ket{\Psi(\dots,z_i,z_{i+1},\dots)}=b_i(z_1,\dots,z_L)\ket{\Psi(\dots,z_{i+1},z_i,\dots)}. \]
It is not hard to show that $b$ is either $1$ or $\frac{r(z_{i+1}-z_i)}{r(z_i-z_{i+1})}$, by the fact that the elements of $\ket\Psi$ are coprime polynomials. Similarly, using the rotation property \eqref{eq:Tsigma} of the transfer matrix we find
\[ \sigma \ket{\Psi(z_1,\dots,z_L)}\propto\ket{\Psi(z_2,\dots,z_L,z_1)}, \]
and the proportionality factor must be $1$ by positivity of the ground state when the arguments are all set to $0$. Thus $\ket\Psi$ satisfies \eqref{eq:qKZrotate}.

Acting on $\ket\Psi$ with the scattering matrix $S_1$, we thus find
\[ S_1(z_1,\dots,z_L)\ket{\Psi(z_1,\dots,z_L)}=\prod_{j=1}^{L-1} b_j(z_1,\dots,z_L)\ket{\Psi(z_1,\dots,z_L)}. \]
But it is not hard to show that
\begin{equation}
\label{eq:SisT}
S_i(z_1,\dots,z_L)=T(w=z_i|z_1,\dots,z_L),
\end{equation}
which means that the product of $b_j$s should be equal to one, indicating that $b_j=1$ for all $j$ and showing that $\ket\Psi$ satisfies \eqref{eq:qKZbulk}.
\end{proof}
\begin{proof}[Proof for type $\hat{\mathrm C}$]
To show \eqref{eq:SisT} for the identified and open cases we must note that while $S_i$ uses the algebraic $K$-matrices defined in \eqref{eq:checkK}, $T$ uses the graphical versions defined in \eqref{eq:K}. At the right boundary, we have $\check K_L(w)=K_L(w)$, but at the left boundary we need the relation
\begin{equation}
\label{eq:KKcheck}
\check K_0(w)\;=\;
\begin{tikzpicture}[baseline=(current  bounding  box.center),>=stealth,scale=0.75]
\usetikzlibrary{arrows}
\draw (0,2) -- (0,0) -- (1,1) -- cycle;
\draw[thick,<-<] (1,1.5) to[out=180,in=90] (0,1) to[out=270,in=180] (1,0.5);
\node[right] at (1,1.5) {${\scriptstyle w}$};
\node[right] at (1,0.5) {${\scriptstyle -w}$};
\node[circle,fill=black,draw=black,inner sep=0pt,minimum size=0.15cm] at (0,1) {};
\end{tikzpicture}
\;=\;
\begin{tikzpicture}[baseline=(current  bounding  box.center),>=stealth,scale=0.75]
\usetikzlibrary{arrows}
\draw (0,2) -- (0,0) -- (2,2) -- (3,1) -- (2,0) -- cycle;
\draw[thick,>->,smooth] (2.8,0.2) to [out=135,in=-45] (1.5,1.5) to[out=135,in=90] (0,1) to[out=270,in=225] (1.5,0.5) to[out=45,in=225] (2.8,1.8);
\node[right] at (2.8,1.8) {${\scriptstyle w}$};
\node[right] at (2.8,0.2) {${\scriptstyle -w}$};
\node[circle,fill=white,draw=black,inner sep=0pt,minimum size=0.15cm] at (0,1) {};
\end{tikzpicture}\;.
\end{equation}

By applying the boundary interlacing conditions \eqref{eq:bdinterlace} to the eigenvector one can show that it satisfies the boundary exchange relations up to a proportionality factor, in the same way as for the bulk. Again we can act the scattering matrix $S_1$ on $\ket\Psi$ and show that the proportionality factors must all be $1$. Thus $\ket\Psi$ satisfies \eqref{eq:qKZCR}--\eqref{eq:qKZCKL}. When the $K$-matrix is the identity the proof is trivial.
\end{proof}
When $s=0$ the qKZ equation becomes an invariance equation,
\begin{equation}
\label{eq:qKZs0}
S_i^{(L)}(z_1,\dots,z_L)\ket{\Phi_L(z_1,\dots,z_L)}=\ket{\Phi_L(z_1,\dots,z_L)} \qquad \forall i.
\end{equation}
\begin{prop}
\label{prop:qkzevec}
Let $\ket{\Phi_L}$ and $\ket{\Phi_{L-2}}$ be solutions of \eqref{eq:qKZs0} for sizes $L$ and $L-2$ respectively, such that
\[ \ket{\Phi_L(\dots,z_i,A+z_i,\dots)}=p_i\ \varphi_i\ket{\Phi_{L-2}(\dots,\hat z_i,\hat z_{i+1},\dots)}, \qquad \forall i, \]
and such that $\ket{\Phi_{L-2}}$ has no overall symmetric factor. Then $\ket{\Phi_{L-2}}$ is a ground state eigenvector of the Brauer transfer matrix of size $L-2$.
\end{prop}
\begin{proof}
We consider the action of $\varphi_{i-1}^\dag$ on the scattering matrix when $z_{i-1}=-A+z_i$. For type $\hat{\mathrm A}$, by the property of the $R$-matrix \eqref{eq:Rpoint}, we have
\[ \varphi_{i-1}^\dag S_i(\dots,-A+z_i,z_i,\dots)\varphi_{i-1}=T(z_i|\dots,\hat z_{i-1},\hat z_i,\dots). \]
For type $\hat{\mathrm C}$, this statement is also true, but to prove it we must again use \eqref{eq:KKcheck}. This is only necessary of course in the identified and open cases, where the left $K$-matrix is non-trivial.

Acting $\varphi_{i-1}^\dag S_i(\dots,-A+z_i,z_i,\dots)$ on $\ket{\Phi(\dots,-A+z_i,z_i,\dots)}$ gives
\[ T(z_i|\dots,\hat z_{i-1},\hat z_i,\dots)\ket{\Phi_{L-2}(\dots,\hat z_{i-1},\hat z_i,\dots)}=\ket{\Phi_{L-2}(\dots,\hat z_{i-1},\hat z_i,\dots)}, \]
so that
\[ \ket{\Phi_{L-2}(\dots,\hat z_{i-1},\hat z_i,\dots)}\propto\ket{\Psi_{L-2}(\dots,\hat z_{i-1},\hat z_i,\dots)}, \]
by uniqueness of the ground state. Since both $\ket\Phi$ and $\ket\Psi$ have no overall symmetric factors, and $\ket\Psi$ satisfies the qKZ system, they must be proportional by a constant.
\end{proof}
\subsection{Solution}
\label{sec:soln}
The expressions for the maximally factorized components given in \eqref{eq:psis} apply to the Brauer loop model simply by setting $s=0$. In addition, for $\eps=0$ \cite{PDF-open} gives a non-recursive expression for the component corresponding to the maximally crossing closed link pattern, which has $\pi(i)=i+L/2$. We will not need this expression.

\subsubsection{Sum rules}
\label{sec:normzns}
Finally, we define the sum rule of the ground state as the sum of all its entries, noting again that the entries have been defined to be coprime (up to a constant factor that will be explained in \secref{mdeg}),
\[ Z_L^{\mathrm{a}}:=\sum_\pi \psi_\pi^{\mathrm{a}}. \]

Because we have set $\epsilon=0$, $Z^{\mathrm a}_L$ is a symmetric polynomial of the arguments $z_1,\ldots,z_L$, and in type $\hat{\mathrm C}$, an even polynomial in these variables. The proof is standard and we describe it briefly. First we write $Z^{\mathrm a}_L=\braket{v}{\Psi^{\mathrm a}}$ where $v$ is the covector with constant entries $1$ in the basis of link patterns. Then we note that at $\epsilon=0$, from \eqref{eq:checkR}, $\bra{v}\check R_i(z)=\bra{v}$, and similarly, from \eqref{eq:checkK}, $\bra{v}\check K_{0,L}(z)=\bra{v}$. We now apply \eqref{eq:qKZbulk} and conclude that $\tau_i$ leaves $\braket{v}{\Phi^{\mathrm a}}$ invariant for all $i=0,\ldots,L$, which is the desired symmetry.

This means that the recurrence relations satisfied by the components of the ground state extend to many more recurrence relations for $Z_L$:
\begin{align}
\label{eq:recsum} Z^{\mathrm{p}}_L(z_j=A+z_i)&=p^{\mathrm{p}}(z_i|\hat z_i,\hat z_{j})\ Z^{\mathrm{p}}_{L-2}(\hat z_i,\hat z_j),\hspace{-1cm} & &\forall i\neq j,\hspace{-1cm} & &\\
\nn Z^{\mathrm{a}}_L(z_j=A\pm z_i)&=p^{\mathrm{a}}(\pm z_i|\hat z_i,\hat z_{j})\ Z^{\mathrm{a}}_{L-2}(\hat z_i,\hat z_j),\hspace{-1cm} & &\forall i\neq j,\hspace{-1cm} & &\mathrm{a}\in\{\mathrm{i},\mathrm{c},\mathrm{o},\mathrm{m}\},\\
\nn Z^{\mathrm{a}}_L(z_j=-A\pm z_i)&=p^{\mathrm{a}}(\mp z_i|\hat z_i,\hat z_{j})\ Z^{\mathrm{a}}_{L-2}(\hat z_i,\hat z_j),\hspace{-1cm} & &\forall i\neq j,\hspace{-1cm} & &\mathrm{a}\in\{\mathrm{i},\mathrm{c},\mathrm{o},\mathrm{m}\},\\
\nn Z^{\mathrm{a}}_L(z_i=\pm A/2)&=p_L^{\mathrm{a}}(\hat z_i)\ Z^{\mathrm{a}}_{L-1}(\hat z_i),\hspace{-1cm} & &\forall i,\hspace{-1cm} & &\mathrm{a}\in\{\mathrm{i},\mathrm{o},\mathrm{m}\}.
\end{align}

One of the results of this paper is explicit expressions for the $\mathrm i$,$\mathrm o$ and $\mathrm m$ sum rules. These are given in \secref{loc} and \secref{conc}, along with the $\mathrm p$ and $\mathrm c$ cases that have been done before \cite{PDF-open,artic32}.


\section{The Brauer loop schemes}\label{sec:brauer}
Following \cite{artic33}, there are essentially two ways to define the Brauer loop schemes in various types: either in terms of infinite periodic matrices (i.e., loop algebras); or as flat limits of certain nilpotent orbit closures. We provide both interpretations below.

\subsection{The infinite strip picture}
\subsubsection{Definitions}
Fix a positive integer $N$. Consider the algebra
$
R:=\{
M=(M_{ij})_{i,j\in\Z}\}
$
of complex {\em upper triangular}\/ matrices that are infinite in both directions, and the subalgebra $R_{\ZN}$ of the $(N,N)$-periodic ones:
\[
R_{\ZN}
:=\{ M\in R\ |\ MS^N=S^N M\},
\]
where $S=(\delta_{i,j-1})\in R_{\ZN}$ is the shift operator.
Then we define the algebra $\M_N$ to be the quotient of $R_{\ZN}$ by the ideal generated by $S^N$:
\[
\M_N:=
R_{\ZN}/\left< S^N\right>.
\]
$\M_N$ is of dimension $N^2$. A fundamental domain for $M\in \M_N$ is
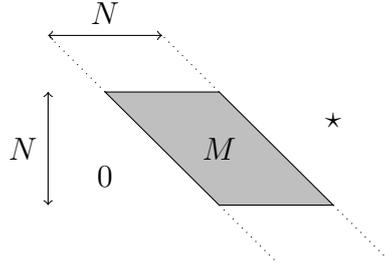
\begin{center}
\begin{tikzpicture}[scale=1.5]
\draw[fill=lightgray] (0,0) -- (1,-1) -- (2,-1) -- (1,0) -- cycle; \node at (1,-0.5) {$M$};
\draw[dotted] (0,0) -- (-0.5,0.5) (1,0) -- (0.5,0.5) (1,-1) -- (1.5,-1.5) (2,-1) -- (2.5,-1.5);
\node at (0,-0.75) {$0$};
\node at (2,-0.25) {$\star$};
\draw[<->] (-0.5,0.5) -- (0.5,0.5) node[above,pos=0.5] {$N$};
\draw[<->] (-0.5,0) -- (-0.5,-1) node[left,pos=0.5] {$N$};
\end{tikzpicture}
\end{center}
where the left diagonal is the main diagonal, and everything left of it is zero, while the right diagonal is the $N^{\rm th}$ diagonal, and everything on it or right of it is ignored. There is some freedom in choosing which $N$ rows we
put in the fundamental domain, i.e., in sliding the latter along the diagonal.
In what follows, we identify an element of $\M_N$
and any of its representatives
when there is no risk of confusion.

The definitions above are directly relevant to type $\hat{\mathrm A}$
(periodic boundary conditions),
and so we shall also write $R_{{\ZN}}=
R^{\mathrm p}_{{\ZN}}$, $\M_N = \M^{\mathrm p}_N$.

Assume now that $N=2n$ is even.
The new ingredient we introduce for type $\hat{\mathrm C}$ is the antidiagonal
symplectic form $J$:
\[
J_{ij}:= \delta_{i+j,N+1} \,\varepsilon_i
\qquad
i,j\in\Z,\
\varepsilon_i:=
\begin{cases}
1&i=1,\ldots,n\pmod N,\\
-1&i=n+1,\ldots,N\pmod N.
\end{cases}
\]
Note that $S^n J S^{n}=-J$.
Define the adjoint of $M$ to be $M^\dagger:=J^{-1}M^T J$ where $J^{-1}=-J$;
explicitly,
\[
M^\dagger_{ij}=\varepsilon_i\varepsilon_j M_{N-j+1,N-i+1}.
\]
Note that $R_{{\ZN}}=R_{{\ZN}}^\dagger$, and the same for $\left<S^N\right>$.
We thus define
\begin{align*}
R_{{\ZN}}^{\mathrm{i}}&:=\{M\in R_{{\ZN}}\ |\ M=-M^\dagger\}, && \M_N^{\mathrm{i}}:=R_{{\ZN}}^{\mathrm{i}}/\left(\left<S^N\right>\cap R_{{\ZN}}^{\mathrm{i}}\right),\\
R_{{\ZN}}^{\mathrm{c}}&:=\{M\in R_{{\ZN}}\ |\ M=M^\dagger\},
&& \M_N^{\mathrm{c}}:=R_{{\ZN}}^{\mathrm{c}}/\left(\left<S^N\right>\cap R_{{\ZN}}^{\mathrm{c}}\right).
\end{align*}
(Note that the definitions above would be unaffected
by the change $J\to S^n J S^{-n}$.) $R_{{\ZN}}^{\mathrm{i}}$ is a Lie subalgebra
of $R_{{\ZN}}$, and $R_{{\ZN}}=R_{{\ZN}}^{\mathrm{i}}\oplus R_{{\ZN}}^{\mathrm{c}}$ as a $R_{{\ZN}}^{\mathrm{i}}$-module.

A fundamental domain is (assuming that in the previous picture the chosen
rows are from $1$ to $N$)
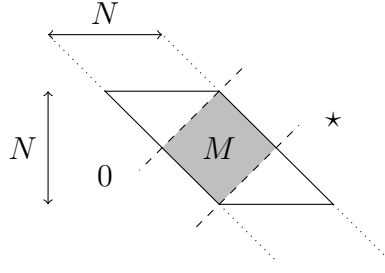
\begin{center}
\begin{tikzpicture}[scale=1.5]
\fill[lightgray] (1,0) -- (0.5,-0.5) -- (1,-1) -- (1.5,-0.5);
\draw (0,0) -- (1,-1) -- (2,-1) -- (1,0) -- cycle; \node at (1,-0.5) {$M$};
\draw[dashed] (1.2,0.2) -- (0.3,-0.7) (0.8,-1.2) -- (1.7,-0.3);
\draw[dotted] (0,0) -- (-0.5,0.5) (1,0) -- (0.5,0.5) (1,-1) -- (1.5,-1.5) (2,-1) -- (2.5,-1.5);
\node at (0,-0.75) {$0$};
\node at (2,-0.25) {$\star$};
\draw[<->] (-0.5,0.5) -- (0.5,0.5) node[above,pos=0.5] {$N$};
\draw[<->] (-0.5,0) -- (-0.5,-1) node[left,pos=0.5] {$N$};
\end{tikzpicture}
\end{center}
where the dashed lines are symmetry axes (i.e., entries that are mirror
images w.r.t.\ one of the axes are either equal or opposite). In
$R_{\ZN}^{\mathrm c}$, the entries on the symmetry axes are zero, whereas in
$R_{\ZN}^{\mathrm i}$, they are free.

Assuming that $N$ is a multiple of $4$, i.e., that $n=2m$,
we can introduce a {\em second}\/ symplectic form
$J'=S^{m} J S^{-m}$ and a second notion of adjoint
$M^\ddagger_{ij}=\varepsilon_{i+m}\varepsilon_{j+m} M_{n+1-j,n+1-i}$. This leads to
more definitions:
\begin{align*}
R_{{\ZN}}^{\mathrm{o}}&:=\{M\in R_{{\ZN}}\ |\ M=-M^\dagger=-M^\ddagger\},
&& \M_N^{\mathrm{o}}:=R_{{\ZN}}^{\mathrm{o}}/\left(\left<S^N\right>\cap R_{{\ZN}}^{\mathrm{o}}\right),\\
R_{{\ZN}}^{\mathrm{m}}&:=\{M\in R_{{\ZN}}\ |\ M=M^\dagger=-M^\ddagger\},
&& \M_N^{\mathrm{m}}:=R_{{\ZN}}^{\mathrm{m}}/\left(\left<S^N\right>\cap R_{{\ZN}}^{\mathrm{m}}\right),
\end{align*}
i.e., a fundamental domain of the form
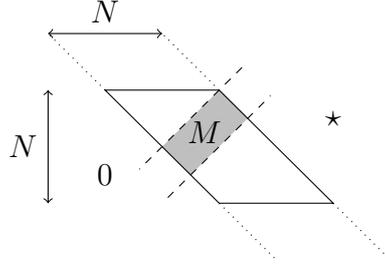
\begin{center}
\begin{tikzpicture}[scale=1.5]
\fill[lightgray] (1,0) -- (0.5,-0.5) -- (0.75,-0.75) -- (1.25,-0.25);
\draw (0,0) -- (1,-1) -- (2,-1) -- (1,0) -- cycle; \node at (0.87,-0.37) {$M$};
\draw[dashed] (1.2,0.2) -- (0.3,-0.7) (0.55,-0.95) -- (1.45,-0.05);
\draw[dotted] (0,0) -- (-0.5,0.5) (1,0) -- (0.5,0.5) (1,-1) -- (1.5,-1.5) (2,-1) -- (2.5,-1.5);
\node at (0,-0.75) {$0$};
\node at (2,-0.25) {$\star$};
\draw[<->] (-0.5,0.5) -- (0.5,0.5) node[above,pos=0.5] {$N$};
\draw[<->] (-0.5,0) -- (-0.5,-1) node[left,pos=0.5] {$N$};
\end{tikzpicture}
\end{center}

We finally define in each type the (unreduced) Brauer loop scheme to be
\[
\tilde E^{\mathrm{a}}_N:=\{ M\in \M^{\mathrm{a}}_N\ |\ M^2=0\},\qquad \mathrm{a}\in\{\mathrm{p},\mathrm{i},\mathrm{c},\mathrm{o},\mathrm{m}\}.
\]
As in \cite{artic33}, noting that among the equations of the scheme are $M_{ii}^2=0$, we prefer to define the (generically reduced) Brauer loop scheme as
\[
E^{\mathrm{a}}_N:=\{ M\in \M^{\mathrm{a}}_N\ |\ M^2=0\text{ and } M_{ii}=0\ \forall i\},\qquad \mathrm{a}\in\{\mathrm{p},\mathrm{i},\mathrm{c},\mathrm{o},\mathrm{m}\}.
\]
As sets, $\tilde E^{\mathrm{a}}_N$ and $E^{\mathrm{a}}_N$ are identical, but as schemes,
the latter is generically reduced (as we shall prove), and conjecturally reduced, whereas the former is neither.
The distinction is rather inessential in type $\hat{\mathrm A}$
(see however \cite[Section~7]{artic33}), but less so in type $\hat{\mathrm C}$, see the discussion before \eqref{eq:defpsi}. Note that $E^{\mathrm a}_N\subset
(\M^{\mathrm a}_N)_{\Delta=0}$ with the notation
$(\M^{\mathrm a}_N)_{\Delta=0}=\{M\in \M^{\mathrm a}_N\ |\ M_{ii}=0\ \forall i\}$.

\subsubsection{Group action and multidegrees}\label{sec:mdeg}
Invertible elements of $R$ act by conjugation on $R$, and among them,
the subgroup
\[
B_{\ZN}:=\{
M\in R^\times\ |\ \exists\lambda\in\C^\times,\ S^N M=\lambda\,M S^N
\},
\]
leaves $R_{{\ZN}}$ and $\left<S^N\right>$ invariant, thus acts on $\M_N$.

A maximal torus $T_{{\ZN}}$ of $B_{{\ZN}}$ consists of diagonal matrices with
the same property:
\[
T_{\ZN}:=B_{\ZN}\cap\{\text{diagonal matrices}\}.
\]
It is of dimension $N+1$. Note however that scalar matrices act trivially
by conjugation, so only $T_{\ZN}/\C^\times$, of dimension $N$, acts on $\M_N^{\mathrm p}$.

We add to $T_{\ZN}$ an additional $\C^\times\ni q$ which acts on $M\in\M_N^{\mathrm p}$ by scaling:
$M\to qM$.

The corresponding Lie algebra ${\mathfrak t}_{{\ZN}}$ has elements of the form
$\text{diag}(z_i)_{i\in\Z}$, where $z_{i+N}=z_i+\epsilon$ for all $i$,
where $\lambda=\exp\epsilon$. Also introduce
the generator $A$ (where $q=\exp A$) of the extra $\C^\times$.
Then the group of characters of $T_{\ZN}\times\C^\times$ (viewed as a lattice in ${\mathfrak t}_{\ZN}\oplus\C$)
is the abelian group
generated by
the $z_i$, $i\in\Z$, $\epsilon$ and $A$ with relations $z_{i+N}=z_i+\epsilon$.
Furthermore, the commutative ring they generate is the equivariant
cohomology ring of a point, or of $\M^{\mathrm p}_N$:
\[
H^\ast_{T_{\ZN}\times\C^\times}(\M_N^{\mathrm p})\cong
\Z[(z_i)_{i\in\Z},A,\epsilon]/\left<z_i+\epsilon-z_{i+N},\ i\in\Z\right>.
\]
Comparing with the notations of \secref{dynkin}, we find that we must identify $L=N$, $s=\epsilon$,
and then $H^\ast_{T_{\ZN}\times\C^\times}$ is the embedding ring for the root lattice of type $\hat{\mathrm A}$, with one additional
variable $A$ (corresponding to the extra circle).

A convenient algebraic framework for computations in equivariant cohomology
of a 
vector space endowed with a linear group action is to use {\em multidegrees};
we refer to \cite{MS-book} for details.
To any subvariety $X$ of $\M_N^{\mathrm p}$ which is invariant by action of $T_{\ZN}\times\C^\times$,
we can thus associate its multidegree $\mdeg X$
in $H^\ast_{T_{\ZN}\times\C^\times}(\M_N^{\mathrm p})$.
Because the real action is $(T_{\ZN}/\C^\times)\times \C^\times$, all our multidegrees
depend only on $z_i-z_j$ (and more precisely, are sums of products
of the weights $A+z_i-z_j$, $i\le j<i+N$).

We now discuss type $\hat{\mathrm C}$. We define
\[
\tilde B_{\ZN}:=\{ M\in B_{\ZN}\ |\ \exists\zeta\in\C^\times,\ MM^\dagger=\zeta\},
\qquad
\tilde T_{\ZN}:=\tilde B_{\ZN}\cap T_{\ZN}.
\]
We could set the scalar $\zeta$ to $1$ because conjugation by a scalar is trivial,
but we prefer to keep it for reasons which will become clear.

There are corresponding Lie algebras for which $M+M^\dagger=u$,
$\zeta=\exp u$.
In particular,
\[
\tilde{\mathfrak t}_{\ZN}:=\{M=\text{diag}(z_i)\in {\mathfrak t}_{\ZN}\ |\ z_{N-i+1}=u-z_i\}.
\]
Then we have
\begin{align*}
H^\ast_{\tilde T_{\ZN}\times\C^\times}(\M_N^{\mathrm i})&\cong H^\ast_{\tilde T_{\ZN}\times\C^\times}(\M_N^{\mathrm c})\\
&\cong \Z[(z_i)_{i\in\Z},A,\epsilon,\alpha]/\left<z_i+\epsilon-z_{i+N},\ z_i+z_{N-i+1}-u,\ i\in\Z\right>.
\end{align*}

We can as before define multidegrees of subvarieties of $\M_N^{\mathrm i,c}$;
because of the trivial conjugaton by scalar matrices,
the parameter $u$ is redundant, and will be set to $0$ in what follows
(it can be recovered by substituting $z_i \mapsto z_i-u/2$). We recover at this stage the embedding ring
for the root lattice of type $\hat{\mathrm C}$ of \secref{dynkin} with the following correspondence:
$L=N/2=n$, $s=\epsilon$.

Similarly, define
\begin{align*}
\ttilde B_{\ZN}&:=\{ M\in B_{\ZN}\ |\ \exists\zeta,\xi\in\C^\times,\ MM^\dagger=\zeta,\ MM^\ddagger=\xi\},\\
\ttilde T_{\ZN}&:=\ttilde B_{\ZN}\cap T_{\ZN}.
\end{align*}
Here the scalar factors become relevant: indeed, it is easy to see
by combining the various equations (in particular,
using $(J'J)^2=S^N$) that $\zeta^2=\lambda\xi^2$, so that, for
$\lambda\ne1$, one cannot set simultaneously $\zeta$ and $\xi$ to $1$.
Writing $\xi=\exp v$,
with the relation $2v=2u-\epsilon$,
the Lie algebra of the corresponding maximal torus is
\[
\ttilde{\mathfrak t}_{\ZN}:=\{M=\text{diag}(z_i)\in {\mathfrak t}_{\ZN}\ |\ z_{N-i+1}=u-z_i,\ z_{n-i+1}=v-z_i\}.
\]

Finally, we have for $\mathrm a\in \{\mathrm o,\mathrm m\}$
\begin{multline*}
H^\ast_{\ttilde T_{\ZN}\times\C^\times}(\M_N^{\mathrm{a}})
\\
\cong\Z[(z_i)_{i\in\Z},A,\epsilon/2,v]/\left<z_i+\epsilon-z_{i+N},\ z_i+z_{N-i+1}-v-\epsilon/2,\ z_i+z_{n-i+1}-v,\
i\in\Z\right>,
\end{multline*}
where we have replaced $u$
with $v+\epsilon/2$.
The parameter $v$ is redundant
and will be set to $0$ (it can be recovered
by $z_i \mapsto z_i-v/2$).
We also recover the embedding ring
for the root lattice of type $\hat{\mathrm C}$ of \secref{dynkin}, but with a different correspondence:
$L=N/4=m$, $s=\epsilon/2$.

At this stage, one would like to introduce the multidegrees of the components of the Brauer loop scheme.
We note that $\tilde E^{\mathrm a}_N$ and $E^{\mathrm a}_N$ are invariant by action of $T_{\ZN}\times\C^\times$ for $\mathrm a=\mathrm p$,
$\tilde T_{\ZN}\times\C^\times$ for $\mathrm a\in\{\mathrm i,\mathrm c\}$,
$\ttilde T_{\ZN}\times\C^\times$ for $\mathrm a\in\{\mathrm o,\mathrm m\}$, and therefore so are their irreducible components.

The most natural quantities are the multidegrees of the primary
top-dimensional components of the scheme $\tilde E_N^{\mathrm a}$
(we conjecture equidimensionality, in which case ``top-dimensional'' can be omitted).
However for practical reasons, it is much easier to deal with (reduced) varieties. Let us therefore define the $E_\pi^{\mathrm a}$ to be the (reduced) irreducible top-dimensional components of $\tilde E_N^{\mathrm a}$ or $E_N^{\mathrm a}$,
where the indexing set for $\pi$ will be determined to be the Brauer link patterns in \secref{irredcomp}.
Then we write
\begin{equation}\label{eq:defpsi}
\phi^{\mathrm a}_\pi:=m_\pi \mdeg E^{\mathrm a}_\pi,\qquad \mathrm a\in\{\mathrm p,\mathrm i,\mathrm c,\mathrm o,\mathrm m\},
\end{equation}
where the multidegree is relative to $\M^{\mathrm a}_N$,
and $m_\pi=2^{\#\{\text{chords}(\pi)\}+\#\{\text{fixed points}(\pi)\}}$ (the number
of fixed points is one if $L$ is odd and $\mathrm a\in\{\mathrm p,\mathrm c\}$, zero otherwise).\footnote{In \secref{irredcomp}, we shall introduce another ``periodic'' diagram
for $\pi$. It is important to note that in the definition of $m_\pi$ we mean the number of chords and fixed points of the ordinary, nonperiodic (for $\mathrm a\ne\mathrm p$) diagram of $\pi$.} We shall show
(see \appref{geomstuff}) that $m_\pi$
is the multiplicity of $E_\pi^{\mathrm a}$ in $\tilde E_N^{\mathrm a}$, i.e., that
$\phi^{\mathrm a}_\pi$ is the multidegree
of the primary component of $\tilde E^{\mathrm a}_N$ associated to $E^{\mathrm a}_\pi$.


\subsubsection{Relation to loop algebras}
Since all matrices we consider commute with $S^N$, it is natural
to consider $t=S^N$ as a scalar. We immediately conclude that
$R_{\ZN}\cong R_N\oplus t\,\g_N\oplus t^2\g_N\oplus\cdots$,
where $\g_N=\g_N^{\mathrm p}$  is the space of $N\times N$ matrices, i.e.,
the Lie algebra $\gl_N(\C)$, and $R_N$
is the space of $N\times N$ upper triangular matrices;
thus identifying $R_{\ZN}$ with the Borel subalgebra
of the loop algebra $\gl_N(\C[t,t^{-1}])$.
Then $\M_N\cong R_{\ZN}/t R_{\ZN}$.

Similarly, denote by $\g^{\mathrm i}_N$
the space of matrices $M\in \gl_N$ that satisfy $JM+M^T J=0$, where by abuse
of notation we use the same letter $J$ for the finite matrix
\begin{equation}\label{eq:Jfin}
J:=
\begin{pmatrix}
0 & \cdots & & & & 1\\
\vdots & & & & \iddots &\\
& & & 1 & &\\
& & -1 & & &\\
& \iddots & & & & \vdots\\
-1 & & & & \cdots & 0
\end{pmatrix},
\end{equation}
and its infinite periodic counterpart defined earlier.

$\g^{\mathrm i}_N$ is the symplectic Lie algebra $\sp_N(\C)$.
Now define $\g^{\mathrm c}_N$ to be the space of matrices satisfying $JM-M^T J=0$.
One has
$\g_N=\g_N^{\mathrm i}\oplus\g_N^{\mathrm c}$
as a $\sp_N(\C)$-module by conjugation.

Then $R_{\ZN}^{\mathrm{a}}\cong R^{\mathrm a}_N\oplus t\, \g^{\mathrm a}_N\oplus t^2 \g^{\mathrm a}_N\oplus\cdots$ for $\mathrm a\in \{\mathrm i,\mathrm c\}$ (where $R^{\mathrm a}_N = R_N \cap \g_N^{\mathrm a}$),
which identifies $R_{\ZN}^{\mathrm i}$ with the Borel subalgebra
of the loop algebra $\sp_N(\C[t,t^{-1}])$.

Finally, define $\sp'_N(\C)$ to be the Lie algebra of matrices
satisfying $J'M+M^T J'=0$, where
\[
J':=\begin{pmatrix}
0&&1\\
&\iddots\\
-1
\\
&&&0&&&1\\
&&&&&\iddots\\
&&&&-1&&0
\end{pmatrix},
\]
As a Lie algebra, $\sp_N(\C)\cap\sp'_N(\C) \cong \sp_n(\C) \oplus\sp_n(\C)$,
i.e., two copies of the symplectic Lie algebra.
However, $R_N\cap\sp_N(\C)\cap\sp'_N(\C)$
is {\em not}\/ its Borel subalgebra.
Define the spaces
$\g^{\mathrm o}_N=\g_N^{\mathrm i}\cap \sp'_N(\C)$,
$\g^{\mathrm m}_N=\g_N^{\mathrm c}\cap \sp'_N(\C)$,
and $R^{\mathrm a}_N = R_N \cap \g_N^{\mathrm a}$,
so that $R_{\ZN}^{\mathrm{a}}\cong R^{\mathrm a}_N\oplus t\, \g_N^{\mathrm a}\oplus t^2 \g_N^{\mathrm a}\oplus\cdots$ for $\mathrm a\in \{\mathrm o,\mathrm m\}$.

\subsection{The Brauer loop schemes as a flat limit}
\subsubsection{Orbit closures and their flat degeneration}\label{sec:orbclos}
Let us define the map from $\g_N$ to $\M_N$ that
takes $M$ to $M_{\le}+t M_>$,
where $M_\le$ (resp.\ $M_>$) indicates the upper (resp.\ strict lower)
triangle of $M$.
(Equivalently, in terms of the strip picture, this amounts to
the parameterization
\begin{center}
\begin{tikzpicture}[scale=1.5]
\draw[fill=lightgray] (0,0) -- (1,-1) -- (2,-1) -- (1,0) -- cycle;
\draw (1,-1) -- (1,0);
\node at (0.7,-0.35) {$M_\le$};
\node at (1.3,-0.65) {$M_>$};
\draw[dotted] (0,0) -- (-0.5,0.5) (1,0) -- (0.5,0.5) (1,-1) -- (1.5,-1.5) (2,-1) -- (2.5,-1.5);
\node at (0,-0.75) {$0$};
\node at (2,-0.25) {$\star$};
\draw[<->] (-0.5,0.5) -- (0.5,0.5) node[above,pos=0.5] {$N$};
\draw[<->] (-0.5,0) -- (-0.5,-1) node[left,pos=0.5] {$N$};
\end{tikzpicture}
\end{center}
of the fundamental domain).

The connection to loop algebras suggests that we should think of $t$ as a numerical parameter which provides a one-parameter family of products on $\g_N$. By varying the value of $t$, one interpolates between ordinary matrix product on $\g_N$ (at $t=1$, a generic fiber) and a degenerate product at the special fiber $t=0$ (denoted $\bullet$ in \cite{artic33}), which is the one on $\M_N$, which makes $\M_N$ isomorphic to the semi-direct product $R_N \times (\g_N/R_N)$, with multiplication $(\U,\La)(\U',\La')=(\U\U',\U\La'+\U'\La)$.

More explicitly, define for any $\mathrm{a}\in\{\mathrm{p},\mathrm{i},\mathrm{c},\mathrm{o},\mathrm{m}\}$
\[
D_{N;t}^{\mathrm a} := \{M\in\g^{\mathrm a}_N\ |\ (M_{\le}+t M_>)^2=0\},\qquad t\ne 0,
\]
and $D_{N;0}^{\mathrm a}=\lim_{t\to 0} D_{N;t}^{\mathrm a}$ to be the flat limit as $t\to 0$.

If $\mathrm a\in\{\mathrm p,\mathrm i,\mathrm c\}$, this is equivalent to saying that $D_{N;0}^{\mathrm a}$ is the {\em normal cone}\/ of $D_N^{\mathrm a}\cap R_N$ inside $D_N^{\mathrm a}:=D_{N;1}^{\mathrm a}$. This is not so for $\mathrm a\in\{\mathrm o,\mathrm m\}$ (note that it is $M$, not $M_\le+t M_>$, that is in $\g^{\mathrm a}_N$).

\begin{prop}\label{prop:normalcone}
$D_{N;0}^{\mathrm a}\subset \tilde E_{N}^{\mathrm a}$ as schemes.
\end{prop}
\begin{proof}
First, one checks that in each case $\mathrm a\in\{\mathrm i,\mathrm c,\mathrm o,\mathrm m\}$, the symmetry of $M\in \mathfrak{g}_N^{\mathrm a}$ turns into the symmetry of $M_\le + t M_> \in \M^{\mathrm a}_N$. It then follows from the above (see also \cite[Section~2.3]{artic33}) that in the limit $t\to0$, the equation $(M_\le+t M_>)^2=0$ in $\mathfrak{g}^{\mathrm a}$ becomes, essentially tautologically, the equation $M^2=0$ in $\M^{\mathrm a}_N$. This implies the inclusion of schemes.
\end{proof}
In principle there may be more equations in the flat limit of the ideal generated
by $(M_\le+t M_>)^2=0$ as $t\to 0$.
In fact, we conjecture that there are not, so that the inclusion of \propref{normalcone} is an
equality. It will be a consequence of the results below that $D_{N;0}^{\mathrm a}=E^{\mathrm a}_N$ as sets,
so that the Brauer loop schemes can be defined alternatively as the normal cones $D_{N;0}^{\mathrm a}$.
We shall also prove (\appref{geomstuff})
that the multiplicities of the $E_\pi^{\mathrm a}$ inside $D_{N;0}^{\mathrm a}$ and $\tilde E_N^{\mathrm a}$ are equal.
Note that some of these results are new even in type $\hat{\mathrm A}$, proving some conjectures of \cite{artic33}.

The group $B_{\ZN}$ (or $\tilde B_{\ZN}$, $\ttilde B_{\ZN}$) does not
act on $\g_N^{\mathrm a}$; instead, we have an action of $GL_N$
(or $Sp_N$, or $Sp_N \cap Sp'_N$). At the level of the torus action, it
is easy to see that the Cartan tori of the latter identify with subgroups of codimension one inside
the Cartan tori of the former. In terms of Lie algebras, it
means that we must restrict to the subalgebra given
by $\epsilon=0$. The degeneration respects that torus action, and therefore
\begin{equation}\label{eq:flatlimmdeg}
\mdeg D_{N;0}^{\mathrm a}|_{\epsilon=0}=\mdeg D_N^{\mathrm a}.
\end{equation}
By standard arguments,
$D_N^{\mathrm a}$ is an orbit closure,
and therefore irreducible. Furthermore, we check smoothness at
a specific point
(with the assumption that $n=N/2$ is even if $\mathrm a=\mathrm c$;
see the discussion in the next section) by an elementary Zariski tangent space computation,
and conclude that $D^{\mathrm a}_N$ is generically reduced.
For $\mathrm a\in\{\mathrm p,\mathrm i,\mathrm c\}$, the smooth point is
\begin{align*}
M&=\left(\begin{array}{ccc|ccc}
&&&&&1
\\
&0&&&\iddots
\\
&&&1
\\
\hline
&&&\\
&0&&&0
\\
&&&
\end{array}
\right),
\qquad \mathrm a\in\{\mathrm p,\mathrm i\},
\\
M&=\left(\begin{array}{ccc|ccc}
&&&&&\begin{array}{cc}\ss 1&\ss 0\\\ss 0&\ss -1\end{array}
\\
&0&&&\iddots
\\
&&&\begin{array}{cc}\ss 1&\ss 0\\\ss 0&\ss -1\end{array}
\\
\hline
&&&\\
&0&&&0
\\
&&&
\end{array}
\right),
\qquad \mathrm a=\mathrm c,\ N=0\pmod{4},
\end{align*}
where blocks are $n\times n$.

For $\mathrm a\in\{\mathrm o,\mathrm m\}$, it is easy to see that
there is a decomposition $\mathbb{C}^N=W_+\oplus W_-$ which is orthogonal w.r.t.\ both $J$ and $J'$,
and such that $J'=\pm i J$ when restricted to $W_\pm\otimes W_\pm$.
With respect to this decomposition, $M\in D_N^{\mathrm o}$ is block diagonal, so that
one simply has $D_N^{\mathrm{o}}\cong D_{N/2;1}^{\mathrm{i}}\times D_{N/2;1}^{\mathrm{i}}$ (as schemes), and no further check is necessary.
Finally, for $\mathrm a=\mathrm m$,
using the same decomposition, one finds this time that $M\in D_N^{\mathrm m}$ has off-diagonal blocks
$X\in \text{Hom}(W_+,W_-), X^\dagger\in \text{Hom}(W_-,W_+)$:
\[
D_N^{\mathrm m}\cong\left\{
\begin{pmatrix}0&X^\dagger\\X&0
\end{pmatrix}\ \middle|\ XX^\dagger=X^\dagger X=0\right\}.
\]
The smooth point is then
\[
X=\begin{pmatrix}
1\\
&\ddots\\
&&1\\
&&&0\\
&&&&\ddots\\
&&&&&0
\end{pmatrix},
\]
where the number of $1$'s is $m=N/4$.

We now compute the multidegree of $D_N^{\mathrm a}$
using localization.

\subsubsection{Localization}\label{sec:loc}
We now wish to use equivariant 
localization techniques to compute the multidegree (i.e., equivariant
cohomology class) of the orbit closure $D_N^{\mathrm a}$.
$D_N^{\mathrm a}$ is a conical affine variety, with unique fixed point
$0$, which is of course singular, so we cannot directly apply
localization to it. Instead, we find a resolution of singularity
$Q^{\mathrm a}_N$ of $D_N^{\mathrm a}$, then express $1$ as a linear combination
of fixed points in the appropriately localized equivariant cohomology of
$Q^{\mathrm a}_N$, and then finally push forward using the resolution map:
by definition the pushforward of $1$
is $\mdeg D_N^{\mathrm a}$ (viewed as an element of the equivariant
cohomology of $\mathfrak{g}^{\mathrm a}_N$, which is the same as that of a point),
whereas each fixed point is sent to $0$, whose class is the product of weights inside $\mathfrak{g}^{\mathrm a}_N$. 

We skip the detailed proofs (see also \cite[Proposition~7]{artic33}) 
and simply provide the formulae in each case:
\begin{itemize}
\item $\mathrm a=\mathrm p$.
This is the case considered in \cite{artic33}. Write $N=2n+r$ with $r\in\{0,1\}$.
$Q_N^{\mathrm p}$ is
the cotangent bundle of the (type A) Grassmannian:
\[
Q_N^{\mathrm{p}}=\{(V,M)\ |\ M\in \g_N^{\mathrm p},\ \dim V=n,\ \Im M \subset V \subset \Ker M\}.
\]

Fixed points are coordinate subspaces $V_I=\text{span}\{\mathrm e_i, i\in I\}$
(where $\mathrm e_1,\ldots,\mathrm e_N$ is the standard basis of $\C^N$),
indexed by $n$-subsets $I$ of $\{1,\ldots,N\}$, and localization gives
\[
\mdeg D_N^{\mathrm{p}}=2^r\prod_{i,j=1}^{N} (A+z_i-z_j) \sum_{\substack{I\subset \{1,\ldots,N\}\\|I|=n}} \prod_{i\in I,j\not\in I}
\frac{1}{(z_j-z_i)(A+z_i-z_j)}.
\]

\item $\mathrm a=\mathrm i$.
Define the cotangent bundle of the Lagrangian (type C) Grassmannian:
\[
Q_N^{\mathrm{i}}=\{(V,M)\ |\ M\in \g_N^{\mathrm i},\ V^\perp=V,\ \Im M \subset V \subset \Ker M\}.
\]
Lagrangian coordinate subspaces are indexed by signs $\varepsilon=(\varepsilon_1,\ldots,\varepsilon_n)\in \{+1,-1\}^n$: explicitly,
\[
V_\varepsilon:=\text{span}\left(
\{
\mathrm e_i, 1\le i\le n, \varepsilon_i=-1
\}
\cup
\{
\mathrm e_{N-i+1}, 1\le i\le n, \varepsilon_i=+1
\}
\right).
\]
Then,
\[
\mdeg D_N^{\mathrm{i}}=A^{-n}\prod_{1\le i\le j\le n} (A\pm z_i\pm z_j)
\sum_{\varepsilon\in\{+1,-1\}^n}
\prod_{\substack{i,j=1\\i\le j}}^n
\frac{1}{(\varepsilon_i z_i+\varepsilon_j z_j)(A-\varepsilon_i z_i-\varepsilon_j z_j)}.
\]

\item $\mathrm a=\mathrm c$.
For $n=N/2$ even, the situation is similar to the identified case:
\[
Q_N^{\mathrm{c}}=\{(V,M)\ |\ M\in \g_N^{\mathrm c},\ V^\perp=V,\ \Im M \subset V \subset \Ker M\}.
\]

\[
\mdeg D_N^{\mathrm{c}}=A^n\prod_{1\le i<j\le n} (A\pm z_i\pm z_j)
\sum_{\varepsilon\in\{+1,-1\}^n}
\prod_{\substack{i,j=1\\i\le j}}^n
\frac{1}{\varepsilon_i z_i+\varepsilon_j z_j}
\prod_{\substack{i,j=1\\i<j}}^n
\frac{1}{A-\varepsilon_i z_i-\varepsilon_j z_j}.
\]

For $n$ odd, the map $(V,M)\mapsto M$ is not generically one-to-one since the rank of a generic element of $D^{\mathrm c}_N$ is $n-1$ (not $n$). The resolution of singularities is slightly more complicated and we shall skip the details, noting that it is simpler to use the recurrence relation \eqref{eq:recsum} to deduce the sum rule in odd size from that in even size.

\item $\mathrm a=\mathrm o$.
Recall that $D_N^{\mathrm{o}}=D_{N/2;1}^{\mathrm{i}}\times D_{N/2;1}^{\mathrm{i}}$, so the resolution of singularities is simply $Q_N^{\mathrm{o}}=Q_{N/2}^{\mathrm{i}}\times Q_{N/2}^{\mathrm{i}}$. We conclude immmediately
\[
\mdeg D_N^{\mathrm{o}}=(\mdeg D_{N/2}^{\mathrm{i}})^2.
\]

\item $\mathrm a=\mathrm m$.
We use again the decomposition discussed at the end of last section, i.e.,
$M\sim\left(\begin{smallmatrix}0&X^\dagger\\X&0
\end{smallmatrix}\right)$
with $X\in \text{Hom}(W_+,W_-)$.
The resolution of singularities is given by
\[
Q_N^{\mathrm{m}}=\{(V_+,V_-,X)\ |\ V_+^\perp=V_+,\ V_-^\perp=V_-,\
\Im X^\dagger \subset V_+ \subset \Ker X,\
\Im X \subset V_- \subset \Ker X^\dagger
\},
\]
and by localization one gets, writing $n=2m$,
\[
\mdeg D_N^{\mathrm{m}}
=\prod_{1\le i, j\le m} (A\pm z_i\pm z_j)
\sum_{\substack{\varepsilon\in\{+1,-1\}^m\\\varepsilon'\in\{+1,-1\}^m}}
\prod_{\substack{i,j=1\\i\le j}}^m
\frac{1}{(\varepsilon_i z_i+\varepsilon_j z_j)(\varepsilon'_i z_i+\varepsilon'_j z_j)}
\prod_{i,j=1}^{m}
\frac{1}{A-\varepsilon_i z_i-\varepsilon'_j z_j}.
\]
\end{itemize}

\begin{remark}
It was shown in \cite[section 7]{artic33} that the type $\hat{\mathrm A}$ localization formula can also be derived via an integral over the unitary group. Similar results can be obtained in type $\hat{\mathrm C}$, with integrals over the compact symplectic group.
\end{remark}


\subsection{Irreducible components}
\label{sec:irredcomp}
This section will outline the relationship between the irreducible components of the scheme $E_N^{\mathrm a}$ and the link patterns of the Brauer loop model with boundary condition $\mathrm{a}$. We recall the notion of link patterns introduced in
\secref{qKZ}, see Figures~\ref{fig:Alp} and \ref{fig:Clp}.
We now describe a map from link patterns of type $\hat{\mathrm C}$ to link patterns of type $\hat{\mathrm A}$ with certain symmetries.

Given a link pattern $\pi\in \mathrm{LP}_L^{\mathrm a}$, we define $\tilde \pi$ according to the following simple symmetry rules:
\begin{enumerate}
\item For a=i,c, we have $L=n=N/2$ and:
\begin{itemize}
\item If $\pi(j)=k\neq b$, then $\tilde\pi(j)=k$ and $\tilde\pi(N-j+1)=N-k+1$.
\item If $\pi(j)=b$, then $\tilde\pi(j)=N-j+1$.
\end{itemize}
\item For a=o,m, we have $L=m=N/4$ and:
\begin{itemize}
\item If $\pi(j)=k\neq l,r$, then $\tilde\pi(j)=k$, $\tilde\pi(n+j)=n+k$, $\tilde\pi(N-j+1)=N-k+1$, and $\tilde\pi(n-j+1)=n-k+1$.
\item If $\pi(j)=l$, then $\tilde\pi(j)=N-j+1$ and $\tilde\pi(n-j+1)=n+j$.
\item If $\pi(j)=r$, then $\tilde\pi(j)=n-j+1$ and $\tilde\pi(N-j+1)=n+j$.
\end{itemize}
\end{enumerate}
For example:
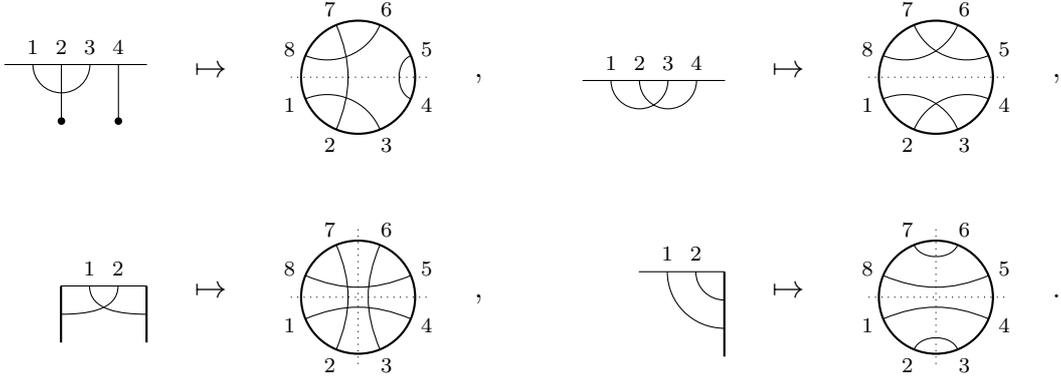
\begin{align*}
\begin{tikzpicture}[baseline=(current  bounding  box.center),scale=0.75]
\draw (0,1.5)--(2.5,1.5);
\draw[smooth] (0.5,1.5) to[out=270,in=180] (1,1) to[out=0,in=270] (1.5,1.5);
\draw (1,1.5) -- (1,0.5);\node[circle,fill=black,inner sep=0pt,minimum size=0.1cm] at (1,0.5) {};
\draw (2,1.5) -- (2,0.5);\node[circle,fill=black,inner sep=0pt,minimum size=0.1cm] at (2,0.5) {};
\node[above] at (0.5,1.5) {${\scriptstyle 1}$};
\node[above] at (1,1.5) {${\scriptstyle 2}$};
\node[above] at (1.5,1.5) {${\scriptstyle 3}$};
\node[above] at (2,1.5) {${\scriptstyle 4}$};
\end{tikzpicture}
\hspace{0.5cm}&\mapsto\hspace{0.5cm}
\begin{tikzpicture}[baseline=(current  bounding  box.center),scale=0.75]
\draw[thick] (0,0) circle(1);
\draw[smooth] ({5*360/8-360/16}:1) to[out={1*360/8-360/16},in={3*360/8-360/16}] ({7*360/8-360/16}:1);
\draw[smooth] ({6*360/8-360/16}:1) to[out={2*360/8-360/16},in={7*360/8-360/16}] ({3*360/8-360/16}:1);
\draw[smooth] ({8*360/8-360/16}:1) to[out={4*360/8-360/16},in={5*360/8-360/16}] ({1*360/8-360/16}:1);
\draw[smooth] ({2*360/8-360/16}:1) to[out={6*360/8-360/16},in={8*360/8-360/16}] ({4*360/8-360/16}:1);
\draw[dotted] (0:1.2) -- (180:1.2);
\node at ({5*360/8-360/16}:1.3) {${\scriptstyle 1}$};
\node at ({6*360/8-360/16}:1.3) {${\scriptstyle 2}$};
\node at ({7*360/8-360/16}:1.3) {${\scriptstyle 3}$};
\node at ({8*360/8-360/16}:1.3) {${\scriptstyle 4}$};
\node at ({1*360/8-360/16}:1.3) {${\scriptstyle 5}$};
\node at ({2*360/8-360/16}:1.3) {${\scriptstyle 6}$};
\node at ({3*360/8-360/16}:1.3) {${\scriptstyle 7}$};
\node at ({4*360/8-360/16}:1.3) {${\scriptstyle 8}$};
\end{tikzpicture}\quad,
&
\begin{tikzpicture}[baseline=(current  bounding  box.center),scale=0.75]
\draw (0,1.5)--(2.5,1.5);
\draw[smooth] (0.5,1.5) to[out=270,in=180] (1,1) to[out=0,in=270] (1.5,1.5);
\draw[smooth] (1,1.5) to[out=270,in=180] (1.5,1) to[out=0,in=270] (2,1.5);
\node[above] at (0.5,1.5) {${\scriptstyle 1}$};
\node[above] at (1,1.5) {${\scriptstyle 2}$};
\node[above] at (1.5,1.5) {${\scriptstyle 3}$};
\node[above] at (2,1.5) {${\scriptstyle 4}$};
\end{tikzpicture}
\hspace{0.5cm}&\mapsto\hspace{0.5cm}
\begin{tikzpicture}[baseline=(current  bounding  box.center),scale=0.75]
\draw[thick] (0,0) circle(1);
\draw[smooth] ({5*360/8-360/16}:1) to[out={1*360/8-360/16},in={3*360/8-360/16}] ({7*360/8-360/16}:1);
\draw[smooth] ({6*360/8-360/16}:1) to[out={2*360/8-360/16},in={4*360/8-360/16}] ({8*360/8-360/16}:1);
\draw[smooth] ({3*360/8-360/16}:1) to[out={7*360/8-360/16},in={5*360/8-360/16}] ({1*360/8-360/16}:1);
\draw[smooth] ({2*360/8-360/16}:1) to[out={6*360/8-360/16},in={8*360/8-360/16}] ({4*360/8-360/16}:1);
\draw[dotted] (0:1.2) -- (180:1.2);
\node at ({5*360/8-360/16}:1.3) {${\scriptstyle 1}$};
\node at ({6*360/8-360/16}:1.3) {${\scriptstyle 2}$};
\node at ({7*360/8-360/16}:1.3) {${\scriptstyle 3}$};
\node at ({8*360/8-360/16}:1.3) {${\scriptstyle 4}$};
\node at ({1*360/8-360/16}:1.3) {${\scriptstyle 5}$};
\node at ({2*360/8-360/16}:1.3) {${\scriptstyle 6}$};
\node at ({3*360/8-360/16}:1.3) {${\scriptstyle 7}$};
\node at ({4*360/8-360/16}:1.3) {${\scriptstyle 8}$};
\end{tikzpicture}\quad,\\[5mm]
\begin{tikzpicture}[baseline=(current  bounding  box.center),scale=0.75]
\draw (0,1)--(1.5,1);
\draw[thick] (1.5,1)--(1.5,0);
\draw[thick] (0,1)--(0,0);
\draw[smooth] (0.5,1) to[out=270,in=180] (1.5,0.5);
\draw[smooth] (1,1) to[out=270,in=0] (0,0.5);
\node[above] at (0.5,1) {${\scriptstyle 1}$};
\node[above] at (1,1) {${\scriptstyle 2}$};
\end{tikzpicture}
\hspace{0.5cm}&\mapsto\hspace{0.5cm}
\begin{tikzpicture}[baseline=(current  bounding  box.center),scale=0.75]
\draw[thick] (0,0) circle(1);
\draw[smooth] ({5*360/8-360/16}:1) to[out={1*360/8-360/16},in={4*360/8-360/16}] ({8*360/8-360/16}:1);
\draw[smooth] ({6*360/8-360/16}:1) to[out={2*360/8-360/16},in={7*360/8-360/16}] ({3*360/8-360/16}:1);
\draw[smooth] ({4*360/8-360/16}:1) to[out={8*360/8-360/16},in={5*360/8-360/16}] ({1*360/8-360/16}:1);
\draw[smooth] ({2*360/8-360/16}:1) to[out={6*360/8-360/16},in={3*360/8-360/16}] ({7*360/8-360/16}:1);
\draw[dotted] (0:1.2) -- (180:1.2);
\draw[dotted] (90:1.2) -- (270:1.2);
\node at ({5*360/8-360/16}:1.3) {${\scriptstyle 1}$};
\node at ({6*360/8-360/16}:1.3) {${\scriptstyle 2}$};
\node at ({7*360/8-360/16}:1.3) {${\scriptstyle 3}$};
\node at ({8*360/8-360/16}:1.3) {${\scriptstyle 4}$};
\node at ({1*360/8-360/16}:1.3) {${\scriptstyle 5}$};
\node at ({2*360/8-360/16}:1.3) {${\scriptstyle 6}$};
\node at ({3*360/8-360/16}:1.3) {${\scriptstyle 7}$};
\node at ({4*360/8-360/16}:1.3) {${\scriptstyle 8}$};
\end{tikzpicture}\quad,
&
\begin{tikzpicture}[baseline=(current  bounding  box.center),scale=0.75]
\draw (0,1.5)--(1.5,1.5);
\draw[thick] (1.5,1.5)--(1.5,0);
\draw[smooth] (0.5,1.5) to[out=270,in=180] (1.5,0.5);
\draw[smooth] (1,1.5) to[out=270,in=180] (1.5,1);
\node[above] at (0.5,1.5) {${\scriptstyle 1}$};
\node[above] at (1,1.5) {${\scriptstyle 2}$};
\end{tikzpicture}
\hspace{0.5cm}&\mapsto\hspace{0.5cm}
\begin{tikzpicture}[baseline=(current  bounding  box.center),scale=0.75]
\draw[thick] (0,0) circle(1);
\draw[smooth] ({5*360/8-360/16}:1) to[out={1*360/8-360/16},in={4*360/8-360/16}] ({8*360/8-360/16}:1);
\draw[smooth] ({6*360/8-360/16}:1) to[out={2*360/8-360/16},in={3*360/8-360/16}] ({7*360/8-360/16}:1);
\draw[smooth] ({4*360/8-360/16}:1) to[out={8*360/8-360/16},in={5*360/8-360/16}] ({1*360/8-360/16}:1);
\draw[smooth] ({2*360/8-360/16}:1) to[out={6*360/8-360/16},in={7*360/8-360/16}] ({3*360/8-360/16}:1);
\draw[dotted] (0:1.2) -- (180:1.2);
\draw[dotted] (90:1.2) -- (270:1.2);
\node at ({5*360/8-360/16}:1.3) {${\scriptstyle 1}$};
\node at ({6*360/8-360/16}:1.3) {${\scriptstyle 2}$};
\node at ({7*360/8-360/16}:1.3) {${\scriptstyle 3}$};
\node at ({8*360/8-360/16}:1.3) {${\scriptstyle 4}$};
\node at ({1*360/8-360/16}:1.3) {${\scriptstyle 5}$};
\node at ({2*360/8-360/16}:1.3) {${\scriptstyle 6}$};
\node at ({3*360/8-360/16}:1.3) {${\scriptstyle 7}$};
\node at ({4*360/8-360/16}:1.3) {${\scriptstyle 8}$};
\end{tikzpicture}\quad.
\end{align*}
It is $\tilde\pi$, rather than $\pi$, that will appear naturally in the geometry, e.g.,
for defining the irreducible components,
and as the distinction between $\pi$ and $\tilde \pi$
is either irrelevant or obvious from context, we will drop the $\ \tilde{}\ $ notation.
As to the pictorial description,
we shall simply call this new representation the periodic diagram of the link pattern $\pi$.

Define $s_i(M)=\sum_{j=i}^{i+N} M_{ij}M_{j,i+N}$ for $M\in E^{\mathrm p}_N$ (note that this is well-defined despite the quotient by $\left<S^N\right>$).

Given a matrix $M$ of size $N$, we define the so-called rank matrix $\rkm(M)$ as
\[ \rkm(M)_{ij}:=\rk{M_{i:N,1:j}}, \]
where $M_{i:N,1:j}$ denotes the submatrix south-west of entry $(i,j)$.

We recall from now on that detailed proofs are only provided for $\mathrm a\in\{\mathrm i,\mathrm c\}$, and occasionally for $\mathrm a=\mathrm p$ when it is not already present in \cite{artic33,artic39}.

\subsubsection{Irreducible components of \texorpdfstring{$E^{\mathrm a}_N$}{scheme} for identified and closed boundaries}\label{sec:irrcomp}
In type $\hat{\mathrm A}$, it is known that $E_N^{\mathrm{p}}$ is equidimensional of dimension $2n^2$, and that $E_N^{\mathrm p}$ decomposes into its irreducible components $E_\pi^{\mathrm p}$ indexed by periodic link patterns \cite{artic33,Rothbach}.

In this section we will prove a similar statement: that the top-dimensional irreducible components of $E_N^{\mathrm{i}}$ and $E_N^{\mathrm{c}}$ are indexed by link patterns in $\mathrm{LP}^{\mathrm i}_n$ and $\mathrm{LP}^{\mathrm c}_n$ respectively. (As mentioned previously, equidimensionality is only conjectured for these two cases.) We find that the dimensions of the two schemes are
\begin{align*}
\dim(E_N^{\mathrm i})&=n(n+1),\\
\dim(E_N^{\mathrm c})&=2\left\lfloor\frac{n^2}{2}\right\rfloor.
\end{align*}
We also show that the top-dimensional components are generically reduced.

We conjecture that the same statement is true for $E_N^{\mathrm{o}}$ and $E_N^{\mathrm{m}}$, and that their respective dimensions are
\begin{align*}
\dim(E^{\mathrm o}_N)=2m(m+1),\\
\dim(E^{\mathrm m}_N)=m(2m+1).
\end{align*}
However the inductive proof of \thmref{Borbits} (using an appropriately defined $\ttilde B_N$) presented technical challenges that obstructed the proofs of \thmref{maxF} and \thmref{genred}. Thus for the rest of this section $\mathrm a\in\{\mathrm p,\mathrm i,\mathrm c\}$ unless stated otherwise.

If $M\in E^{\mathrm a}_N$, recall from \secref{orbclos} that we can write $M$ as a pair $(\U,\La)$, where $\U$ belongs to $\mathcal{O}^{\mathrm a}_N=\left\{ \U\in R^{\mathrm a}_N\ |\ \U^2=0 \right\}$, and $\La\in\g_N^{\mathrm a}/R_N^{\mathrm a}$ such that $\La\U+\U\La$ (i.e., its strict lower triangular part) is $0$. We will assume the diagonal of $M$ to be zero (i.e., consider the reduced scheme $E^{\mathrm a}_N$).

We define the Borel subgroups
\begin{align*}
B_N&:=\{ M\in GL_N\ |\ M_{ij}=0,\ j<i \},\\
\tilde B_N&:=\{ M\in B_N\ |\ M^{-1}=M^\dag \},
\end{align*}
where $M^\dag=J^{-1}M^T J$ with the symplectic form as given in \eqref{eq:Jfin}. Acting by conjugation with $B_N$ leaves $R^{\mathrm p}_N$ and therefore $\mathcal{O}^{\mathrm p}_N$ invariant, and similarly for $\tilde B_N$, $R^{\mathrm i,c}_N$ and $\mathcal{O}^{\mathrm i,c}_N$. We will use the notation $\cdot$ for conjugation.

\begin{defn}
\begin{align*}
\mathrm{Inv}^{\mathrm i}_N &:= \{ \pi\text{ involution of } \{1,\ldots,N\}\ |\ \pi(i)=N-\pi_{N-i+1}+1 \},\\
\mathrm{Inv}^{\mathrm c}_N &:= \{ \pi\in \mathrm{Inv}^{\mathrm i}_N \ |\ \pi(i)\neq N-i+1\ \forall i \}.
\end{align*}
\end{defn}
\begin{exmp}
\begin{align*}
\mathrm{Inv}^{\mathrm i}_4&=\left\{\ \begin{tikzpicture}[baseline=-3pt,scale=0.5]
\draw[thick] (0,0) circle(1);
\node at (225:1) {${\scriptstyle \bullet}$};
\node at (315:1) {${\scriptstyle \bullet}$};
\node at (45:1) {${\scriptstyle \bullet}$};
\node at (135:1) {${\scriptstyle \bullet}$};
\node at (225:1.4) {${\scriptstyle 1}$};
\node at (315:1.4) {${\scriptstyle 2}$};
\node at (45:1.4) {${\scriptstyle 3}$};
\node at (135:1.4) {${\scriptstyle 4}$};
\end{tikzpicture},\;
\begin{tikzpicture}[baseline=-3pt,scale=0.5]
\draw[thick] (0,0) circle(1);
\node at (315:1) {${\scriptstyle \bullet}$};
\node at (45:1) {${\scriptstyle \bullet}$};
\node at (225:1.4) {${\scriptstyle 1}$};
\node at (315:1.4) {${\scriptstyle 2}$};
\node at (45:1.4) {${\scriptstyle 3}$};
\node at (135:1.4) {${\scriptstyle 4}$};
\draw[smooth] (225:1) to[out=45,in=315] (135:1);
\end{tikzpicture},\;
\begin{tikzpicture}[baseline=-3pt,scale=0.5]
\draw[thick] (0,0) circle(1);
\node at (225:1) {${\scriptstyle \bullet}$};
\node at (135:1) {${\scriptstyle \bullet}$};
\node at (225:1.4) {${\scriptstyle 1}$};
\node at (315:1.4) {${\scriptstyle 2}$};
\node at (45:1.4) {${\scriptstyle 3}$};
\node at (135:1.4) {${\scriptstyle 4}$};
\draw[smooth] (315:1) to[out=135,in=225] (45:1);
\end{tikzpicture},\;
\begin{tikzpicture}[baseline=-3pt,scale=0.5]
\draw[thick] (0,0) circle(1);
\node at (225:1.4) {${\scriptstyle 1}$};
\node at (315:1.4) {${\scriptstyle 2}$};
\node at (45:1.4) {${\scriptstyle 3}$};
\node at (135:1.4) {${\scriptstyle 4}$};
\draw[smooth] (225:1) to[out=45,in=315] (135:1);
\draw[smooth] (315:1) to[out=135,in=225] (45:1);
\end{tikzpicture},\;
\begin{tikzpicture}[baseline=-3pt,scale=0.5]
\draw[thick] (0,0) circle(1);
\node at (225:1.4) {${\scriptstyle 1}$};
\node at (315:1.4) {${\scriptstyle 2}$};
\node at (45:1.4) {${\scriptstyle 3}$};
\node at (135:1.4) {${\scriptstyle 4}$};
\draw[smooth] (225:1) to[out=45,in=135] (315:1);
\draw[smooth] (135:1) to[out=315,in=225] (45:1);
\end{tikzpicture},\;
\begin{tikzpicture}[baseline=-3pt,scale=0.5]
\draw[thick] (0,0) circle(1);
\node at (225:1.4) {${\scriptstyle 1}$};
\node at (315:1.4) {${\scriptstyle 2}$};
\node at (45:1.4) {${\scriptstyle 3}$};
\node at (135:1.4) {${\scriptstyle 4}$};
\draw (225:1) -- (45:1);
\draw (315:1) -- (135:1);
\end{tikzpicture}\ \right\},\\
\mathrm{Inv}^{\mathrm c}_4&=\left\{\ \begin{tikzpicture}[baseline=-3pt,scale=0.5]
\draw[thick] (0,0) circle(1);
\node at (225:1) {${\scriptstyle \bullet}$};
\node at (315:1) {${\scriptstyle \bullet}$};
\node at (45:1) {${\scriptstyle \bullet}$};
\node at (135:1) {${\scriptstyle \bullet}$};
\node at (225:1.4) {${\scriptstyle 1}$};
\node at (315:1.4) {${\scriptstyle 2}$};
\node at (45:1.4) {${\scriptstyle 3}$};
\node at (135:1.4) {${\scriptstyle 4}$};
\end{tikzpicture},\;
\begin{tikzpicture}[baseline=-3pt,scale=0.5]
\draw[thick] (0,0) circle(1);
\node at (225:1.4) {${\scriptstyle 1}$};
\node at (315:1.4) {${\scriptstyle 2}$};
\node at (45:1.4) {${\scriptstyle 3}$};
\node at (135:1.4) {${\scriptstyle 4}$};
\draw[smooth] (225:1) to[out=45,in=135] (315:1);
\draw[smooth] (135:1) to[out=315,in=225] (45:1);
\end{tikzpicture},\;
\begin{tikzpicture}[baseline=-3pt,scale=0.5]
\draw[thick] (0,0) circle(1);
\node at (225:1.4) {${\scriptstyle 1}$};
\node at (315:1.4) {${\scriptstyle 2}$};
\node at (45:1.4) {${\scriptstyle 3}$};
\node at (135:1.4) {${\scriptstyle 4}$};
\draw (225:1) -- (45:1);
\draw (315:1) -- (135:1);
\end{tikzpicture}\ \right\}.
\end{align*}

\end{exmp}
The following definitions give unique matrix representations of the involutions defined above.
\begin{defn}
For $\pi\in\mathrm{Inv}^{\mathrm i}_N$ we define the matrix $\pi^{\mathrm i}_<\in \mathcal{O}_N^{\mathrm i}$ as
\[ \left(\pi^{\mathrm i}_<\right)_{ij}:=\begin{cases}
                             0 & j\leq i,\\
                             -\delta_{i,\pi(j)} & n<i<j,\\
                             \delta_{i,\pi(j)} & \text{else}.
                             \end{cases}
\]

For $\pi\in\mathrm{Inv}^{\mathrm c}_N$ we define $\pi^{\mathrm c}_<\in \mathcal{O}_N^{\mathrm c}$ as
\[ \left(\pi^{\mathrm c}_<\right)_{ij}:=\begin{cases}
                             0 & j\leq i,\\
                             0 & j+i=N+1,\\
                             -\delta_{i,\pi(j)} & N-j+1<i\leq n.\\
                             \delta_{i,\pi(j)} & \text{else}.
                             \end{cases}
\]
\end{defn}
\begin{exmp}
\[ \qquad (351624)^{\mathrm i}_< = \begin{pmatrix}
                 0 & 0 & 1 & 0 & 0 & 0 \\
                 0 & 0 & 0 & 0 & 1 & 0 \\
                 0 & 0 & 0 & 0 & 0 & 0 \\
                 0 & 0 & 0 & 0 & 0 & -1 \\
                 0 & 0 & 0 & 0 & 0 & 0 \\
                 0 & 0 & 0 & 0 & 0 & 0 \\
                 \end{pmatrix},\quad (563421)^{\mathrm c}_< = \begin{pmatrix}
                 0 & 0 & 0 & 0 & 1 & 0 \\
                 0 & 0 & 0 & 0 & 0 & -1 \\
                 0 & 0 & 0 & 0 & 0 & 0 \\
                 0 & 0 & 0 & 0 & 0 & 0 \\
                 0 & 0 & 0 & 0 & 0 & 0 \\
                 0 & 0 & 0 & 0 & 0 & 0 \\
                 \end{pmatrix}.
                 \]
\end{exmp}

\begin{thm}
\label{thm:Borbits}
For $\mathrm a\in \{\mathrm i,\mathrm c\}$, each $\tilde B_N$-orbit of $\mathcal{O}_N^{\mathrm a}$ contains exactly one $\pi^{\mathrm a}_<$ where $\pi\in\mathrm{Inv}^{\mathrm a}_N$. Orbits are thus naturally labelled by these involutions.
\end{thm}
\begin{remark}
This is a modification of the main theorem in \cite{Meln}, which applies to $\mathrm a=\mathrm p$, and the proof follows that given there.
\end{remark}
\begin{proof}
We use induction from size $N-2$ to size $N$. For $N=2$ we could have that $\mathcal{O}^{\mathrm a}_2 \ni X=0$ (which is trivial), or, for $\mathrm{a}=\mathrm{i}$, the upper-right entry of $X$ is nonzero. The latter case is conjugate to $(2,1)^{\mathrm i}_<$ by a diagonal matrix.

We now consider general $N$. For any matrix $X\in \mathcal{O}^{\mathrm a}_N$, we can form the matrix $\hat X\in \mathcal{O}^{\mathrm a}_{N-2}$ by truncating the first and last row and column of $X$. Assuming that the claim is true for $N-2$, there is a unique $\hat\pi\in\mathrm{Inv}^{\mathrm a}_{N-2}$ such that $\hat w:=\hat\pi^{\mathrm a}_<=\hat U\hat X\hat U^{-1}$ for some $\hat U\in \tilde B_{N-2}$. Define $U_0\in \tilde B_N$ as
\[ U_0:=\begin{pmatrix}
     1 &        & \\
       & \hat U & \\
       &        & 1
     \end{pmatrix}.
\]
The matrix $Y=U_0XU_0^{-1}$ has its middle $N-2\times N-2$ block equal to $\hat w$. Its first row is free (not including entry $(1,1)$, which equals 0), and its last column is decided from the first row by the symmetry of $\mathcal{O}^{\mathrm a}_N$. Our aim is to find a transformation matrix $T\in \tilde B_N$ so that $TYT^{-1}=\pi^{\mathrm a}_<=:w$ for a unique $\pi\in\mathrm{Inv}^{\mathrm a}_N$. We note that rank is preserved by conjugation, so $\rk{Y}=\rk{X}=\rk{w}$.

We now define $U_1\in \tilde B_N$ with first row
\[ \begin{pmatrix}
   1 & \left[\displaystyle{-\sum_{s=2}^{N-1} \hat w_{j-1,s-1}\ Y_{1,s}}\right]_{2<j<N-1} & 0
   \end{pmatrix},
\]
middle $N-2\times N-2$ block equal to the $N-2$ identity matrix, and other entries decided by the symmetry of $\tilde B_N$. Let $Z=U_1YU_1^{-1}$. If $\rk{Y}=\rk{\hat X}$, then $\rk{w}=\rk{\hat w}$, so the first and last row and column of $w$ must be zero. Then $T=U_1$, and $Z=w$.

If $\rk{Y}=\rk{\hat X}+1$ (this is only possible for $\mathrm{a}=\mathrm{i}$), then $\rk{w}=\rk{\hat w}+1$, so $w_{1N}=1$ and all other extra entries must be zero. We define $U_2\in \tilde B_N$ as
\[ U_2:=\begin{pmatrix}
     \frac1{\sqrt{Z_{1N}}} &         & \\
                           & I_{N-2} & \\
                           &         & \sqrt{Z_{1N}}\\
     \end{pmatrix},
\]
then $w=U_2ZU_2^{-1}$ and $T=U_2U_1$.

Finally, if $\rk{Y}=\rk{\hat X}+2$, we define $k$ to be the column index of the first nonzero entry in the first row of $Z$. We define $U_3\in \tilde B_N$ as having diagonal $1$ except for
\[ (U_3)_{11}=\frac1{Z_{1k}},\qquad (U_3)_{NN}=Z_{1k}, \]
and $k$th row given by
\[ (U_3)_{kj}=\frac{Z_{1j}}{Z_{1k}},\quad k<j<N,\qquad (U_3)_{kN}=\frac{Z_{1N}}{2Z_{1k}}. \]
By the symmetry of $\tilde B_N$ the $(N-k+1)$th column is also nontrivial:
\[ (U_3)_{ik}=\begin{cases}
              \frac{-Z_{1N}}{2Z_{1k}^2} & k>n,\ i=1,\\
              \frac{-Z_{1,N-i+1}}{Z_{1k}} & k>n,\ 1<i<N-k+1,\\
              \frac{Z_{1N}}{2Z_{1k}^2} & k\leq n,\ i=1,\\
              \frac{Z_{1,N-i+1}}{Z_{1k}} & k\leq n,\ 1<i\leq n,\\
              \frac{-Z_{1,N-i+1}}{Z_{1k}} & k\leq n,\ n<i<N-k+1,
              \end{cases}
\]
and all other entries of $U_3$ are zero. Then $w=U_3ZU_3^{-1}$ and $T=U_3U_1$.
\end{proof}

As a consequence, $\tilde B_N\cdot E^{\mathrm a}_N$ for $\mathrm{a}\in\{\mathrm{i},\mathrm{c}\}$ breaks into disjoint components, labelled by involutions. We denote these by $F^{\mathrm a}_\pi$:
\begin{defn}
\[ F^{\mathrm a}_\pi:=\left\{ M=(\U,\La)\in E^{\mathrm a}_N\ \middle|\ \exists U\in \tilde B_N : U\U U^{-1}=\pi^{\mathrm a}_< \right\}. \]
\end{defn}

\begin{thm}
\label{thm:maxF}
The sets $F^{\mathrm a}_\pi$ with the highest dimension have $\pi$ corresponding to a link pattern.
\end{thm}
\begin{remark}
This is the analog of \cite[Theorem~3]{artic33} for $\mathrm a=\mathrm p$.
\end{remark}
\begin{proof}
We have
\[ \dim(F^{\mathrm a}_\pi)=\dim(\tilde B_N\cdot \pi^{\mathrm a}_<)+\dim\{\La\in\g_N^{\mathrm a}/R_N^{\mathrm a}\ |\ (\pi^{\mathrm a}_< \La+\La\pi^{\mathrm a}_<)_>=0\}. \]
To obtain the dimension of the $\tilde B_N$-orbit, we calculate the dimension of the tangent space at $\pi^{\mathrm a}_<$. This is $\{(U\pi^{\mathrm a}_<-\pi^{\mathrm a}_<U)\}$, where $U$ is in the Lie algebra of $\tilde B_N$; that is, $U$ is weakly upper-triangular with $U=-U^\dag$.

The second term is
\[ \dim\{\La\in\g_N^{\mathrm a}/R_N^{\mathrm a}\ |\ (\pi^{\mathrm a}_< \La+\La\pi^{\mathrm a}_<)_>=0\}=\dim(\g_N^{\mathrm a}/R_N^{\mathrm a})-\dim\{(\pi^{\mathrm a}_< \La+\La\pi^{\mathrm a}_<)_>\ |\ \La\in\g_N^{\mathrm a}/R_N^{\mathrm a}\}. \]
The dimension of $\g_N^{\mathrm a}/R_N^{\mathrm a}$ is $n^2$ for $\mathrm{a}=\mathrm{i}$ and $n(n-1)$ for $\mathrm{a}=\mathrm{c}$.

To calculate the dimensions of $\{(U\pi^{\mathrm a}_<-\pi^{\mathrm a}_<U)\}$ and $\{(\pi^{\mathrm a}_< \La+\La\pi^{\mathrm a}_<)_>\}$, we note that no more than two entries of $\pi^{\mathrm a}_<$ will ever be involved in calculating a single entry of the matrices. Thus we can do the calculations for $N=2,4,6,8$ by brute force, and the results will extend easily to larger sizes. We find

\begin{align*}
\dim(F^{\mathrm i}_\pi)&=n^2
+\#\begin{tikzpicture}[baseline=(current  bounding  box.center),scale=0.5]
\draw[thick] (0,0) circle(1);
\draw[smooth] (90:1) to[out=270,in=90] (270:1);
\node at (90:1.4) {$a'$};
\node at (270:1.4) {$a$};
\end{tikzpicture}
+\#\begin{tikzpicture}[baseline=(current  bounding  box.center),scale=0.5]
\draw[thick] (0,0) circle(1);
\draw[smooth] (45:1) to[out=225,in=45] (225:1);
\draw[smooth] (135:1) to[out=315,in=135] (315:1);
\node at (135:1.4) {$a'$};
\node at (225:1.4) {$a$};
\node at (45:1.4) {$b'$};
\node at (315:1.4) {$b$};
\end{tikzpicture}
+2\#\begin{tikzpicture}[baseline=(current  bounding  box.center),scale=0.5]
\draw[thick] (0,0) circle(1);
\draw[smooth] (45:1) to[out=225,in=315] (135:1);
\draw[smooth] (225:1) to[out=45,in=135] (315:1);
\node at (135:1.4) {$a'$};
\node at (225:1.4) {$a$};
\node at (45:1.4) {$b'$};
\node at (315:1.4) {$b$};
\end{tikzpicture}\ ,\\
\dim(F^{\mathrm c}_\pi)&=n(n-1)
+\#\begin{tikzpicture}[baseline=(current  bounding  box.center),scale=0.5]
\draw[thick] (0,0) circle(1);
\draw[smooth] (45:1) to[out=225,in=45] (225:1);
\draw[smooth] (135:1) to[out=315,in=135] (315:1);
\node at (135:1.4) {$a'$};
\node at (225:1.4) {$a$};
\node at (45:1.4) {$b'$};
\node at (315:1.4) {$b$};
\end{tikzpicture}
+2\#\begin{tikzpicture}[baseline=(current  bounding  box.center),scale=0.5]
\draw[thick] (0,0) circle(1);
\draw[smooth] (45:1) to[out=225,in=315] (135:1);
\draw[smooth] (225:1) to[out=45,in=135] (315:1);
\node at (135:1.4) {$a'$};
\node at (225:1.4) {$a$};
\node at (45:1.4) {$b'$};
\node at (315:1.4) {$b$};
\end{tikzpicture}\ .
\end{align*}
The highest dimension is thus
\begin{align*}
\max(\dim(F^{\mathrm i}_{\pi}))&=n(n+1),\\
\max(\dim(F^{\mathrm c}_{\pi}))&=2\floor*{\frac{n^2}{2}},
\end{align*}
which occurs if $\pi$ has as few fixed points as possible and no instances of $\begin{tikzpicture}[baseline=(current  bounding  box.center),scale=0.5]
\draw[thick] (0,0) circle(1);
\draw[smooth] (45:1) to[out=225,in=45] (225:1);
\draw[smooth] (135:1) to[out=315,in=135] (315:1);
\node at (135:1.4) {$a'$};
\node at (225:1.4) {$a$};
\node at (45:1.4) {$b'$};
\node at (315:1.4) {$b$};
\end{tikzpicture}$. These $\pi$ correspond to link patterns.
\end{proof}
\begin{cor}\label{cor:dimoneless}
The sets $F^{\mathrm a}_\pi$ with highest dimension minus one, i.e.~$\dim(F^{\mathrm p}_{\pi})=2n^2-1$, $\dim(F^{\mathrm i}_{\pi})=n(n+1)-1$, $\dim(F^{\mathrm c}_{\pi})=2\floor*{\frac{n^2}{2}}-1$, have $\pi$ being an involution that looks like a link pattern except for:
\begin{align*}
\mathrm a=\mathrm p:&\qquad \text{one pair of fixed points,}\\
\mathrm a=\mathrm i:&\qquad \text{one instance of} \begin{tikzpicture}[baseline=(current  bounding  box.center),scale=0.5]
\draw[thick] (0,0) circle(1);
\draw[smooth] (45:1) to[out=225,in=45] (225:1);
\draw[smooth] (135:1) to[out=315,in=135] (315:1);
\node at (135:1.4) {$a'$};
\node at (225:1.4) {$a$};
\node at (45:1.4) {$b'$};
\node at (315:1.4) {$b$};
\end{tikzpicture}
\text{, or one pair of fixed points at $a$ and $a'$,}\\
\mathrm a=\mathrm c:&\qquad \text{one instance of} \begin{tikzpicture}[baseline=(current  bounding  box.center),scale=0.5]
\draw[thick] (0,0) circle(1);
\draw[smooth] (45:1) to[out=225,in=45] (225:1);
\draw[smooth] (135:1) to[out=315,in=135] (315:1);
\node at (135:1.4) {$a'$};
\node at (225:1.4) {$a$};
\node at (45:1.4) {$b'$};
\node at (315:1.4) {$b$};
\end{tikzpicture}.
\end{align*}
\end{cor}

\begin{defn}
For $\pi$ a link pattern $\in\mathrm{LP}_L^{\mathrm a}$, we define $E^{\mathrm a}_\pi$ as the closure of $F^{\mathrm a}_\pi$.
\end{defn}
The projection $(\U,\La)\mapsto \U$ makes $F_\pi$ a vector bundle over the orbit $\tilde B_N\cdot \pi^{\mathrm a}_<$.
As the closure of a vector bundle over the orbit of a connected group,
$E^{\mathrm a}_\pi$ is irreducible. \thmref{maxF} says that the $E^{\mathrm a}_\pi$ are the
top-dimensional components of $E^{\mathrm a}_N$ (we conjecture
that there are no other components; this can be probably be proved
in a similar way as in the periodic case \cite{Rothbach}).

\begin{thm}
\label{thm:genred}
Each $E^{\mathrm a}_\pi$ is generically reduced (as a component of $E^{\mathrm a}_N$).
\end{thm}
\begin{remark}
This proof is similar to that of \cite[Theorem~4]{artic33}.
\end{remark}
\begin{proof}
We need to show that the Zariski tangent space at a generic point has the same dimension as $E^{\mathrm a}_N$. The tangent space is given by the set of all matrices $P\in\left(\M_N^{\mathrm{a}}\right)_{\Delta=0}$ that satisfy the derivative of the defining equation $M^2=0$:
\[ P\underline{\pi}t+\underline{\pi}tP=0, \]
where $\underline{\pi}$ is the matrix representation of $\pi$ (with diagonal zeroed out) that belongs to $E^{\mathrm a}_N$, and $\underline{\pi}t$ is $\underline{\pi}$ multiplied by a generic diagonal matrix with restrictions necessary for the result to belong to $\M_N^{\mathrm{a}}$. We note that both $P$ and $\underline\pi t$ have zero diagonal. To compute the dimension of this set we consider the individual matrix components of the above equation:
\begin{equation}
\label{eq:Ppit0}
0=\left(P\underline{\pi}t+\underline{\pi}tP\right)_{ij}=P_{i\pi(j)}t_j[i<\pi(j)<j<i+N]+P_{\pi(i)j}t_{\pi(i)}[i<\pi(i)<j<i+N],
\end{equation}
where $[a]$ stands for $1$ if $a$ is true and $0$ if $a$ is false. In this equation, only two loops ever interact, so we only need to consider small size examples (up to $N=8$). For $\mathrm{a}=\mathrm{i}$ there are $7$ base cases that need to be considered. We will give $4$ example calculations for $\mathrm{a}=\mathrm{i}$ here, the rest are similar.

We first note that due to symmetry, the RHS of \eqref{eq:Ppit0} is the same for $(i,j)$ as for $(j',i')$, meaning that only one of these will contribute to the dimension. We also note that if $j=i'$, the RHS is automatically zero due to symmetry. Thus we will only give nontrivial equations where $j<i'$.
\begin{enumerate}
\item \[\pi=\begin{tikzpicture}[baseline=(current  bounding  box.center),scale=0.6]
\draw[thick] (0,0) circle(1);
\draw[smooth] (45:1) to[out=225,in=135] (315:1);
\draw[smooth] (225:1) to[out=45,in=315] (135:1);
\node at (225:1.3) {${\scriptstyle a}$};
\node at (315:1.3) {${\scriptstyle b}$};
\node at (45:1.3) {${\scriptstyle b'}$};
\node at (135:1.3) {${\scriptstyle a'}$};
\end{tikzpicture} \qquad\qquad\qquad
\begin{tabular}{lll}
$i$  & $j$  & \\
$a$  & $b'$\hspace{1cm} & $0=P_{ab}t_{b'}$\\
$b$  & $a$             & $0=P_{ba'}t_{a}+P_{b'a}t_{b'}$\\
$b'$ & $a$              & $0=P_{b'a'}t_{a}$
\end{tabular}
\]
After applying the known symmetries of $P_{ij}$ and $t_j$, there are only $2$ independent equations in this list.
\item \[\pi=\begin{tikzpicture}[baseline=(current  bounding  box.center),scale=0.6]
\draw[thick] (0,0) circle(1);
\draw[smooth] (45:1) to[out=225,in=315] (135:1);
\draw[smooth] (315:1) to[out=135,in=45] (225:1);
\node at (225:1.3) {${\scriptstyle a}$};
\node[right] at (303:1.1) {${\scriptstyle \pi(a)}$};
\node[right] at (57:1.1) {${\scriptstyle \pi(a')}$};
\node at (135:1.3) {${\scriptstyle a'}$};
\end{tikzpicture} \qquad\qquad\qquad
\begin{tabular}{lll}
$i$  & $j$  & \\
$a$  & $\pi(a')$\hspace{1cm} & $0=P_{\pi(a)\pi(a')}t_{\pi(a)}$\\
$\pi(a')$ & $a$              & $0=P_{a'a}t_{a'}$
\end{tabular}
\]
These $2$ equations are independent.
\item \[\pi=\begin{tikzpicture}[baseline=(current  bounding  box.center),scale=0.6]
\draw[thick] (0,0) circle(1);
\draw[smooth] (30:1) to[out=210,in=330] (150:1);
\draw[smooth] (90:1) to[out=270,in=90] (270:1);
\draw[smooth] (210:1) to[out=30,in=150] (330:1);
\node at (210:1.3) {${\scriptstyle a}$};
\node at (270:1.3) {${\scriptstyle b}$};
\node[right] at (320:1) {${\scriptstyle \pi(a)}$};
\node[right] at (40:1) {${\scriptstyle \pi(a')}$};
\node at (90:1.3) {${\scriptstyle b'}$};
\node at (150:1.3) {${\scriptstyle a'}$};
\end{tikzpicture} \qquad\qquad\qquad
\begin{tabular}{lll}
$i$       & $j$  & \\
$a$       & $b'$                  & $0=P_{ab} t_{b'} +P_{\pi(a)b'} t_{\pi(a)}$\\
$b$       & $a$                   & $0=P_{b\pi(a)} t_a +P_{b'a} t_{b'}$\\
$\pi(a)$  & $b$                   & $0=P_{\pi(a)b'} t_{b} +P_{ab} t_{a}$\\
$\pi(a')$ & $b$                   & $0=P_{\pi(a')b'} t_{b} +P_{a'b} t_{a'}$\\
\end{tabular}
\]
Here we have only included those equations that don't already appear in the previous example, thus for every pair of chords that are mirror images we must add $2$ equations. These $4$ equations are independent, so in total there are $6$ equations in this example.
\item \[ \pi=\begin{tikzpicture}[baseline=(current  bounding  box.center),scale=0.6]
\draw[thick] (0,0) circle(1);
\draw[smooth] (337.5:1) to[out=157.5,in=67.5] (247.5:1);
\draw[smooth] (292.5:1) to[out=112.5,in=22.5] (202.5:1);
\draw[smooth] (157.5:1) to[out=337.5,in=247.5] (67.5:1);
\draw[smooth] (22.5:1) to[out=202.5,in=292.5] (112.5:1);
\node at (202.5:1.3) {${\scriptstyle a}$};
\node at (247.5:1.3) {${\scriptstyle b}$};
\node at (302.5:1.5) {${\scriptstyle \pi(a)}$};
\node[right] at (332.5:1) {${\scriptstyle \pi(b)}$};
\node[right] at (27.5:1) {${\scriptstyle \pi(b')}$};
\node at (57.5:1.4) {${\scriptstyle \pi(a')}$};
\node at (112.5:1.3) {${\scriptstyle b'}$};
\node at (157.5:1.3) {${\scriptstyle a'}$};
\end{tikzpicture} \qquad\qquad\qquad
\begin{tabular}{lll}
$i$       & $j$  & \\
$a$       & $\pi(b)$              & $0=P_{ab} t_{\pi(b)} +P_{\pi(a)\pi(b)} t_{\pi(a)}$ \\
$a$       & $\pi(b')$             & $0=P_{\pi(a)\pi(b')} t_{\pi(a)}$ \\
$a$       & $b'$                  & $0=P_{a\pi(b')} t_{b'} +P_{\pi(a)b'} t_{\pi(a)}$ \\
$b$       & $a$                   & $0=P_{b\pi(a)} t_{a} +P_{\pi(b)a} t_{\pi(b)}$ \\
$\pi(a)$  & $b'$                  & $0=P_{\pi(a)\pi(b')} t_{b'}$ \\
$\pi(a)$  & $b$                   & $0=P_{\pi(a)\pi(b)} t_{b} +P_{ab} t_{a}$ \\
$\pi(b)$  & $\pi(a)$              & $0=P_{\pi(b)a} t_{\pi(a)} +P_{b\pi(a)} t_{b}$ \\
$\pi(b')$ & $a$                   & $0=P_{b'a} t_{b'}$ \\
$\pi(b')$ & $\pi(a)$              & $0=P_{\pi(b')a} t_{\pi(a)} +P_{b'\pi(a)} t_{b'}$ \\
$b'$      & $\pi(a)$              & $0=P_{b'a} t_{\pi(a)}$
\end{tabular}
\]
There are $8$ independent equations in this list, and as before we must add $4$ equations, giving a total of $12$.
\end{enumerate}
The other cases can be treated in the same way, and we find that there are $2$ equations for every pair of chords in the {\em periodic}\/ diagram of $\pi$, which comes to $n(n-1)$ equations. The dimension of the larger space $\left(\M_N^{\mathrm{i}}\right)_{\Delta=0}$ is $2n^2$, thus the dimension of the Zariski tangent space is $2n^2-n(n-1)=n(n+1)$, the same as $F^{\mathrm i}_\pi$.

For $\mathrm{a}=\mathrm{c}$ there are $6$ base cases to consider. Recall that the symmetry implies that $P_{ii'}=0$ and $t_i=0$ if $\pi(i)=i'$. We give here two examples.
\begin{enumerate}
\item \[ \pi=\begin{tikzpicture}[baseline=(current  bounding  box.center),scale=0.6]
\draw[thick] (0,0) circle(1);
\draw[smooth] (30:1) to[out=210,in=330] (150:1);
\draw[smooth] (210:1) to[out=30,in=150] (330:1);
\node at (90:1) {$\bullet$};
\node at (270:1) {$\bullet$};
\node at (210:1.3) {${\scriptstyle a}$};
\node at (270:1.4) {${\scriptstyle b}$};
\node[right] at (320:1) {${\scriptstyle \pi(a)}$};
\node[right] at (40:1) {${\scriptstyle \pi(a')}$};
\node at (90:1.4) {${\scriptstyle b'}$};
\node at (150:1.3) {${\scriptstyle a'}$};
\end{tikzpicture} \qquad\qquad\qquad
\begin{tabular}{lll}
$i$       & $j$  & \\
$a$       & $b'$\hspace{1cm}        & $0=P_{\pi(a)b'} t_{\pi(a)}$\\
$b$       & $a$                   & $0=P_{b\pi(a)} t_a$\\
$\pi(a)$  & $b$                   & $0=P_{ab} t_{a}$\\
$\pi(a')$ & $b$                   & $0=P_{a'b} t_{a'}$\\
\end{tabular}
\]
These $4$ equations are all independent.
\item \[ \pi=\begin{tikzpicture}[baseline=(current  bounding  box.center),scale=0.6]
\draw[thick] (0,0) circle(1);
\draw[smooth] (337.5:1) to[out=157.5,in=67.5] (247.5:1);
\draw[smooth] (292.5:1) to[out=112.5,in=22.5] (202.5:1);
\draw[smooth] (157.5:1) to[out=337.5,in=247.5] (67.5:1);
\draw[smooth] (22.5:1) to[out=202.5,in=292.5] (112.5:1);
\node at (202.5:1.3) {${\scriptstyle a}$};
\node at (247.5:1.3) {${\scriptstyle b}$};
\node at (302.5:1.5) {${\scriptstyle \pi(a)}$};
\node[right] at (332.5:1) {${\scriptstyle \pi(b)}$};
\node[right] at (27.5:1) {${\scriptstyle \pi(b')}$};
\node at (57.5:1.4) {${\scriptstyle \pi(a')}$};
\node at (112.5:1.3) {${\scriptstyle b'}$};
\node at (157.5:1.3) {${\scriptstyle a'}$};
\end{tikzpicture} \qquad\qquad\qquad
\begin{tabular}{lll}
$i$       & $j$  & \\
$a$       & $\pi(b)$\hspace{1cm}  & $0=P_{ab} t_{\pi(b)} +P_{\pi(a)\pi(b)} t_{\pi(a)}$ \\
$a$       & $\pi(b')$             & $0=P_{\pi(a)\pi(b')} t_{\pi(a)}$ \\
$a$       & $b'$                  & $0=P_{a\pi(b')} t_{b'} +P_{\pi(a)b'} t_{\pi(a)}$ \\
$b$       & $a$                   & $0=P_{b\pi(a)} t_{a} +P_{\pi(b)a} t_{\pi(b)}$ \\
$\pi(a)$  & $b'$                  & $0=P_{\pi(a)\pi(b')} t_{b'}$ \\
$\pi(a)$  & $b$                   & $0=P_{\pi(a)\pi(b)} t_{b} +P_{ab} t_{a}$ \\
$\pi(b)$  & $\pi(a)$              & $0=P_{\pi(b)a} t_{\pi(a)} +P_{b\pi(a)} t_{b}$ \\
$\pi(b')$ & $a$                   & $0=P_{b'a} t_{b'}$ \\
$\pi(b')$ & $\pi(a)$              & $0=P_{\pi(b')a} t_{\pi(a)} +P_{b'\pi(a)} t_{b'}$ \\
$b'$      & $\pi(a)$              & $0=P_{b'a} t_{\pi(a)}$
\end{tabular}
\]
There are $8$ independent equations in this list.
\end{enumerate}
We find that there are $2$ equations for every pair of chords that are not mirror images, as well as $1$ for every fixed point-chord pair, in total $n(n-1)-2\floor*{\frac{n}{2}}$ equations. The dimension of the larger space $\left(\M_N^{\mathrm{c}}\right)_{\Delta=0}$ is $2n(n-1)$, so the dimension of the Zariski tangent space is $2\floor*{\frac{n^2}{2}}$, the same as $F^{\mathrm a}_\pi$.
\end{proof}

\subsubsection{Defining Equations}

We first find another characterization of the $E^{\mathrm a}_\pi$. From its definition, $E^{\mathrm a}_N$, and therefore its irreducible components, are invariant by conjugation
and scaling, i.e., under the action of the group $B_{\ZN}\times \C^{\times}$ (resp.\ $\tilde B_{\ZN}\times \C^\times$, $\ttilde B_{\ZN}\times \C^\times$).
The latter is a semi-direct product of $T_{\ZN}\times \C^\times$ and $U_{\ZN}$,
where
\[
U_{\ZN} := \{ M\in B_{\ZN}\ |\ M_{ii}=1\ \forall i\},
\]
and similarly for $\tilde B_{\ZN}$ and $\ttilde B_{\ZN}$, with $\tilde U_{\ZN} = U_{\ZN} \cap \tilde B_{\ZN}$ and $\ttilde U_{\ZN} = U_{\ZN} \cap \ttilde B_{\ZN}$.
Since the use of the full group does not significantly simplify the orbit structure, we investigative below dense orbits under $\tilde U_{\ZN}$ in the $E_\pi^{\mathrm a}$, $\mathrm a\in\{\mathrm i,\mathrm c\}$.

\begin{thm}\label{thm:denseorbit}
\[ E^{\mathrm a}_\pi=\overline{\tilde U_{\ZN}\cdot\{\underline{\pi} t\ |\ t\textnormal{ diagonal }\in\M_N^{\mathrm{a}}\}}. \]
\end{thm}
\begin{proof}
First we compute the dimension of $\tilde U_{\ZN}\cdot\{\underline{\pi}t\}$.
It is not hard to show that each $\tilde U_{\ZN}$-orbit contains only one $\underline{\pi}t$. Given this, we have
\[ \dim\big(\tilde U_{\ZN}\cdot\{\underline{\pi}t\}\big)=\dim\big(\{\underline{\pi}t\}\big)+\dim\big(\tilde U_{\ZN}\cdot(\underline{\pi}t)\big), \]
where the second term, the dimension of a $\tilde U_{\ZN}$-orbit for a generic choice of $t$, is equal to the number of equations defining the infinitesimal stabilizer
\[ \{P\in\tilde{\mathfrak u}_{\ZN}\ |\ P\underline{\pi}t-\underline{\pi}tP=0\}. \]
where $\tilde{\mathfrak u}_{\ZN}$ is the Lie algebra of $\tilde U_{\ZN}$. We call this number $d^{\mathrm a}_P$.

Unlike the periodic case, we cannot calculate the number of equations defining the infinitesimal stabilizer in the same way as for \thmref{genred}. Instead, we have
\begin{align*}
d^{\mathrm i}_P&=n^2-\#(a\sim a'),\\
d^{\mathrm c}_P&=n(n-1),
\end{align*}
and
\begin{align*}
\dim\big(\{\underline{\pi^{\mathrm i}}t\}\big)&=n+\#(a\sim a'),\\
\dim\big(\{\underline{\pi^{\mathrm c}}t\}\big)&=2\floor*{\frac{n}{2}},
\end{align*}
so
\begin{align*}
\dim\big(\tilde U_{\ZN}\cdot\{\underline{\pi}t\}\big)&=n(n+1),\\
\dim\big(\tilde U_{\ZN}\cdot\{\underline{\pi}t\}\big)&=2\floor*{\frac{n^2}{2}}.
\end{align*}

For $\pi$ a link pattern, we note that $\tilde U_{\ZN}\cdot\{\underline{\pi} t\ |\ t\text{ invertible}\} \subset F^{\mathrm a}_\pi$, because the upper triangular part of any matrix in the former is $\tilde B_N$-conjugate to $\pi_<$.
Therefore, $\overline{\tilde U_{\ZN}\cdot\{\underline{\pi} t\ |\ t\text{ invertible}\}}=\overline{\tilde U_{\ZN}\cdot\{\underline{\pi} t\}}\subset \overline{F^{\mathrm a}_\pi}=E^{\mathrm a}_\pi$.
Since $\overline{\tilde U_{\ZN}\cdot\{\underline{\pi}t\}}$ has the same dimension as $E^{\mathrm a}_\pi$, and the latter is irreducible, they must be equal.
\end{proof}

A similar statement holds for $\mathrm a\in\{\mathrm o,\mathrm m\}$,
i.e.,
$E^{\mathrm a}_\pi=\overline{\ttilde U_{\ZN}\cdot\{\underline{\pi} t\}}$.

\begin{thm}\label{thm:defeq}
Any $M\in E^{\mathrm a}_\pi$ satisfies the following equations:
\begin{enumerate}
  \item $M\in \M^{\mathrm a}_N$;
  \item $M^2 = 0$;
  \item $s_k(M)=s_l(M)$ when $l\in\cl(k)\cup\cl\left(\pi(k)\right)$;
  \item $\rkm(M)\leq \rkm(\underline\pi)$.\footnote{Note that the symmetries in condition (1) implies $\rkm(M)_{ij}=\rkm(M)_{kl}$ when $(k,l)\in\cl{(i,j)}$.}
\end{enumerate}
\end{thm}
\begin{proof}
For $\mathrm{a}=\mathrm{p}$ this was proved in \cite{artic33}. For $\mathrm{a}\in\{\mathrm{i},\mathrm{c}\}$ only equations (3) and (4) are new; one easily checks that they are satisfied by $\underline{\pi}t$, and that they are invariant by conjugation
by $\tilde B_{\ZN}$.
\end{proof}
We conjecture that these are the defining equations of $E^{\mathrm a}_\pi$. At least, we know that these equations define a set that is the union of $E^{\mathrm a}_\pi$ and of lower dimensional pieces, because the other top-dimensional components contain matrices that do not satisfy the equations.
If this conjecture is true, it implies the following:
\begin{conj}\label{conj:fromp}
For any link pattern $\pi\in\mathrm{LP}^{\mathrm a}_L$ and its associated periodic link pattern $\tilde\pi$ as described at the start of \secref{irredcomp}, we have
\[ E^{\mathrm p}_{\tilde\pi} \cap\M_N^{\mathrm a} = E^{\mathrm a}_{\pi}. \]
\end{conj}
Once again, one can prove the slightly weaker statement that $E^{\mathrm p}_{\tilde\pi} \cap\M_N^{\mathrm a}$ has $E^{\mathrm a}_{\pi}$ as its unique top-dimensional component, because intersecting with any other component of $E^{\mathrm a}_N$ reduces its dimension, some equation of \thmref{defeq} for $E^{\mathrm p}_{\pi}$ being violated.

\subsection{The permutation sector}
For simplicity we assume in this section that $N=2n=4m$ (the case $n=2m+1$ can
be treated analogously).

Define the permutation subspace $\M^{\mathrm{perm}}_{N}$ to be the linear subspace of $\M_N$
\[
\M_{N}^{\mathrm{perm}}:=
\{M\in \M_N\ |\ M_{ij}=0\text{ for } m<i\le j\le 3m\text{ or } 3m<i<j\le 5m\}.
\]
(Compared to the definition in \cite[Section~5]{artic33}, we have
shifted by $m$ along the diagonal in order for the subspace to be
invariant under the symmetry of type $\hat{\mathrm C}$.)

In the strip picture, choosing the fundamental domain to be between rows $m+1$ and $5m$, we have
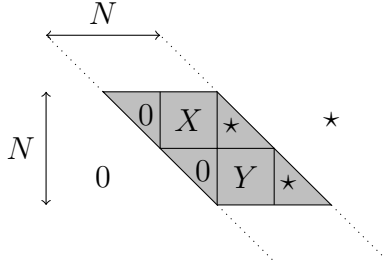
\begin{center}
\begin{tikzpicture}[scale=1.5]
\draw[fill=lightgray] (0,0) -- (1,-1) -- (2,-1) -- (1,0) -- cycle;
\draw (0.5,0) -- (0.5,-0.5) -- (1,-0.5) -- (1,-1);
\draw (1,0) -- (1,-0.5) -- (1.5,-0.5) -- (1.5,-1);
\node at (0.75,-0.25) {$X$};
\node at (1.25,-0.75) {$Y$};
\node at (0.375,-0.2) {$0$};
\node at (0.875,-0.7) {$0$};
\node at (1.125,-0.3) {$\star$};
\node at (1.625,-0.8) {$\star$};
\draw[dotted] (0,0) -- (-0.5,0.5) (1,0) -- (0.5,0.5) (1,-1) -- (1.5,-1.5) (2,-1) -- (2.5,-1.5);
\node at (0,-0.75) {$0$};
\node at (2,-0.25) {$\star$};
\draw[<->] (-0.5,0.5) -- (0.5,0.5) node[above,pos=0.5] {$N$};
\draw[<->] (-0.5,0) -- (-0.5,-1) node[left,pos=0.5] {$N$};
\end{tikzpicture}
\end{center}
where we label the two $n\times n$ submatrices $X$ and $Y$ for convenience.

Now define
\[
E^{\mathrm{a,perm}}_{N}:=E^{\mathrm a}_{N}\cap \M_{N}^{\mathrm{perm}},
\qquad \mathrm a\in\{\mathrm p,\mathrm i,\mathrm c\}.
\]

In \cite{artic33}, it is explained how the equations satisfied by $M$ in $E_N^{\mathrm{p,perm}}$ only involve $X$ and $Y$, and are:
\[
E_{N}^{\mathrm{p,perm}}=\{ XY\text{ and }YX\text{ upper triangular}\},
\]
(so that it is isomorphic to
the so-called ``upper-upper scheme''
$\{ X,Y\in\gl_n\ |\ XY\text{ and }YX$ $\text{upper triangular}\}$ \cite{Kn-uu} times some irrelevant vector space),
and that $E_{N}^{\mathrm{p,perm}}$ is a complete intersection, allowing us to compute
its multidegree:
\[
\mdeg E_{N}^{\mathrm{p,perm}}
=A^N\prod_{\substack{m<i<j\le 3m\\\text{or}\\3m<i<j\le 5m}}
(A+z_i-z_j)(2A+z_j-z_i-\epsilon).
\]

The same argument works for $\mathrm a\in\{\mathrm i,\mathrm c\}$.
The symmetry axes are
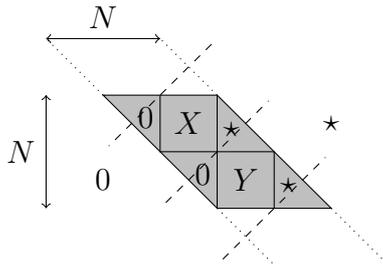
\begin{center}
\begin{tikzpicture}[scale=1.5]
\draw[fill=lightgray] (0,0) -- (1,-1) -- (2,-1) -- (1,0) -- cycle;
\draw (0.5,0) -- (0.5,-0.5) -- (1,-0.5) -- (1,-1);
\draw (1,0) -- (1,-0.5) -- (1.5,-0.5) -- (1.5,-1);
\draw[dashed] (0.95,0.45) -- (0.05,-0.45) (0.55,-0.95) -- (1.45,-0.05) (1.05,-1.45) -- (1.95,-0.55);
\node at (0.75,-0.25) {$X$};
\node at (1.25,-0.75) {$Y$};
\node at (0.375,-0.2) {$0$};
\node at (0.875,-0.7) {$0$};
\node at (1.125,-0.3) {$\star$};
\node at (1.625,-0.8) {$\star$};
\draw[dotted] (0,0) -- (-0.5,0.5) (1,0) -- (0.5,0.5) (1,-1) -- (1.5,-1.5) (2,-1) -- (2.5,-1.5);
\node at (0,-0.75) {$0$};
\node at (2,-0.25) {$\star$};
\draw[<->] (-0.5,0.5) -- (0.5,0.5) node[above,pos=0.5] {$N$};
\draw[<->] (-0.5,0) -- (-0.5,-1) node[left,pos=0.5] {$N$};
\end{tikzpicture}
\end{center}
so that we find
\begin{align*}
E_{N}^{\mathrm{i,perm}}&=\{ Y=X^\dagger, XY\text{ and }YX\text{ upper triangular}\},\\
E_{N}^{\mathrm{c,perm}}&=\{ Y=-X^\dagger, XY\text{ and }YX\text{ upper triangular}\},
\end{align*}
where $X^\dagger=J^{-1}X^T J$, and
$J$ denotes the $n\times n$ skew-symmetric matrix of the type of \eqref{eq:Jfin}.
That is, $E_N^{\mathrm{i,perm}}$ and $E_N^{\mathrm{c,perm}}$ are isomorphic to
the ``symplectic upper-upper scheme''
$\{X\in\gl_n\ |\ XX^\dagger\text{ and }X^\dagger X\text{ upper triangular}\}$
times some irrelevant vector spaces (the latter being due to the $\star$ entries, being careful that the symmetry imposes
linear relations between them, and in particular imposes zeroes on the symmetry axis in $E_N^{\mathrm{c,perm}}$).

The counting of equations goes as follows:
taking into account the symmetry of $M$,
there are $2m(m+1)$ (resp.\ $2m^2$) linear equations defining
$E_{N}^{\mathrm{i,perm}}$ (resp.\ $E_{N}^{\mathrm{c,perm}}$). In both cases,
similarly taking into account the symmetry of $XX^\dagger$ and $X^\dagger X$, there are $2m(m-1)$
quadratic equations.
The total number of equations is therefore
equal to $4m^2$ (resp.\ $2m(2m-1)$), which is
the codimension of $E_N^{\mathrm a}\supset E_{N}^{\mathrm{a,perm}}$; therefore
$E_{N}^{\mathrm{a,perm}}$ is a complete intersection, and its multidegree is the product of
the weights of its equations:
{\small
\begin{align}
\label{eq:mdegps} &\mdeg E_{N}^{\mathrm{i,perm}}=\prod_{1\le i\le j\le m} (A+z_i-z_j) (A-z_i-z_j-\epsilon) \prod_{1\le i<j\le m}(2A+z_j-z_i-\epsilon)(2A+z_i+z_j)\\
\nn &\qquad\times \prod_{m+1\le i\le j\le 2m} (A+z_i-z_j)(A+z_i+z_j) \prod_{m+1\le i<j\le 2m}(2A+z_j-z_i-\epsilon)(2A-z_i-z_j-\epsilon),\\
\nn &\mdeg E_{N}^{\mathrm{c,perm}}=A^{2m} \prod_{1\le i<j\le m} (A+z_i-z_j) (A-z_i-z_j-\epsilon) (2A+z_j-z_i-\epsilon)(2A+z_i+z_j) \\
\nn &\qquad\times \prod_{m+1\le i<j\le 2m} (A+z_i-z_j)(A+z_i+z_j)(2A+z_j-z_i-\epsilon)(2A-z_i-z_j-\epsilon).
\end{align}
}

As a complete intersection $E_{N}^{\mathrm{a,perm}}$ is equidimensional (of the same dimension
as $E_N^{\mathrm a}$), and therefore a union of top-dimensional components
of $E_N^{\mathrm a}$. In order to find which, we simply test whether $\underline\pi t$ belongs to $\M_N^{\mathrm{perm}}$.
We easily find
\[
E_{N}^{\mathrm{a,perm}}=\bigcup_{\pi: \pi(\{1,\ldots,m\})=\{m+1,\ldots,2m\}} E_\pi^{\mathrm a}.
\]
Such link patterns are in bijection with permutations of $\{1,\ldots,m\}$.

Considering $m_\pi=2^n$ for all such link patterns, we can also write
\[
\sum_{\pi: \pi(\{1,\ldots,m\})=\{m+1,\ldots,2m\}} \phi_\pi^{\mathrm a} = 2^n \mdeg E_N^{\mathrm{a,perm}},\qquad
\mathrm a\in\{\mathrm i,\mathrm c\},
\]
where the RHS is given explicitly by \eqref{eq:mdegps}.

Note that we have
$E_\pi^{\mathrm{i}}\cong E_\pi^{\mathrm{c}}\times \C^{n}$ for all such $\pi$, or
\[
\phi_\pi^{\mathrm i}=\prod_{i=1}^{m} (A-2z_i-\epsilon)\prod_{i=m+1}^{2m} (A+2z_i) \ \phi_\pi^{\mathrm c},\qquad \pi(\{1,\ldots,m\})=\{m+1,\ldots,2m\}.
\]
(The prefactor is due to the different embedding space.)


\subsection{Commuting varieties}
In \cite{Kn-uu}, it is shown that one particular component of the upper-upper
scheme is the singular fiber of a one-parameter flat family whose generic
fiber is the {\em commuting variety}
\[
C_n:=\{ X, Y\in\gl_n\ |\ XY=YX\}.
\]

In \cite{artic33}, this was used to provide a formula for the
degree of the commuting variety; in our notations, one has
\[
\deg C_n = \deg E^{\mathrm p}_{(2n,2n-1,\ldots,1)},
\]

\[ \pi=\begin{tikzpicture}[baseline=(current  bounding  box.center),scale=0.75]
\draw[thick] (0,0) circle(1);
\draw[smooth] (15:1) to[out=195,in=165] (345:1);
\draw[smooth] (45:1) to[out=225,in=135] (315:1);
\draw[smooth] (165:1) to[out=345,in=15] (195:1);
\draw[smooth] (135:1) to[out=315,in=45] (225:1);
\node at (195:1.3) {${\scriptstyle 1}$};
\node at (225:1.3) {${\scriptstyle 2}$};
\node at (165:1.4) {${\scriptstyle 2n}$};
\node at (135:1.4) {${\scriptstyle 2n-1}$};
\node at (0:0.05) {${\scriptstyle \dots}$};
\end{tikzpicture}\ , \]

(and more generally, equality of multidegrees with the appropriate
correspondence of torus actions).

Assume now $n$ even. Using the exact same argument, one can show
that a particular component of the ``symplectic upper-upper scheme''
(see previous section) is the singular fiber
of a one-parameter flat family whose generic
fiber is the ``symplectic commuting variety''
\[
\tilde C_n:=\{ X\in\gl_n\ |\ XX^\dagger=X^\dagger X\}.
\]
This implies that
\begin{align*}
\deg \tilde C_n &= \deg E^{\mathrm i}_{(n,n-1,\ldots,1)} = \deg E^{\mathrm c}_{(n,n-1,\ldots,1)}
\\
&=1,11,1583,3186265,92351668113
\ldots\ ,\qquad n=2,4,\ldots\ ,
\end{align*}

\[ \pi=\begin{tikzpicture}[baseline=(current  bounding  box.center),scale=0.75]
\draw[thick] (0,0) circle(1);
\draw[smooth] (15:1) to[out=195,in=345] (165:1);
\draw[smooth] (45:1) to[out=225,in=315] (135:1);
\draw[smooth] (345:1) to[out=165,in=15] (195:1);
\draw[smooth] (315:1) to[out=135,in=45] (225:1);
\draw[dotted] (0:1.2) -- (180:1.2);
\node at (195:1.3) {${\scriptstyle 1}$};
\node at (225:1.3) {${\scriptstyle 2}$};
\node at (345:1.3) {${\scriptstyle n}$};
\node at (315:1.4) {${\scriptstyle n-1}$};
\node at (270:0.6) {${\scriptstyle \vdots}$};
\node at (90:0.6) {${\scriptstyle \vdots}$};
\end{tikzpicture}\ . \]

In principle, an explicit formula for the (multi)degree of $\tilde C_n$
can be obtained by repeated application of divided difference operators
and by using some formulae of \cite{PDF-open} in type $\mathrm a=\mathrm c$.
We shall not reproduce them here because they are rather cumbersome.

\section{From the Brauer loop schemes to the loop model}\label{sec:brauertoloop}
We now provide the link between the geometric construction of \secref{brauer} to the loop model of Sections~\ref{sec:qKZ} and \ref{sec:loopmodel}. As explained in \secref{qKZsoln} and \secref{mdeg}, the correspondence of parameters is as follows:
\begin{itemize}
\item The length of the loop model $L$ is related to the size $N$ of the matrices by:
$L=N$ for $\mathrm a=\mathrm p$,
$L=N/2$ for $\mathrm a\in\{\mathrm i,\mathrm c\}$,
$L=N/4$ for $\mathrm a\in\{\mathrm o,\mathrm m\}$.
\item The shift $s$ of the qKZ equation is related to the equivariance parameter $\epsilon$ by:
$\epsilon=s$ for $\mathrm a\in\{\mathrm p,\mathrm i,\mathrm c\}$,
$\epsilon=s/2$ for $\mathrm a\in\{\mathrm o,\mathrm m\}$.
\end{itemize}

The precise theorem, as advertised in the introduction is:
\begin{thm}\label{thm:main}
In all types $\mathrm a\in\{\mathrm p,\mathrm i,\mathrm c,\mathrm o,\mathrm m\}$,
the vector $\ket\Phi=\sum_{\pi\in\mathrm{LP}^{\mathrm a}_L} \phi^{\mathrm a}_\pi\ket\pi$ of multidegrees $\phi^{\mathrm a}_\pi = m_\pi \mdeg E_\pi^{\mathrm a}$ of the irreducible components
of the Brauer loop scheme $E_N^{\mathrm a}$ satisfies the qKZ system (\ref{eq:qKZbulk}--\ref{eq:qKZrotate}) or
(\ref{eq:qKZCR}--\ref{eq:qKZCKL}), as well as the recurrence relations (\ref{eq:psirecur}), (\ref{eq:lrecur}) and (\ref{eq:rrecur}) (up to normalization), thus identifying it with the unique (up to normalization) minimal degree polynomial solution of the qKZ system.
\end{thm}
The rest of this section is dedicated to the proof of this theorem.

\subsection{L\'evy subgroups}\label{sec:levy}
The geometric interpretation of the quantum Knizhnik--Zamolodchikov equation
follows the same general philosophy that was outlined in \cite{artic32} and then developed in \cite{artic33,artic39}.
It is based on a combination of ``cutting'' -- intersecting with hypersurfaces -- and ``sweeping'' -- taking the image
under L\'evy subgroups, similarly to the pullbacks/pushforwards in convolution actions \cite{CG-book}.

Let $B\subset SL_2$ be the group of $2\times 2$ invertible upper triangular matrices inside the group of $2\times2$ matrices
of determinant $1$.
We start with the following standard lemma: (see also \cite[lemma 8]{artic39} and \cite[lemma 1]{artic33}; the use
of $GL_2$ instead of $SL_2$ makes no difference)
\begin{lm}\label{lm:divdiff}
  Let $X$ be a variety in a vector space $V$ equipped with a $SL_2$-representation, such that $X$ is $B$-invariant and conical.
If the generic fiber of the map
  \begin{equation*}
    \mu: SL_2 \times_B X \to V
  \end{equation*}
  is finite over ${\rm Image}\ \mu$,
  call its cardinality $k$; otherwise let $k=0$.
  (The latter occurs iff $X$ is $SL_2$-invariant.) Then
  \begin{equation*}
    k\,\mdeg ({\rm Image}\ \mu) = -\partial_i \ \mdeg X,
  \end{equation*}
where $\partial_i$ is the divided difference operator $\partial_i f = (f - \tau f)/\alpha$ as defined in \secref{dynkin}, and $\alpha$ is the root of $SL_2$.
\end{lm}
The multidegree is w.r.t.\ the Cartan torus of $SL_2$.

To each node $i$ in the Dynkin diagram of
$\hat{\mathrm A}$ or $\hat{\mathrm C}$, we can associate groups $B^{(i)}\subset SL_2^{(i)}$
(isomorphic to $B\subset SL_2$) defined by:
\begin{multline*}
SL_2^{(i)}=\Big\{M=(M_{jk})_{j,k\in\Z}\ \Big|\ M S^N = S^N M,\ MM^\dagger=1\text{ if }\mathrm a\in\{\mathrm i,\mathrm c,\mathrm o,\mathrm m\},\ MM^\ddagger=1\text{ if }\mathrm a\in\{\mathrm o,\mathrm m\},\\
 M_{jk}=\delta_{jk}\text{ unless }(j,k)\in \cl\left({\scriptstyle\left\{\stack{i}{i+1}\right.},{\scriptstyle\left\{\stack{i}{i+1}\right.}\right),\
\begin{vmatrix} M_{ii} & M_{i,i+1} \\ M_{i+1,i} & M_{i+1,i+1}
\end{vmatrix}=1
\Big\},
\end{multline*}
and $B^{(i)}= SL_2^{(i)}\cap B_{\ZN}$.

Note that in all types, the isomorphism from $SL_2^{(i)}$ to $SL_2$ consists in extracting the $2\times 2$ submatrix at rows
and columns $i,i+1$. When there is no risk of confusion we shall identify $SL_2^{(i)}$ and $SL_2$ via this isomorphism.

Next, given $P\in SL_2^{(i)}$ and $M\in R^{\mathrm a}_{\ZN}$, one can consider conjugating: $P M P^{-1}$. Two problems arise at this stage.
Firstly, the result has entries below the diagonal. There are various ways to deal with this: the one we choose here is to
restrict ourselves to matrices
sitting in the subspace $R^{\mathrm a}_{\ZN}\cap\{M_{ii}=M_{i+1,i+1}=M_{i,i+1}=0\}$, which is stable under the $SL_2^{(i)}$ action.
Secondly, this {\em does not}\/ descend to an action on (appropriate subspaces of) $\M^{\mathrm a}_N$, because
the action does not preserve the ideal generated by $S^N$. In \cite{artic39}, this difficulty is circumvented by working
inside $R^{\mathrm a}_{\ZN}$, i.e., taking preimages of subvarieties of $\M^{\mathrm a}_N$ before taking their image (``sweeping'' them)
under $SL_2^{(i)}$ and then taking the image again in $\M^{\mathrm a}_N$.
Here, to slightly simplify the discussion,
we shall by abuse of notation identify such a subvariety with its preimage.

Taking into account the fact
that the multidegree depends on the embedding space, we are led to the following modification of the lemma:
\begin{itemize}
\item For ``closed boundaries'', that is, for $\mathrm a=\mathrm c$ and $i=0,L$ or for $\mathrm a=\mathrm m$ and $i=0$,
the lemma applies without any changes to the multidegrees w.r.t.\ $\M^{\mathrm a}_N$, the divided difference operators
being the ones defined in \eqref{eq:divdiffops}.
\item In all other cases, to apply the lemma to multidegrees w.r.t.~$\M^{\mathrm a}_N$,
the divided difference operator of \eqref{eq:divdiffops} has to be conjugated, i.e., replaced with
\[
\partial'_i = (A+\alpha_i) \partial_i \frac{1}{A+\alpha_i},
\]
(the factor $A+\alpha_i$ being the weight of $M_{i,i+1}$). 
\end{itemize}

\subsection{Geometry of the exchange relations}
We start with the exchange relations \eqref{eq:qKZbulk} or \eqref{eq:qKZCR},
$i=1,\ldots,L-1$. Note that in type $\hat{\mathrm A}$, the rotation equation \eqref{eq:qKZrotate} is trivially
satisfied due to the cyclic nature of the Brauer loop scheme (see \cite{artic39}), so the equation \eqref{eq:qKZbulk}
is also valid at $i=L$.

We rewrite this equation here for convenience:
\begin{equation}\label{eq:qKZbulk2}
\check R_i(z_i-z_{i+1})\ket{\Phi(\dots,z_i,z_{i+1},\dots)}=\ket{\Phi(\dots,z_{i+1},z_i,\dots)},
\end{equation}
where $\ket\Phi=\sum_{\pi\in \text{LP}^{\mathrm a}_L} \phi_\pi^{\mathrm a} \ket{\pi}$.

As explained in \secref{qKZfacsyms}, when writing \eqref{eq:qKZbulk2} in components, there are two cases to consider, depending on whether $\pi(i)\ne i+1$ or $\pi(i)=i+1$. We treat them separately below.

\subsubsection{The \texorpdfstring{$f_i$}{f} action}\label{sec:fi}
We assume that $\pi \in \mathrm{LP}^{\mathrm a}_L$ is such that $\pi(i)\ne i+1$. Our goal is to prove \eqref{eq:qKZf}
for the multidegrees $\phi^{\mathrm a}_\pi$. The geometric procedure is as follows:
\begin{itemize}
\item ``Sweep'' $E^{\mathrm a}_\pi$ with $SL_2^{(i)}$.
\item ``Cut'' the result with $(M^2)_{i+1,i+N}=0$, and show that it produces $E^{\mathrm a}_\pi \cup E^{\mathrm a}_{f_i\pi}$.
\end{itemize}
\eqref{eq:qKZf} will then be a translation into multidegrees of this construction.

We shall need the following
\begin{lm}\label{lm:basicgeom}
\begin{enumerate}
\item
If $M,M'$ are generic elements of components $E_\pi^{\mathrm a}$ and $E_{\pi'}^{\mathrm a}$ with $\pi(i)\ne i+1$, $\pi'(i)\ne i+1$, such that $M'=PMP^{-1}$, $P\in SL_2^{(i)}$, then $s_j(M)=s_j(M')$ if $j\in\cl(i),\cl(i+1)$, and $\{s_i(M),s_{i+1}(M)\}=\{s_i(M'),s_{i+1}(M')\}$; and for a fixed $M'$ (resp.\ $M$), the set of possible cosets of $P$ in $SL_2^{(i)}/B^{(i)}$ (resp.\ $B^{(i)}\backslash SL_2^{(i)}$) consists of exactly two points, corresponding to whether these $s_i,s_{i+1}$ are in the same order, or reversed.
\item In the particular case where $M=\underline \pi t$, $t$ generic diagonal, then the two classes of $P$ have representatives $\left(\begin{smallmatrix}1&0\\0&1\end{smallmatrix}\right)$ and $i\left(\begin{smallmatrix}0&1\\1&0\end{smallmatrix}\right)$; in the latter case,
\[
P \underline\pi t P^{-1}= \underline{f_i\pi}\, t',
\]
with $t'$ the diagonal matrix obtained from $t$ by switching diagonal entries at $j\in\cl(i),\cl(i+1)$. $M$ and $M'$ playing symmetric roles, an analogous result holds for $M'=\underline\pi t$.
\end{enumerate}
\end{lm}
\begin{proof}
$P$ conjugates the matrices $M^2$, $M'{}^2$, and in particular (restricting to rows $i,i+1$ and columns $i+N,i+N+1$),
their $2\times 2$ submatrices
around the $N^{\text{th}}$ diagonal,
which are upper triangular with eigenvalues $s_i(M),s_{i+1}(M)$ and $s_i(M'),s_{i+1}(M')$.
Generically these eigenvalues are distinct because $\pi(i)\ne i+1$ or $\pi'(i)\ne i+1$,
so that $P=b'{}^{-1} P_0 b$ where $b$, $b'$ are upper triangular matrices which diagonalize these $2\times2$
submatrices,
and $P_0$ is as in the second part of the lemma. The rest is a direct computation.
\end{proof}

\subsubsection*{The degree of the sweeping map}
We compute the cardinality of the generic fiber of the map
$SL_2^{(i)}\times_{B^{(i)}} E^{\mathrm a}_\pi\to SL_2^{(i)}\cdot E_\pi^{\mathrm a}$. Since the $SL_2^{(i)}$ action is a group action,
we may assume that the fiber $\{(P,M)\ |\ PMP^{-1}=M'\}$
is that of an element $M'$ of $E^{\mathrm a}_\pi$, and furthermore that it is of the form $M'=\underline\pi t$.
We are in the situation of \lmref{basicgeom}
with $M,M'\in E^{\mathrm a}_\pi$. We conclude that there are two possibilities:
\begin{itemize}
\item If $f_i \pi \ne \pi$, (i.e.,
if at least one of $i$ or $i+1$  is paired in $\pi$ (not connected to the boundary),
or in type $\mathrm a=\mathrm o$ if they are connected to different boundaries), then only the first coset,
namely, $P\in B^{(i)}$, leads to
$M'\in E^{\mathrm a}_\pi$, and therefore
the fiber (in $SL_2^{(i)}\times_{B^{(i)}} E^{\mathrm a}_\pi$) is a point.
\item If $f_i\pi = \pi$ (i.e., if both $i$ and $i+1$ are connected to the (same) boundary),
then both cosets lead to $M'\in E^{\mathrm a}_\pi$, and the fiber consists of two points.
\end{itemize}
In conclusion, we find that
\begin{equation}\label{eq:cardfib}
\text{cardinality of a generic fiber of }SL_2^{(i)}\times_{B^{(i)}} E^{\mathrm a}_\pi \to SL_2^{(i)}\cdot E^{\mathrm a}_\pi=
\begin{cases}
1&\text{if }f_i\pi\ne\pi,\\
2&\text{if }f_i\pi=\pi.
\end{cases}
\end{equation}

\subsubsection*{Determination of the result of sweeping and cutting}
The generic fiber being finite, the image has same dimension as the source, i.e.,
$\dim(SL_2^{(i)}\cdot E^{\mathrm a}_\pi) = \dim E^{\mathrm a}_\pi + 1$. An elementary calculation shows that the only equation of $E^{\mathrm a}_N$
that $SL_2^{(i)}\cdot E^{\mathrm a}_\pi$ violates is $(M^2)_{i+1,i+N}=0$. Noting that $\{(M^2)_{i+1,i+N}=0\}$ is a Cartier divisor in the (irreducible) variety
$SL_2^{(i)}\cdot E^{\mathrm a}_\pi$,
we conclude that $(SL_2^{(i)}\cdot E^{\mathrm a}_\pi)\cap \{ (M^2)_{i+1,i+N}\}$ is a subscheme
of $E^{\mathrm a}_N$ of pure dimension $\dim E^{\mathrm a}_N$, therefore a union of its top-dimensional components $E^{\mathrm a}_\pi$.

In order to determine which, we apply again \lmref{basicgeom}.
We first compute the image of $M=\underline\pi t$:
\[
(SL_2^{(i)}\cdot \{\underline\pi t\})\cap \{ (M^2)_{i+1,i+N}=0\} = B^{(i)}\cdot\{\underline\pi t\} \cup B^{(i)}\cdot\{ \underline{f_i\pi}\,t' \},
\]
with $t'$ as in the lemma.
Finally, taking the union over $t$ and the closure of the $B_{\ZN}$ (resp.\ $\tilde B_{\ZN}$, $\ttilde B_{\ZN}$) orbit,
we obtain:
\begin{equation}\label{eq:figeom}
(SL_2^{(i)}\cdot E^{\mathrm a}_\pi)\cap \{ (M^2)_{i+1,i+N}=0\} = E^{\mathrm a}_\pi \cup E^{\mathrm a}_{f_i\pi}.
\end{equation}
The equation above is only an equality of sets;
however since $E^{\mathrm a}_N$ is generically reduced in top dimension (\thmref{genred}),
both sides of the equation are generically reduced (which is all that matters for multidegree purposes).

\subsubsection*{Multidegree equality}
Using \eqref{eq:cardfib} and applying \lmref{divdiff},
we have
\[
\mdeg SL_2^{(i)}\cdot E^{\mathrm a}_\pi=
2^{\delta_{\pi,f_i\pi}}
(-\divdif_i) \mdeg E_\pi^{\mathrm a}.
\]

Next we intersect the variety $SL_2^{(i)}\cdot E^{\mathrm a}_\pi$
with the hypersurface $\{(M^2)_{i+1,i+N}=0\}$; by the properties
of multidegrees, this multiplies its multidegree with
$\mdeg \{(M^2)_{i+1,i+N}=0\}=2A+z_{i+1}-z_{i+N}=2A+z_{i+1}-z_i-\epsilon$.
Finally, we apply \eqref{eq:figeom}, noting that the factor
of $2$ when $\pi=f_i\pi$
is compensated by the fact that $E_\pi=E_{f_i\pi}$,
and find
\[
(2A+z_{i+1}-z_i-\epsilon) (-\divdif_i)
\mdeg  E^{\mathrm a}_\pi = \mdeg E^{\mathrm a}_\pi + \mdeg E^{\mathrm a}_{f_i\pi}.
\]

Equivalently, using \eqref{eq:defpsi} and
noting that $\pi$ and $f_i\pi$ have the same number of chords, we find
\begin{equation}\label{eq:fi}
(2A+z_{i+1}-z_i-\epsilon) (-\divdif_i)
\phi_\pi^{\mathrm a} = \phi_\pi^{\mathrm a} + \phi_{f_i\pi}^{\mathrm a}.
\end{equation}

\subsubsection{The \texorpdfstring{$e_i$}{e} action}
We now assume that $\pi \in \mathrm{LP}^{\mathrm a}_L$ is such that $\pi(i)= i+1$, and we wish to prove that
\eqref{eq:qKZi} is satisfied by the multidegrees $\phi^{\mathrm a}_\pi$.

Fully interpreting geometrically the $e_i$ equation \eqref{eq:qKZi}
is rather complicated (see \cite{hdr} and \cite[arXiv v1]{artic39} for the case of type $\mathrm{\hat A}$).
Given such a link pattern $\pi$, the geometric construction is:
\begin{itemize}
\item Cut $E^{\mathrm a}_\pi$ with $M_{i,i+1} = 0$, producing $F_1$;
\item Throw away the $SL_2^{(i)}$-invariant components (giving
  $\bigcup_{\rho\in\varepsilon(\pi,i)} X^{\mathrm a}_{\rho,i}$, to be defined below), and
then sweep with $SL_2^{(i)}$, producing $F_2$;
\item Finally, cut with $(M^2)_{i+1,i+N}=0$, producing $F_3$, and show that
  $F_3=\bigcup_{\rho\ne\pi: e_i\rho=\pi} E^{\mathrm a}_\rho \cap \{s_i(M)=s_{i+1}(M)\}$.
\end{itemize}
The multidegree of $F_3$ is the desired expression.

Here we shall only provide a semi-geometric proof of \eqref{eq:qKZf},
as in \cite{artic39}: we shall stop at the first stage in the construction, i.e., only analyze $F_1$ above,
and then use the (already proven) $f_i$ equation \eqref{eq:fi} to conclude.

\subsubsection*{The auxiliary varieties $X_{\rho,i}^{\mathrm a}$}
Denote by $|\rho|$ the number of crossings of $\rho$.
\begin{prop}\label{prop:defX}
Given $\rho$ a link pattern such that $\rho(i)\ne i+1$ and $|\rho|\ge|f_i\rho|$:
\begin{itemize}
\item If $f_i\rho \ne \rho$, then
$E^{\mathrm a}_\rho\cap \{s_i(M)=s_{i+1}(M)\}$ has a single geometric component.
Call it $X^{\mathrm a}_{\rho,i}$.
\item If $f_i\rho = \rho$,
$E^{\mathrm a}_\rho\cap \{s_i(M)=s_{i+1}(M)\}$
has two geometric components, one of which is $SL^{(i)}_2$-invariant.
Call $X^{\mathrm a}_{\rho,i}$ the other one. 
\end{itemize}
In both cases
$E^{\mathrm a}_\rho\cap \{s_i(M)=s_{i+1}(M)\}$ is generically reduced at $X^{\mathrm a}_{\rho,i}$.
\end{prop}
\begin{proof}
(We only give the full proof for $\mathrm a\in\{\mathrm p,\mathrm i,\mathrm c\}$.)
This is similar to \cite[Proposition~9]{artic39}. Since $\rho(i)\ne i+1$,
$E^{\mathrm a}_\rho \not\subset \{s_i(M)=s_{i+1}(M)\}$ (as can be checked on say $\underline\rho t$),
so all the geometric components of $E^{\mathrm a}_\rho \cap \{s_i(M)=s_{i+1}(M)\}$ have dimension
$\dim E^{\mathrm a}_N -1$. We use the decomposition
\[
E^{\mathrm a}_\rho \cap \{s_i(M)=s_{i+1}(M)\}
=\bigsqcup_\sigma (E^{\mathrm a}_\rho \cap \{s_i(M)=s_{i+1}(M)\}\cap F_\sigma),
\]
and consider only pieces of dimension $\dim E^{\mathrm a}_N -1$.

We start from $F_\sigma$ itself. For it
to have dimension $\ge\dim E^{\mathrm a}_N -1$, according to \corref{dimoneless},
$\sigma$ can have one pair of the form
$\begin{tikzpicture}[baseline=(current  bounding  box.center),scale=0.5]
\draw[thick] (0,0) circle(1);
\draw[smooth] (45:1) to[out=225,in=45] (225:1);
\draw[smooth] (135:1) to[out=315,in=135] (315:1);
\node at (135:1.4) {$a'$};
\node at (225:1.4) {$a$};
\node at (45:1.4) {$b'$};
\node at (315:1.4) {$b$};
\end{tikzpicture}$ for $\mathrm a\in\{\mathrm i,\mathrm c\}$,
or a pair of fixed points for $\mathrm a\in\{\mathrm p,\mathrm i\}$.

We then intersect with $\{ s_i(M)=s_{i+1}(M)\}$.
There are three possibilities:
\begin{enumerate}
\item
$\sigma(\{i,i+1\})\not\subset \{i,i+1,N-i,N-i+1\}$.
The equation $s_i(M)=s_{i+1}(M)$ being $\tilde B_N$-invariant and linear in $\La$
(in the $M=\U+\La$ decomposition, cf.~\secref{irrcomp}),
$F_\sigma\cap \{s_i(M)=s_{i+1}(M)\}$ is a subvector bundle of $F_\sigma$, where the dimension of the fiber can
be easily evaluated, say at $\U=\sigma^{\mathrm a}_<$, where the extra equation
$\La_{i\leftrightarrow\sigma(i)}=\La_{i+1\leftrightarrow \sigma(i+1)}$ reduces it by one compared to that of $F_\sigma$.
This implies that $\sigma$ must be a link pattern (otherwise the dimension is too low).
\item
$\sigma(i)=N-i$, $\sigma(i+1)=N-i+1$, in which case $F_\sigma \subset \{ s_i(M)=s_{i+1}(M) \}$.
For the dimension to be right, $\sigma$ cannot have any other crossing pairs of the same form or any fixed points.
\item $\sigma(i)=i+1$, $\sigma(i+1)=i$.
\end{enumerate}

Now we want to intersect with $E^{\mathrm a}_\rho$. This immediately excludes case (3),
because if $\sigma(i)=i+1$, the rank condition (equations (4) in \thmref{defeq}) of $E^{\mathrm a}_\rho$ at $(i,i+1)$ is violated, so
$F_\sigma \cap E^{\mathrm a}_\rho$ is empty.
We are left with cases (1) and (2); in both, $F_\sigma\cap \{s_i(M)=s_{i+1}(M)\}$ is a subvector bundle of $F_\sigma$,
so is an open (irreducible) variety of the target dimension.

If $\sigma$ differs from $\rho$ outside $\{i,i+1\}$, then it is easy to see that some of the equations of \thmref{defeq}
of $E^{\mathrm a}_\rho$ are violated by say $\underline \sigma t$ with $t_i\ne 0$,
$t_i t_{\sigma(i)}=t_{i+1}t_{\sigma(i+1)}$. Indeed, the $s_i(\underline \sigma t)$ only have the repeats of the pairings
of $\sigma$, so $\rho$ cannot have more pairings than $\sigma$ (otherwise equations (3) of \thmref{defeq} would be violated).
Inversely, assuming the pairings of $\rho$ are a strict subset of those of $\sigma$,
i.e.,  there exists $i<\sigma(i)=j$, but $\rho(i)\ne j$, then the rank condition
(equation (4) of \thmref{defeq}) of $\rho$ at $(i,j)$ would be violated (and in fact, in that case, the intersection would
be empty).
It follows that the dimension of $F_\sigma\cap \{s_i(M)=s_{i+1}(M)\} \cap E^{\mathrm a}_\rho$ is less than the target.
Therefore the two possibilities reduce to:
\begin{enumerate}
\item $\sigma=\rho$, in which case $E^{\mathrm a}_\rho \cap \{s_i(M)=s_{i+1}(M)\} \cap F_\sigma =
F_\rho \cap \{s_i(M)=s_{i+1}(M)\}$, of course.

If $f_i\rho=\rho$, we show in \appref{inv} that $\overline{F_\rho\cap \{s_i(M)=s_{i+1}(M)\}}$ is $SL_2^{(i)}$-invariant.

If $f_i\rho\ne\rho$,
we call $X^{\mathrm a}_{\rho,i}$ the closure of $F_\rho \cap \{ s_i(M)=s_{i+1}(M)\}$
(with its reduced structure); or,
\item $\sigma(i)=N-i$, $\sigma(i+1)=N-i+1$ and is identical to $\rho$ elsewhere, i.e.,
\[ \rho =\;\begin{tikzpicture}[baseline=(current  bounding  box.center),scale=0.75]
\draw[thick] (0,0) circle(1);
\draw[smooth] (112.5:1) to[out=292.5,in=67.5] (247.5:1);
\draw[smooth] (67.5:1) to[out=247.5,in=112.5] (292.5:1);
\node at (247.5:1.3) {${\scriptstyle i}$};
\node at (292.5:1.3) {${\scriptstyle i+1}$};
\node at (117:1.4) {${\scriptstyle N-i+1}$};
\node at (63:1.4) {${\scriptstyle N-i}$};
\node at (0:0.8) {.};
\node at (20:0.8) {.};
\node at (-20:0.8) {.};
\node at (180:0.8) {.};
\node at (200:0.8) {.};
\node at (160:0.8) {.};
\end{tikzpicture}\ ,\qquad\qquad
\sigma =\;\begin{tikzpicture}[baseline=(current  bounding  box.center),scale=0.75]
\draw[thick] (0,0) circle(1);
\draw[smooth] (112.5:1) to[out=292.5,in=112.5] (292.5:1);
\draw[smooth] (67.5:1) to[out=247.5,in=67.5] (247.5:1);
\node at (247.5:1.3) {${\scriptstyle i}$};
\node at (292.5:1.3) {${\scriptstyle i+1}$};
\node at (117:1.4) {${\scriptstyle N-i+1}$};
\node at (63:1.4) {${\scriptstyle N-i}$};
\node at (0:0.8) {.};
\node at (20:0.8) {.};
\node at (-20:0.8) {.};
\node at (180:0.8) {.};
\node at (200:0.8) {.};
\node at (160:0.8) {.};
\end{tikzpicture}\ . \]
This situation can only occur when $f_i\rho=\rho$. In this case, we claim that
$\overline{F_\sigma}=\overline{E^{\mathrm a}_\rho\cap F_\sigma\cap\{ s_i(M)=s_{i+1}(M)\}}$. That $F_\sigma=F_\sigma\cap\{ s_i(M)=s_{i+1}(M)\}$
is obvious.

Next, we note that $\overline{\tilde B_N\cdot \rho^{\mathrm i}_<}
\supset \tilde B_N\cdot\sigma^{\mathrm i}_<$. Indeed, the matrix $P$
with submatrix
$\left(\begin{smallmatrix}
i/t & 1/t & 0 & 0
\\
0   & t   & 0 & 0
\\
0   & 0   & 1/t & 1/t
\\
0   & 0   & 0   & -i t
\end{smallmatrix}\right)
$
at rows $i,i+1,N-i,N-i+1$ and identity elsewhere is symplectic and
sends $\rho^{\mathrm i}_<$ to $P \rho^{\mathrm i}_< P^{-1} = \sigma^{\mathrm i}_< + O(t^2)$.

This implies that $E_\rho \cap F_\sigma = \overline{F_\rho} \cap F_\sigma$
is a vector bundle over $\tilde B_N\cdot \sigma^{\mathrm i}_<$; the dimension of its fiber
is greater or equal to that of $F_\rho$ by semi-continuity of rank, and less than
or equal to that of $F_\sigma$ by obvious inclusion. But according
to the dimension count in the proof of \thmref{maxF}, the latter two are
equal, and therefore there is equality of dimensions, which implies
$E_\rho\cap F_\sigma=F_\sigma$.

In this case, we call $\overline{F_\sigma}$
(with its reduced structure) $X^{\mathrm a}_{\rho,i}$.
\end{enumerate}

In all cases, note that, as the closure of a vector bundle
over an open (irreducible) variety,
$X^{\mathrm a}_{\rho,i}$ is irreducible.
Generic reducedness in $E^{\mathrm a}_\rho \cap \{s_i(M)=s_{i+1}(M)\}$
is shown in \appref{geomstuff}.
\end{proof}
\begin{remark}
One can also analyze the case $|\rho|<|f_i\rho|$, cf.~\cite[Appendix~B]{artic39} in type $\hat{\mathrm A}$, with similar conclusions as when $\rho=f_i\rho$, but we shall not need it here.
\end{remark}


\subsubsection*{Determination of the result of cutting}
We recall that a link pattern $\pi$ such that $\pi(i)=i+1$ is fixed.
Define
\[
\varepsilon(\pi,i):=\{\rho\ne\pi\ |\ e_i\rho=\pi, |\rho|\ge|f_i\rho|\},
\]
e.g.,
\begin{align*}
\mathrm a&=\mathrm i: &
\pi&=
\begin{tikzpicture}[baseline=(current  bounding  box.center),scale=0.6]
\draw[thick] (0,0) circle(1);
\draw[smooth] (22.5:1) to[out=202.5,in=157.5] (337.5:1);
\draw[smooth] (67.5:1) to[out=247.5,in=112.5] (292.5:1);
\draw[smooth] (112.5:1) to[out=292.5,in=337.5] (157.5:1);
\draw[smooth] (202.5:1) to[out=22.5,in=67.5] (247.5:1);
\node at (202.5:1.3) {${\scriptstyle 1}$};
\node at (247.5:1.3) {${\scriptstyle 2}$};
\node at (292.5:1.3) {${\scriptstyle 3}$};
\node at (337.5:1.3) {${\scriptstyle 4}$};
\end{tikzpicture}\ ,
&
\varepsilon(\pi,1)&=\left\{
\begin{tikzpicture}[baseline=(current  bounding  box.center),scale=0.6]
\draw[thick] (0,0) circle(1);
\draw[smooth] (22.5:1) to[out=202.5,in=157.5] (337.5:1);
\draw[smooth] (67.5:1) to[out=247.5,in=337.5] (157.5:1);
\draw[smooth] (112.5:1) to[out=292.5,in=67.5] (247.5:1);
\draw[smooth] (202.5:1) to[out=22.5,in=112.5] (292.5:1);
\node at (202.5:1.3) {${\scriptstyle 1}$};
\node at (247.5:1.3) {${\scriptstyle 2}$};
\node at (292.5:1.3) {${\scriptstyle 3}$};
\node at (337.5:1.3) {${\scriptstyle 4}$};
\end{tikzpicture},
\begin{tikzpicture}[baseline=(current  bounding  box.center),scale=0.6]
\draw[thick] (0,0) circle(1);
\draw[smooth] (22.5:1) to[out=202.5,in=337.5] (157.5:1);
\draw[smooth] (67.5:1) to[out=247.5,in=112.5] (292.5:1);
\draw[smooth] (112.5:1) to[out=292.5,in=67.5] (247.5:1);
\draw[smooth] (202.5:1) to[out=22.5,in=157.5] (337.5:1);
\node at (202.5:1.3) {${\scriptstyle 1}$};
\node at (247.5:1.3) {${\scriptstyle 2}$};
\node at (292.5:1.3) {${\scriptstyle 3}$};
\node at (337.5:1.3) {${\scriptstyle 4}$};
\end{tikzpicture},
\begin{tikzpicture}[baseline=(current  bounding  box.center),scale=0.6]
\draw[thick] (0,0) circle(1);
\draw[smooth] (22.5:1) to[out=202.5,in=157.5] (337.5:1);
\draw[smooth] (67.5:1) to[out=247.5,in=112.5] (292.5:1);
\draw[smooth] (112.5:1) to[out=292.5,in=67.5] (247.5:1);
\draw[smooth] (202.5:1) to[out=22.5,in=337.5] (157.5:1);
\node at (202.5:1.3) {${\scriptstyle 1}$};
\node at (247.5:1.3) {${\scriptstyle 2}$};
\node at (292.5:1.3) {${\scriptstyle 3}$};
\node at (337.5:1.3) {${\scriptstyle 4}$};
\end{tikzpicture}
\right\}\ ,\\
\mathrm a&=\mathrm c: &
\pi&=
\begin{tikzpicture}[baseline=(current  bounding  box.center),scale=0.6]
\draw[thick] (0,0) circle(1);
\draw[smooth] (22.5:1) to[out=202.5,in=337.5] (157.5:1);
\draw[smooth] (67.5:1) to[out=247.5,in=292.5] (112.5:1);
\draw[smooth] (202.5:1) to[out=22.5,in=157.5] (337.5:1);
\draw[smooth] (292.5:1) to[out=112.5,in=67.5] (247.5:1);
\node at (202.5:1.3) {${\scriptstyle 1}$};
\node at (247.5:1.3) {${\scriptstyle 2}$};
\node at (292.5:1.3) {${\scriptstyle 3}$};
\node at (337.5:1.3) {${\scriptstyle 4}$};
\end{tikzpicture}\ ,
&
\varepsilon(\pi,2)&=\left\{
\begin{tikzpicture}[baseline=(current  bounding  box.center),scale=0.6]
\draw[thick] (0,0) circle(1);
\draw[smooth] (22.5:1) to[out=202.5,in=292.5] (112.5:1);
\draw[smooth] (67.5:1) to[out=247.5,in=337.5] (157.5:1);
\draw[smooth] (202.5:1) to[out=22.5,in=112.5] (292.5:1);
\draw[smooth] (337.5:1) to[out=157.5,in=67.5] (247.5:1);
\node at (202.5:1.3) {${\scriptstyle 1}$};
\node at (247.5:1.3) {${\scriptstyle 2}$};
\node at (292.5:1.3) {${\scriptstyle 3}$};
\node at (337.5:1.3) {${\scriptstyle 4}$};
\end{tikzpicture}
\right\}\ .
\end{align*}
Then one has $e_i^{-1}(\pi)-\{\pi\}= \bigcup_{\rho\in \varepsilon(\pi,i)} \{\rho,f_i\rho\}$.

\begin{prop}\label{prop:eicut}
The geometric components of
$E^{\mathrm a}_\pi \cap \{M_{i,i+1}=0\}$ are
the $X^{\mathrm a}_{\rho,i}$, $\rho\in \varepsilon(\pi,i)$, as well
as one extra $SL_2^{(i)}$-invariant component in the cases
$\mathrm a\in\{\mathrm p, \mathrm c\}$.

The multiplicity of $X^{\mathrm a}_{\rho,i}$ in
$E_\pi \cap \{M_{i,i+1}=0\}$ is $1$ except for the single case (in type $\mathrm a=\mathrm o$) of $\rho$ connecting
$i,i+1$ to distinct boundaries,
i.e., $\rho(i)=r$, $\rho(i+1)=\ell$,
in which case the multiplicity is $2$.
\end{prop}
\begin{proof}
The proof is along the same lines as that of \propref{defX}.
Since $\pi(i)=i+1$, $E_\pi^{\mathrm a}\not\subset \{M_{i,i+1}=0\}$
and $E^{\mathrm a}_\pi \cap \{M_{i,i+1}=0\}$ is of dimension $\dim E^{\mathrm a}_N-1$.
Once again we use the decomposition
\[
E^{\mathrm a}_\pi \cap \{M_{i,i+1}=0\}
=\bigsqcup_\sigma (E^{\mathrm a}_\pi \cap \{M_{i,i+1}=0\} \cap F_\sigma).
\]
Intersecting with $\{ M_{i,i+1}=0 \}$ amounts to imposing $\sigma(i)\ne i+1$ (assuming for $\mathrm a=\mathrm p$ that $i\ne N$,
a case we can always avoid by cyclic symmetry),
in which case this equation is automatically satisfied. So we have
\[
E^{\mathrm a}_\pi \cap \{M_{i,i+1}=0\}
=\bigsqcup_{\sigma:\,\sigma(i)\ne i+1} (E^{\mathrm a}_\pi \cap F_\sigma),
\]
and we must consider only pieces of dimension $\dim E^{\mathrm a}_N-1$.

\begin{enumerate}
\item
If $\sigma$ is a link pattern, since among the equations (3) of \thmref{defeq} for $E^{\mathrm a}_\pi$ are $s_i(M)=s_{i+1}(M)$, we have $E^{\mathrm a}_\pi \cap F_\sigma=E^{\mathrm a}_\pi\cap F_\sigma \cap \{s_i(M)=s_{i+1}(M)\}$, where $F_\sigma \cap \{s_i(M)=s_{i+1}(M)\}$ is irreducible and of the target dimension. Now, using the exact same reasoning as in the proof of \propref{defX}, if $\sigma$ differs from $\pi$ outside of $\{i,i+1,\sigma(i),\sigma(i+1)\}$ (and their images under the symplectic symmetry), then an equation among those of \thmref{defeq} for $E^{\mathrm a}_\pi$ is not satisfied, and the intersection has too low dimension.

This implies that $e_i\sigma=\pi$. Furthermore, if $|f_i\sigma|\ge|\sigma|$, i.e., if $i$ and $i+1$ are both connected to the boundary in $\sigma$, or the arches coming from them do not cross, then an additional equation of type (4) of \thmref{defeq}, namely if say $i+1$ is connected to $j$, the rank condition at $(i+1,j)$, is violated, and similarly in other cases. So $|f_i\sigma|<|\sigma|$.

We conclude in the end that $\sigma\in\varepsilon(i,\pi)$, $f_i\sigma\ne\sigma$.
In that case, from \thmref{defeq},
$E^{\mathrm a}_\pi \cap F_\sigma \subset F_\sigma\cap \{s_i(M)=s_{i+1}(M)\}$.
In fact, one can easily show the equality -- the proof is given in type $\hat{\mathrm A}$ in \cite[Appendix~B]{artic39},
but it works in all types, so we shall not repeat it here.
In the proof of \propref{defX} we have seen
that $F_\sigma\cap \{s_i(M)=s_{i+1}(M)\}$ is irreducible and that its closure is $X^{\mathrm a}_{\sigma, i}$.

The other two cases are obtained by assuming that $F_\sigma$ is of dimension $\dim E^{\mathrm a}_N-1$,
applying \corref{dimoneless} again, and eliminating cases by more of the same dimension considerations. In the end we find either:
\item $\sigma$ has a crossing of the form
$\begin{tikzpicture}[baseline=(current  bounding  box.center),scale=0.5]
\draw[thick] (0,0) circle(1);
\draw[smooth] (45:1) to[out=225,in=45] (225:1);
\draw[smooth] (135:1) to[out=315,in=135] (315:1);
\node at (135:1.4) {$a'$};
\node at (225:1.4) {$a$};
\node at (45:1.4) {$b'$};
\node at (315:1.4) {$b$};
\end{tikzpicture}$, which forces it to be $\sigma(i)=N-i$, $\sigma(i+1)=N-i+1$, and
$\sigma(j)=\pi(j)$ for $j\ne i,i+1$. In this case we claim that
$F_\sigma\cap E_\pi = F_\sigma$. This is identical to case (2) in the proof
of \propref{defX}.
First we check that $\overline{\tilde B_N\cdot \pi^{\mathrm i}_<}
\supset \tilde B_N\cdot \sigma^{\mathrm i}_<$. Indeed, the matrix $P$
with submatrix
$\left(\begin{smallmatrix}
t & 0 & 0 & 0
\\
0 &1/t&-1/t&0
\\
0 & 0 & t & 0
\\
0 & 0 & 0 & 1/t
\end{smallmatrix}\right)
$
at rows $i,i+1,N-i,N-i+1$ and identity elsewhere is symplectic and
sends $\pi^{\mathrm i}_<$ to $P \pi^{\mathrm i}_< P^{-1} = \sigma^{\mathrm i}_< + O(t^2)$.
We then conclude that $E_\pi \cap F_\sigma = \overline{F_\pi} \cap F_\sigma$
is a vector bundle over $\tilde B_N\cdot \sigma^{\mathrm i}_<$, and that the dimension
of its fiber matches that of $F_\sigma$, which implies
$E_\pi\cap F_\sigma=F_\sigma$.

If $\mathrm a=\mathrm c$, we show in \appref{inv} that $\overline{F_\sigma}$ is $SL_2^{(i)}$-invariant.

If $\mathrm a=\mathrm i$, $\overline{F_\sigma}=X^{\mathrm a}_{\rho,i}$ with $\rho(i)=N-i+1$, $\rho(i+1)=N-i$, i.e.,
this is the second case considered in the proof of \propref{defX}.
We find this way $X^{\mathrm a}_{\rho,i}$ with $f_i\rho=\rho$; or,

\item $\sigma$ has two fixed points which forces it to be
$\sigma(i)=i$, $\sigma(i+1)=i+1$ and $\sigma(j)=\pi(j)$ for $j\ne i,i+1$.
This is only possible for $\mathrm a=\mathrm p$, and the corresponding component is $SL_2^{(i)}$-invariant, as discussed in \cite[sect.~5.4]{artic39}.

\end{enumerate}

What we have obtained is a set-theoretic decomposition
of $E^{\mathrm a}_\pi \cap \{M_{i,i+1}=0\}$, and we need to calculate multiplicities.
This computation is performed in \appref{geomstuff}.
\end{proof}

\subsubsection*{Multidegree equality}
As usual, intersections of (irreducible) varieties with hypersurfaces result in the multidegree identities
\begin{align*}
\mdeg(E_\pi^{\mathrm a} \cap \{M_{i,i+1}=0\})&=(A+z_i-z_{i+1})\mdeg E_\pi^{\mathrm a},
\\
\mdeg(E_\rho^{\mathrm a} \cap \{s_i(M)=s_{i+1}(M)\})&=(2A-\epsilon)\mdeg E^{\mathrm a}_\rho.
\end{align*}
Finally, the decomposition of \propref{eicut} combined with \propref{defX} translates into
\[
(A+z_i-z_{i+1})\mdeg E^{\mathrm a}_\pi = \sum_{\rho\in \varepsilon(\pi,i)}
2^{\delta_{\rho_i,r}\delta_{\rho_{i+1},\ell}}
(2A-\epsilon)\mdeg E^{\mathrm a}_\rho\pmod{\Ker\divdif_i},
\]
where we have used \lmref{divdiff}
to take care of the $SL_2^{(i)}$-invariant terms.

We now apply $-\divdif_i$ and multiply by $2A+z_{i+1}-z_i-\epsilon$:
\begin{align*}
(2A+z_{i+1}-z_i-\epsilon)&(A+z_i-z_{i+1})(-\partial_i)\mdeg E_\pi
\\
&= (2A-\epsilon)\sum_{\rho\in \varepsilon(\pi,i)}
2^{\delta_{\rho_i,r}\delta_{\rho_{i+1},\ell}}
(2A+z_{i+1}-z_i-\epsilon)
(-\divdif_i)\mdeg E_{\rho}
\\
&=
(2A-\epsilon)\sum_{\rho\in \varepsilon(\pi,i)}
2^{\delta_{\rho_i,r}\delta_{\rho_{i+1},\ell}}
(\mdeg E_{\rho}+\mdeg E_{f_i\rho})
&&\qquad \llap{(using Eq.~\eqref{eq:fi})}
\\
&=
(2A-\epsilon)\sum_{\rho\in e_i^{-1}(\pi)-\{\pi\}}
2^{\delta_{\rho_i,b/\ell/r}\delta_{\rho_{i+1},b/\ell/r}}
\mdeg E_\rho.
\end{align*}
Now note that all the $\rho\in e_i^{-1}(\pi)-\{\pi\}$ such that $i$ and $i+1$ are not both connected to a boundary
have the same number of chords as $\pi$, whereas the ones such that they are have one extra chord, which matches the power of $2$ above;
therefore, using \eqref{eq:defpsi}, we find
\[
(2A+z_{i+1}-z_i-\epsilon)(A+z_i-z_{i+1})(-\partial_i)\phi^{\mathrm a}_\pi
=
(2A-\epsilon)\sum_{\rho\ne\pi: e_i\rho=\pi}
\phi^{\mathrm a}_\rho.
\]

\subsection{Geometry of the boundary exchange relations}

In what follows we are necessarily in type $\hat{\mathrm C}$.

\subsubsection{The invariant components}
\begin{prop}\label{prop:bdryinv}
Given a link pattern $\pi$, $E^{\mathrm a}_\pi$ is $SL_2^{(L)}$-invariant iff $\pi(L)\ne L+1$.
\end{prop}
\begin{proof}
The $SL_2^{(L)}$-invariance of $E_\pi^{\mathrm a}$ when $\pi(L)\ne L+1$ is given in \appref{inv}.

Conversely, if $\pi(L)=L+1$, $SL_2^{(L)}\cdot \underline \pi t$ has entries below the diagonal, so the corresponding component cannot be invariant.
\end{proof}

\subsubsection*{Multidegree equality}
Assume now that $\pi(L)\ne L+1$ (or, using the alternate notation, $\pi(L)\ne r$).
Note that this is necessarily the case if $\mathrm a\in\{\mathrm c,\mathrm m\}$.

We apply \lmref{divdiff}, minding, as explained right below it, the conjugation of the divided difference operator,
and find:
\[
\phi^{\mathrm a}_\pi\text{ is an even polynomial in $z_L$ }\times
\begin{cases}
1 &\mathrm a\in\{\mathrm c,\mathrm m\},
\\
A+2z_L&\mathrm a\in\{\mathrm i,\mathrm o\},\quad
\pi(L)\ne r.
\end{cases}
\]

With the exact same arguments, we find at the left boundary:
\[
\phi^{\mathrm a}_\pi\text{ is an even polynomial in $z_1$ }\times
\begin{cases}
1 &\mathrm a=\mathrm c,
\\
A-2z_1+\epsilon&\mathrm a\in\{\mathrm i,\mathrm o,\mathrm m\},\quad
\pi(1)\ne \ell.
\end{cases}
\]

\subsubsection{The noninvariant components}
We now assume that $\pi(L)=L+1$ (which implies $\mathrm a\in\{\mathrm i,\mathrm o\}$). The geometric construction corresponding to the boundary exchange relation \eqref{eq:qKZCKL} is the following:
\begin{itemize}
\item Cut $E^{\mathrm a}_\pi$ with $M_{L,L+1}=0$.
\item Throw away the $SL_2^{(L)}$-invariant components, and sweep with $SL_2^{(L)}$,
producing
\[ \bigcup_{\rho\ne\pi: e_L \rho = \pi} E^{\mathrm a}_\rho. \]
\end{itemize}

\subsubsection*{Determination of the result of cutting}
Given a link pattern $\rho$ such that $\rho(L)\ne L+1$, define $Y^{\mathrm a}_\rho$ to be the closure of $F_{\rho'}$,
where $\rho'$ is obtained from $\rho$ by permuting the images of $L$ and $L+1$,
i.e.,
\[\rho=\;
\begin{tikzpicture}[baseline=(current  bounding  box.center),scale=0.6]
\draw[thick] (0,0) circle(1);
\draw[smooth] (22.5:1) to[out=202.5,in=247.5] (67.5:1);
\draw[smooth] (157.5:1) to[out=337.5,in=292.5] (112.5:1);
\draw[smooth] (202.5:1) to[out=22.5,in=67.5] (247.5:1);
\draw[smooth] (292.5:1) to[out=112.5,in=157.5] (337.5:1);
\node at (337.5:1.3) {${\scriptstyle L}$};
\node at (17:1.6) {${\scriptstyle L+1}$};
\end{tikzpicture}\ ,
\qquad\qquad
\rho'=\;
\begin{tikzpicture}[baseline=(current  bounding  box.center),scale=0.6]
\draw[thick] (0,0) circle(1);
\draw[smooth] (337.5:1) to[out=157.5,in=247.5] (67.5:1);
\draw[smooth] (157.5:1) to[out=337.5,in=292.5] (112.5:1);
\draw[smooth] (202.5:1) to[out=22.5,in=67.5] (247.5:1);
\draw[smooth] (292.5:1) to[out=112.5,in=202.5] (22.5:1);
\node at (337.5:1.3) {${\scriptstyle L}$};
\node at (17:1.6) {${\scriptstyle L+1}$};
\end{tikzpicture}\ .
\]
Note that $\rho'$ is no longer a link pattern; in fact,
acording to \corref{dimoneless}, $\dim Y^{\mathrm a}_\rho = \dim E_N^{\mathrm a}-1$.

\begin{lm}\label{lm:YEE}
\[
Y^{\mathrm a}_\rho \subset E^{\mathrm a}_\rho \cap E^{\mathrm a}_{e_L\rho}.
\]
\end{lm}
\begin{proof}
This is exactly \cite[lemma 15]{artic39} at $i=L$ intersected
with $\M^{\mathrm a}_N$,
taking into account the remark after \conjref{fromp}.
\end{proof}
In fact, just as in \cite[lemma 15]{artic39}, we conjecture equality.

\begin{prop}\label{prop:bdrycut}
The geometric components of
$E^{\mathrm a}_\pi \cap \{ M_{L,L+1}=0 \}$ are the
$Y^{\mathrm a}_\rho$, $\rho\in e_L^{-1}(\pi)-\{\pi\}$ and some $SL_2^{(L)}$-invariant pieces.
The multiplicity of $Y^{\mathrm a}_\rho$ in $E^{\mathrm a}_\pi \cap \{ M_{L,L+1}=0 \}$ is $1$.
\end{prop}
\begin{proof} (We give the proof for $\mathrm a=\mathrm i$)
The proof is very similar to that of \propref{eicut}. $E^{\mathrm a}_\pi \cap \{ M_{L,L+1}=0 \}$ is of dimension one less than
$E^{\mathrm a}_N$, so we decompose
\[
E^{\mathrm a}_\pi \cap \{ M_{L,L+1}=0 \} = \bigsqcup_{\sigma: \sigma(L)\ne L+1} (E^{\mathrm a}_\pi \cap F_\sigma),
\]
and consider pieces of dimension $\dim E^{\mathrm a}_N-1$.

\begin{enumerate}
\item If $\sigma$ is a link pattern, we show in \appref{inv} that
any irreducible component of $\overline{E^{\mathrm a}_\pi \cap F_\sigma}$ of dimension $\dim E^{\mathrm a}_N-1$ is $SL_2^{(L)}$-invariant.

The other two cases are obtained by assuming $F_\sigma$ is of dimension $\dim E^{\mathrm a}_N-1$,
applying \corref{dimoneless} again, and eliminating cases by more of the same dimension considerations. In the end we find either:
\item $\sigma$ has a crossing of the form
$\begin{tikzpicture}[baseline=(current  bounding  box.center),scale=0.5]
\draw[thick] (0,0) circle(1);
\draw[smooth] (45:1) to[out=225,in=45] (225:1);
\draw[smooth] (135:1) to[out=315,in=135] (315:1);
\node at (135:1.4) {$a'$};
\node at (225:1.4) {$a$};
\node at (45:1.4) {$b'$};
\node at (315:1.4) {$b$};
\end{tikzpicture}$, which forces it to be of the form $\rho'$ described above, where $\rho$ is a link pattern such that $e_L\rho=\pi$, $\rho\ne\pi$.
In this case, according to \lmref{YEE}, $F_\sigma\cap E^{\mathrm a}_\pi=F_\sigma$ and its closure is $Y^{\mathrm a}_\rho$; or,

\item $\sigma$ has two fixed points which forces it to be
$\sigma(L)=L$, $\sigma(L+1)=L+1$ and $\sigma(j)=\pi(j)$ for $j\ne L,L+1$.
The corresponding component is $SL_2^{(L)}$-invariant, as proved in \appref{inv}.
\end{enumerate}

The multiplicity computation is performed in \appref{geomstuff}.
\end{proof}

\subsubsection*{The degree of the sweeping map}
The fiber of the map
$SL_2^{(L)}\times_{B^{(L)}} Y^{\mathrm a}_\rho\to SL_2^{(L)}\cdot Y_\rho^{\mathrm a}$ is more difficult to study than in \secref{fi}
because we have the identity $s_L(M)=s_{L+1}(M)$ (by symplectic symmetry), which means the block of $M^2$ on the $N^{\mathrm{th}}$ diagonal provides us no
useful information. Instead we proceed as follows.

Since the $SL_2^{(L)}$ action is a group action, we can as usual look at the fiber $\{(P,M)\ |\ PMP^{-1}=M'\}$ of an element $M'\in Y^{\mathrm a}_\rho$.
Consider the ranks of successive submatrices of $M'$ southwest of entries $(\rho(L),L-1), (\rho(L),L), (\rho(L),L+1)$, respectively.
Since $M'\in Y^{\mathrm a}_\rho$, these must be of the form $r,r,r+1$, where $r$ is the number of pairings inside $\{\rho(L+1)+1,\ldots,L-1\}$, e.g.,
\[ \rho=\begin{tikzpicture}[baseline=(current  bounding  box.center),scale=0.75,every node/.style={inner sep=0,outer sep=0}]
\draw[thick] (0,0) circle(1);
\draw[smooth] (191.25:1) to[out=11.25,in=348.75] (168.75:1);
\draw[smooth] (213.75:1) to[out=33.75,in=168.75] (348.75:1);
\draw[smooth] (236.25:1) to[out=56.25,in=101.25] (281.25:1);
\draw[smooth] (258.75:1) to[out=78.75,in=281.25] (101.25:1);
\draw[smooth] (303.75:1) to[out=123.75,in=146.25] (326.25:1);
\draw[smooth] (11.25:1) to[out=191.25,in=326.25] (146.25:1);
\draw[smooth] (33.75:1) to[out=213.75,in=236.25] (56.25:1);
\draw[smooth] (78.75:1) to[out=258.75,in=303.75] (123.75:1);
\node[anchor=east] at (146.25:1.2) {${\scriptstyle \rho(L+1)}$};
\node[anchor=west] at (11.25:1.2) {${\scriptstyle L+1}$};
\node[anchor=west] at (348.75:1.2) {${\scriptstyle L}$};
\node[anchor=east] at (213.75:1.2) {${\scriptstyle \rho(L)}$};
\end{tikzpicture}\ ,
\qquad
\rho'=\begin{tikzpicture}[baseline=(current  bounding  box.center),scale=0.75,every node/.style={inner sep=0,outer sep=0}]
\draw[thick] (0,0) circle(1);
\draw[smooth] (191.25:1) to[out=11.25,in=348.75] (168.75:1);
\draw[smooth] (213.75:1) to[out=33.75,in=191.25] (11.25:1);
\draw[smooth] (236.25:1) to[out=56.25,in=101.25] (281.25:1);
\draw[smooth] (258.75:1) to[out=78.75,in=281.25] (101.25:1);
\draw[smooth] (303.75:1) to[out=123.75,in=146.25] (326.25:1);
\draw[smooth] (348.75:1) to[out=168.75,in=326.25] (146.25:1);
\draw[smooth] (33.75:1) to[out=213.75,in=236.25] (56.25:1);
\draw[smooth] (78.75:1) to[out=258.75,in=303.75] (123.75:1);
\node[anchor=east] at (146.25:1.2) {${\scriptstyle \rho(L+1)}$};
\node[anchor=west] at (11.25:1.2) {${\scriptstyle L+1}$};
\node[anchor=west] at (348.75:1.2) {${\scriptstyle L}$};
\node[anchor=east] at (213.75:1.2) {${\scriptstyle \rho(L)}$};
\end{tikzpicture}\ ,
\qquad r=2.\]
Now conjugate $M'$ with $P\in SL_2^{(L)}$; the effect is to mix columns $L$ and $L+1$, and the same ranks for $P^{-1} M' P$ (for generic $P$) are
$r,r+1,r+1$. This violates the rank equations of $M$ unless $P$ does not send column $L+1$ to $L$, i.e., $P\in B^{(L)}$. This is equivalent to saying
that the fiber of the map $SL_2^{(L)}\times_{B^{(L)}} Y^{\mathrm a}_\rho\to SL_2^{(L)}\cdot Y_\rho^{\mathrm a}$ is of cardinality $1$.

\subsubsection*{Determination of the result of sweeping}
According to \lmref{YEE}, $Y^{\mathrm a}_\rho \subset E^{\mathrm a}_\rho$, and $E^{\mathrm a}_\rho$ is $SL_2^{(L)}$-invariant by \propref{bdryinv}.
So $SL_2^{(L)}\cdot Y^{\mathrm a}_\rho\subset E^{\mathrm a}_\rho$, and since the fiber above is finite, and $\dim Y^{\mathrm a}_\rho = \dim E^{\mathrm a}_N-1$,
$SL_2^{(L)}\cdot Y^{\mathrm a}_\rho$ and $E^{\mathrm a}_\rho$ have the same dimension.
We conclude from irreducibility of $E^{\mathrm a}_\rho$ that
\begin{equation}\label{eq:bdrysw}
SL_2^{(L)}\cdot Y_\rho^{\mathrm a} = E^{\mathrm a}_\rho.
\end{equation}

\subsubsection*{Multidegree identity}
\propref{bdrycut} implies that
\[
(A+2z_L) \mdeg E^{\mathrm a}_\pi
=
\sum_{\rho\ne\pi: e_L\rho = \pi}
\mdeg Y^{\mathrm a}_\rho.
\]

We then sweep with $SL_2^{(L)}$, apply \lmref{divdiff} with the generic fiber of cardinality $1$, and obtain using \eqref{eq:bdrysw}
\[
(A+2z_L) (-\partial_L) \mdeg E^{\mathrm a}_\pi
=
\sum_{\rho\ne\pi: e_L\rho = \pi}
\mdeg E^{\mathrm a}_\rho.
\]

Finally, noting that $\pi$ has one more chord than the $\rho\ne\pi$ such that $e_L\rho=\pi$, we have
\[
(A+2z_L) (-\partial_L) \phi^{\mathrm a}_\pi
=
2\sum_{\rho\ne\pi: e_L\rho = \pi}
\phi^{\mathrm a}_\rho.
\]

\subsection{Geometry of the recurrence relations}
This section follows closely \cite[Section~6]{artic33}, which covers the periodic case. Here we consider $\mathrm a\in\{\mathrm i,\mathrm c\}$.
\subsubsection{The bulk recurrence}
\begin{prop}
\label{prop:geomrec}
Fix an $i$, $0<i<n$. For a link pattern $\pi=\varphi_i\hat\pi$,
\[ \phi^{\mathrm a}_\pi(z_{i+1}=A+z_i)=A^2 p_i^{\mathrm a}(z_i|\dots,\hat z_i,\hat z_{i+1},\dots)\phi^{\mathrm a}_{\hat\pi}(\dots,\hat z_i,\hat z_{i+1},\dots), \]
where $p_i^{\mathrm i}$ and $p_i^{\mathrm c}$ are given in \eqref{eq:propfacs}, and $\varphi_i$ was defined in \lmref{RKids}.
%
\end{prop}
\begin{remark}
Note that if $\pi(i)\neq i+1$, $M_{i,i+1}=0$ so the multidegree disappears when $(A+z_i-z_{i+1})=0$.
\end{remark}
\begin{proof}
We define the hyperplane
\[ H^{\mathrm a} = \M^{\mathrm a}_N \cap \{M_{jk}=0\ |\ (j,k)\in \cl(i,i+1) \}, \]
and the linear spaces
\[ L^{\mathrm i}=\mathbb{C}(\mathrm e_{i,i+1}-\mathrm e_{N-i,N-i+1}),\qquad L^{\mathrm c}=\mathbb{C}(\mathrm e_{i,i+1}+\mathrm e_{N-i,N-i+1}), \]
noting $\M^{\mathrm a}_N=H^{\mathrm a}\times L^{\mathrm a}$. The equations defining $E^{\mathrm a}_\pi$ can be written as
\begin{equation}
\label{eq:qMr}
q_j M_{i,i+1}^{d_j}+r_j=0,\qquad j=0,1,\dots\ ,
\end{equation}
where in the $j$th equation, $d_j$ is the highest power of $M_{i,i+1}$, $q_j$ is the coefficient of $M_{i,i+1}^{d_j}$, and $r_j$ is the remainder. Now call $\Theta^{\mathrm a}_\pi$ the scheme defined by the equations $q_j=0$, $\forall j$. This can also be thought of as the result of taking $M_{i,i+1}$ to infinity in $E^{\mathrm a}_\pi$. Then by \cite[Corollary~2.6]{KMY}, we have
\[ \mdeg_{\mathcal M_N^{\mathrm a}}{E^{\mathrm a}_\pi}\big|_{A+z_i-z_{i+1}=0}=\mdeg_{H^{\mathrm a}}{\Theta^{\mathrm a}_\pi}\big|_{A+z_i-z_{i+1}=0}. \]
We can now extract some factors of the RHS by examining the defining equations of $\Theta^{\mathrm a}_\pi$.

Amongst the defining equations of $E^{\mathrm a}_\pi$ are the defining equations of $E^{\mathrm a}_N$:
\[ \left(M^2\right)_{kl}=\sum_{j=k+1}^{l-1}M_{kj}M_{jl}=0, \qquad l-k<N,\quad l\neq N-k+1. \]
Writing these equations in the form \eqref{eq:qMr}, we find that $d_j$ can either equal $1$ or $0$. For $d_j=1$ we must have $k=i$ or $l=i+1$, meaning that the following equations form part of the definition of $\Theta^{\mathrm a}_\pi$:
\begin{align*}
M_{jk}&=0,\quad (j,k)\in \cl(i+1,a),&& a\neq i,i+1,N-i+1,\\
M_{jk}&=0,\quad (j,k)\in \cl(b,i),&& b\neq i,i+1,N-i,
\end{align*}
and with these substituted into the remaining equations we find no dependence on $M_{jk}$ when $(j,k)\in \cl(i,a)\ \forall a\neq i+1$, $\cl(b,i+1)\ \forall b\neq i$, or $\cl(i+1,i)$ (see \figref{recurmat}).
\begin{figure}[ht]
\begin{tikzpicture}[scale=0.25]
\fill[lightgray] (6,-3) -- (7,-3) -- (7,-6) -- (6,-6);
\fill[lightgray] (8,-3) -- (12,-3) -- (12,-4) -- (8,-4);
\fill[lightgray] (12,-4) -- (13,-4) -- (13,-6) -- (12,-6);
\fill[lightgray] (8,-7) -- (8,-8) -- (13,-8) -- (13,-7);
\fill[pattern=north west lines,pattern color=lightgray] (8,-3) -- (8,-6) -- (7,-6) -- (7,-2) -- (12,-2) -- (12,-3);
\fill[pattern=north west lines,pattern color=lightgray] (8,-6) -- (13,-6) -- (13,-4) -- (14,-4) -- (14,-7) -- (8,-7);
\fill[pattern=north west lines,pattern color=lightgray] (12,-3) -- (12,-4) -- (13,-4) -- (13,-3);
\fill[black] (8,-6) -- (8,-7) -- (7,-7) -- (7,-6);
\draw (0,0) -- (10,-10) -- (20,-10) -- (10,0) -- cycle;
\draw[dashed] (10,0) -- (5,-5) (10,-10) -- (15,-5);
\draw[dotted] (0,0) -- (-1,1) (10,0) -- (9,1) (10,-10) -- (11,-11) (20,-10) -- (21,-11);
\draw[gray] (2,0) -- (2,-2) -- (12,-2) -- (12,-10);
\draw[gray] (3,0) -- (3,-3) -- (13,-3) -- (13,-10);
\draw[gray] (4,0) -- (4,-4) -- (14,-4) -- (14,-10);
\draw[gray] (6,0) -- (6,-6) -- (16,-6) -- (16,-10);
\draw[gray] (7,0) -- (7,-7) -- (17,-7) -- (17,-10);
\draw[gray] (8,0) -- (8,-8) -- (18,-8) -- (18,-10);
\node[anchor=east] at (0,-2.6) {${\scriptstyle i}$};
\node[anchor=east] at (0,-3.7) {${\scriptstyle i+1}$};
\node[anchor=east] at (0,-6.6) {${\scriptstyle N-i}$};
\node[anchor=east] at (0,-7.7) {${\scriptstyle N-i+1}$};
\node[anchor=south] at (6.5,0) {\rotatebox{90}{${\scriptstyle N-i}$}};
\node[anchor=south] at (7.5,0) {\rotatebox{90}{${\scriptstyle N-i+1}$}};
\node[anchor=south] at (12.5,0) {\rotatebox{90}{${\scriptstyle N+i}$}};
\node[anchor=south] at (13.5,0) {\rotatebox{90}{${\scriptstyle N+i+1}$}};
\end{tikzpicture}
\caption{The fundamental region of a generic matrix with $\mathrm i$ or $\mathrm c$-type symmetry. The black entry has been taken to infinity. In the defining equations of $\Theta^{\mathrm a}_\pi$, there is no dependence on those entries shaded with slanted lines, and the grey entries are equal to zero.}
\label{fig:recurmat}
\end{figure}
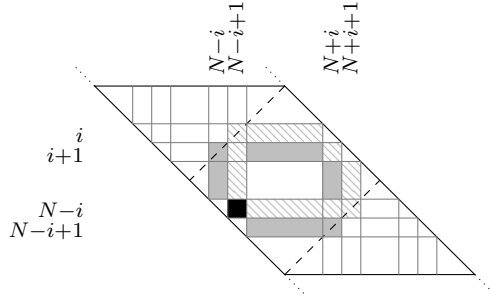

Taking the leading coefficient of $M_{i,i+1}$ in each of the other equations satisfied by $E^{\mathrm a}_\pi$ (as listed in \thmref{defeq}), we find these are also independent of all the matrix elements with an index of $i$, $i+1$, $N-i$, or $N-i+1$. We further find that they are exactly the equations of \thmref{defeq} that are satisfied on $E^{\mathrm a}_{\hat\pi}$, where $\hat\pi$ is the involution of size $N-4$ that is $\pi$ with the links from $i$ to $i+1$ and $N-i$ to $N-i+1$ removed.
Since, as observed after \thmref{defeq}, these equations define $E^{\mathrm a}_{\hat\pi}$ up to lower dimensional pieces,
and the flat limit of $E^{\mathrm a}_\pi$ is equidimensional, we conclude that after removal of rows and columns $i$, $i+1$, $N-i$, and $N-i+1$, we obtain
$E^{\mathrm a}_{\hat\pi}$ (up to embedded components, which are irrelevant for multidegree purposes).

We can therefore relate the multidegree of $\Theta^{\mathrm a}_\pi$ to the multidegree of $E^{\mathrm a}_{\hat\pi}$, by intersecting $\Theta^{\mathrm a}_\pi$ successively with a series of hyperplanes and using the inductive definition of the multidegree. The hyperplanes we use are the ones defined by $M_{jk}=0$ for $j$ or $k$ in $\cl(i)$ or $\cl(i+1)$ (with the exception of $(j,k)\in\cl(i,i+1)$ and, in the closed case, any choice of $j$ and $k$ for which $j=N-k+1$, because the matrix entries on the symmetry axis are already zero by definition). The result of intersecting $H^{\mathrm a}$ with these hyperplanes is $\M^{\mathrm a}_{N-4}$, so we have
\begin{multline*}
\mdeg_{H^{\mathrm i}}{\Theta^{\mathrm i}_\pi}\big|_{A+z_i-z_{i+1}=0}=A^2 (A-2z_i-\eps)(3A+2z_i) \prod_{a\neq i,i+1}^{n} (2A+z_a+z_i)(A-z_a-z_i-\eps) \\
\times \prod_{a=1}^{i-1} (A+z_a-z_i)(2A-z_a+z_i-\eps) \prod_{a=i+2}^{n}(A+z_a-z_i-\eps)(2A-z_a+z_i) \mdeg_{\M^{\mathrm i}_{N-4}}{E^{\mathrm i}_{\hat\pi}},
\end{multline*}
\begin{multline*}
\mdeg_{H^{\mathrm c}}{\Theta^{\mathrm c}_\pi}\big|_{A+z_i-z_{i+1}=0}=A^2 \prod_{a\neq i,i+1}^{n} (2A+z_a+z_i)(A-z_a-z_i-\eps)\\
\times \prod_{a=1}^{i-1} (A+z_a-z_i)(2A-z_a+z_i-\eps) \prod_{a=i+2}^{n}(A+z_a-z_i-\eps)(2A-z_a+z_i) \mdeg_{\M^{\mathrm c}_{N-4}}{E^{\mathrm c}_{\hat\pi}}.
\end{multline*}
Using \eqref{eq:defpsi}, we have the result.
\end{proof}

Similar arguments can be used to reproduce \propref{geomrec} in the cases $\mathrm a\in\{\mathrm o,\mathrm m\}$.

\subsubsection{The boundary recurrence}
\begin{prop}
\label{prop:geombdryrec}
For a link pattern $\pi=\tilde\varphi_n \hat\pi$,
\[ \phi^{\mathrm i}_\pi(z_n=-A/2)=A\ p^{\mathrm i}_n(z_1,\dots,z_{n-1})\ \phi^{\mathrm i}_{\hat\pi}(z_1,\dots,z_{n-1}), \]
and for a link pattern $\pi=\tilde\varphi_0 \hat\pi$,
\[ \phi^{\mathrm i}_\pi(z_1=(A-\eps)/2)=A\ p^{\mathrm i}_0(z_2,\dots,z_n)\ \phi^{\mathrm i}_{\hat\pi}(z_2,\dots,z_n), \]
where $p^{\mathrm i}_n$ and $p^{\mathrm i}_0$ are given in \eqref{eq:bdpropfacs}.
\end{prop}
\begin{remark}
As before, if $\pi(n)\neq n+1$ (resp.~$\pi(1)\neq N$), then $M_{n,n+1}=0$ ($M_{N1}=0$) and the multidegree disappears when $z_n=-A/2$ ($z_1=(A-\eps)/2$). Further, note that the proposition only refers to the identified case; in the closed case the below proof does not work because $M_{n,n+1}$ and $M_{N1}$ are both zero by definition, and thus do not appear in the defining equations. Since we only consider the identified case we will drop the $\mathrm i$ superscript for the proof.
\end{remark}
\begin{proof}
Right boundary: We define
\[ H = \M_N \cap \{M_{n,n+1}=0 \},\quad L=\mathbb{C}\mathrm e_{n,n+1}. \]
As before we write the defining equations of $E_\pi$ in the form $q_j M_{n,n+1}^{d_j}+r_j=0$, and call $\Theta_\pi$ the scheme defined by $q_j=0\ \forall j$. Again by \cite[Corollary~2.6]{KMY},
\[ \mdeg{E_\pi}\big|_{A+2z_n=0}=\mdeg_{H}{\Theta_\pi}\big|_{A+2z_n=0}. \]

From the defining equations of $E_N$ we find in the definition of $\Theta_\pi$ the following equations (see \figref{bdrecurmat}):
\[ M_{jk}=0,\quad (j,k)\in\cl(a,n), \qquad a\neq n,n+1, \]
and we also find no dependence on $M_{jk}$ for $(j,k)\in\cl(n,a)\ \forall a\neq n+1$ or $(j,k)\in\cl(n+1,n)$. We again find that the rest of the equations defining $\Theta_\pi$ are exactly the equations defining $E_{\hat\pi}$,
where $\hat\pi$ is the involution of size $N-2$ that is $\pi$ with the link from $n$ to $n+1$ removed.

\begin{figure}[ht]
\begin{tikzpicture}[scale=0.25]
\fill[lightgray] (11,-4) -- (11,-5) -- (5,-5) -- (5,-4);
\fill[pattern=north west lines,pattern color=lightgray] (11,-4) -- (11,-3) -- (5,-3) -- (5,-4);
\fill[pattern=north west lines,pattern color=lightgray] (12,-5) -- (12,-4) -- (11,-4) -- (11,-5);
\fill[black] (5,-3) -- (5,-4) -- (4,-4) -- (4,-3);
\draw (0,0) -- (8,-8) -- (16,-8) -- (8,0) -- cycle;
\draw[dashed] (8,0) -- (4,-4) (8,-8) -- (12,-4);
\draw[dotted] (0,0) -- (-1,1) (8,0) -- (7,1) (8,-8) -- (9,-9) (16,-8) -- (17,-9);
\draw[gray] (3,0) -- (3,-3) -- (11,-3) -- (11,-8);
\draw[gray] (4,0) -- (4,-4) -- (12,-4) -- (12,-8);
\draw[gray] (5,0) -- (5,-5) -- (13,-5) -- (13,-8);
\node[anchor=east] at (0,-3.6) {${\scriptstyle n}$};
\node[anchor=east] at (0,-4.6) {${\scriptstyle n+1}$};
\node[anchor=south] at (4.5,0) {\rotatebox{90}{${\scriptstyle n+1}$}};
\node[anchor=south] at (11.5,0) {\rotatebox{90}{${\scriptstyle N+n}$}};
\end{tikzpicture}
\caption{The fundamental region of a generic matrix with $\mathrm i$-type symmetry. The black entry has been taken to infinity. In the defining equations of $\Theta_\pi$, there is no dependence on those entries shaded with slanted lines, and the grey entries are equal to zero.}
\label{fig:bdrecurmat}
\end{figure}
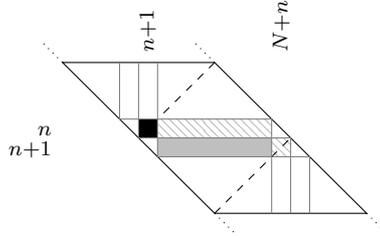

Intersecting $\Theta_\pi$ with hyperplanes defined by $M_{jk}=0$ for $j$ or $k$ in $\cl(n)$, excepting $(j,k)\in\cl(n,n+1)$, we find
\[ \mdeg_{H}{\Theta_\pi}\big|_{A+2z_n=0}=A\prod_{a=1}^{n-1}\frac{(3A+2z_j)(3A-2z_j-2\eps)}{4}\mdeg_{\M^{\mathrm i}_{N-2}}{E_{\hat\pi}}, \]
and again using \eqref{eq:defpsi}, we have the result.

Left boundary: The previous argument can be slightly modified, to obtain
\[ \mdeg{E_\pi}\big|_{A-2z_1-\eps=0}=A\prod_{a=2}^{n}\frac{(3A-2z_j-\eps)(3A+2z_j-\eps)}{4}\mdeg_{\M^{\mathrm i}_{N-2}}{E_{\hat\pi}}, \]
where $\hat\pi$ is the involution of size $N-2$ that is $\pi$ with the link from $1$ to $N$ removed.
\end{proof}
A similar argument can be made to reproduce \propref{geombdryrec} in the case $\mathrm a=\mathrm o$, as well as in the case $\mathrm a=\mathrm m$ where only the right boundary statement applies.

\subsection{The specialization \texorpdfstring{$\epsilon=2A$}{epsilon=2A}}
We mention an interesting specialization of the $\phi_\pi$, which is
$\epsilon=2A$. Note that this corresponds to the loop weight $\beta$ being
infinite.

\begin{prop}\label{prop:specinf}
One has:
\begin{align*}
\phi^{\mathrm p}_\pi|_{\epsilon=2A} &= \displaystyle
A^L (-1)^{|\pi|}\prod_{\substack{1\le i<j\le L\\ j\ne \pi(i)}} (A+z_i-z_j),
\\
\phi^{\mathrm i}_\pi|_{\epsilon=2A} &=\displaystyle
A^L (-1)^{|\pi|}
\prod_{i=1}^L (A+2z_i)\prod_{\substack{1\le i<j\le L\\ j\ne \pi(i)}} (A+z_i-z_j)(A+z_i+z_j),
\\
\phi^{\mathrm c}_\pi|_{\epsilon=2A} &=\displaystyle
A^L (-1)^{|\pi|}\prod_{\substack{1\le i<j\le L\\ j\ne \pi(i)}} (A+z_i-z_j)(A+z_i+z_j),
\\
\phi^{\mathrm o}_\pi|_{\epsilon=2A} &=\displaystyle
A^L
\prod_{1\le i<j\le L} (z_i-z_j)(z_i+z_j)(A+z_i-z_j)(A+z_i+z_j)
\prod_{i=1}^L
\begin{cases}
A+2z_i&\pi(i)=\ell,\\
-2z_i&\pi(i)=r,\\
0&\text{otherwise},
\end{cases}
\\[2mm]
\phi^{\mathrm m}_\pi|_{\epsilon=2A} &=\displaystyle
\begin{cases}
A^L
\prod_{1\le i<j\le L} (z_i-z_j)(z_i+z_j)(A+z_i-z_j)(A+z_i+z_j)
& \pi = (b,\ldots,b),\\
0&\text{otherwise},
\end{cases}
\end{align*}
where $|\pi|$ is defined as the number of crossings of $\pi$,
plus (the location of the unpaired site minus one) in odd size for $\mathrm a\in\{\mathrm p,\mathrm c\}$.
\end{prop}
\begin{proof}
The recurrence relations \eqref{eq:psirecur}, \eqref{eq:lrecur} and \eqref{eq:rrecur}, combined with the qKZ system \eqref{eq:qKZbulk}--\eqref{eq:qKZrotate} or \eqref{eq:qKZCR}--\eqref{eq:qKZCKL},
provide an infinite number of values for the $\phi_\pi$, even after
the specialization $\epsilon=2A$. So we only need to check that
all these equations are satisfied by the expression in the proposition.
We leave it as an exercise to the reader (see also \cite[Lemma 2]{artic32} for
a similar proof).
\end{proof}
It would be interesting to find a geometric interpretation of this
specialization.

\subsection{Conclusion}
\label{sec:conc}
Noting that $A^L|\phi^{\mathrm a}_\pi$ for all $\pi$ and all types $\mathrm a$ because this is nothing but
the product of weights of the equations $M_{ii}=0$, $i=1,\ldots,L$,
it is natural to redefine the $\phi^{\mathrm a}_\pi$
by dividing them by $A^L$;
since $E_N^{\mathrm a}$ actually sits in $(\M_N^{\mathrm a})_{\Delta=0}$,
we redefine
\[
\psi^{\mathrm a}_\pi=m_\pi \mdeg_{(\M_N^{\mathrm a})_{\Delta=0}} E_\pi^{\mathrm a}.
\]

The $\psi^{\mathrm a}_\pi$ satisfy the qKZ system \eqref{eq:qKZbulk}--\eqref{eq:qKZrotate} or \eqref{eq:qKZCR}--\eqref{eq:qKZCKL}, as well as recurrence relations of the form \eqref{eq:psirecur}, \eqref{eq:lrecur} and \eqref{eq:rrecur}. This shows existence of the would-be solution of the qKZ system that was studied in \secref{qKZ}, where its uniqueness was proved.

In particular, by setting $\epsilon=0$, we conclude from \propref{qkzevec} that the $\psi^{\mathrm a}_\pi$ are the entries of the ground state of the Brauer loop model.

We also show:
\begin{prop}\label{prop:gcd}
The greatest common denominator of the $\psi^{\mathrm a}_\pi|_{\epsilon=0}$, and therefore of the $\psi^{\mathrm a}_\pi$, is $1$.
\end{prop}
Here we ignore possible numerical factors, i.e., consider the gcd as polynomials with coefficients in $\mathbb Q$.
\begin{proof}
In types $\mathrm a\in\{\mathrm p,\mathrm c\}$, we use \propref{specinf}: the greatest common denominator of the $\phi^{\mathrm a}_\pi/A^L|_{\epsilon=A=0}$ is clearly $1$.

In other types, we use \propref{evecqkz}: $\ket\Psi/\text{gcd}(\psi_\pi)$ is the ground state eigenvector and therefore a solution of the qKZ system. But so is $\ket\Psi$, so that the gcd must be a Weyl-group invariant polynomial.
We then use the fully factorized components \eqref{eq:psis}: from the degree of $\ket\Psi$, the factors $S$ must
be equal to $1$. Furthermore, by inspection, no nontrivial Weyl-group invariant polynomial divides them.
(Alternatively, we can use the form of the fully factorized components directly at $\epsilon=0$, since
as remarked at the beginning of \secref{soln} they are the same
as those at generic $\epsilon$ in which we set $\epsilon=0$, and conclude by degree).
\end{proof}

Consider now the sum rule $Z^{\mathrm a}_L=\sum_{\pi\in\mathrm{LP}_L^{\mathrm a}} \psi_{\pi}^{\mathrm a}$ at $\eps=0$. It satisfies the recurrence relations given in \eqref{eq:recsum}. These combined with the symmetry properties specify
$Z^{\mathrm a}_L$
as a function of say $z_1$ at a certain number of points.
In \appref{bounddeg}, this number is carefully computed
for each type $\mathrm{a}$ and compared to the degree in each variable of
$Z^{\mathrm a}_L$. The former is found to be
strictly lower than that the latter,
so that $Z^{\mathrm a}_L$
is specified uniquely by these recurrence relations, along
with the initial condition $1$ in size $0$.

With the exception of the mixed case, the result can be written in determinant or pfaffian form. The expression for the periodic case comes from \cite{artic32}, and the closed case comes from \cite{PDF-open}. First define
\[ b(z_i,z_j)=\frac{(A^2-(z_i-z_j)^2)(A^2-(z_i+z_j)^2)}{z_i^2-z_j^2}. \]
The sum rules are
\begin{align}
\label{eq:norms} Z_L^{\mathrm{p}}&=
  \begin{dcases}
  2^{L/2}\prod_{1\leq i<j\leq L} \frac{A^2-(z_i-z_j)^2}{z_i-z_j} \Pf{\left[\frac{z_i-z_j}{A^2-(z_i-z_j)^2}\right]}_{1\leq i,j\leq L} &\; L \text{ even},\\
   (-2)^{(L+1)/2}\prod_{1\leq i<j\leq L}\frac{A^2-(z_i-z_j)^2}{z_i-z_j} \Pf\left[\begin{array}{cc} \left[\frac{z_i-z_j}{A^2-(z_i-z_j)^2}\right]_{1\leq i,j\leq L} & \left[-1\right]_{1\leq j\leq L}\\[2mm] \left[1\right]_{1\leq i\leq L} & 0\end{array}\right] &\; L \text{ odd},
  \end{dcases}\\
\nn Z_{L}^{\mathrm{i}}&=
  \begin{dcases}
  2^{L/2}\prod_{1\leq i<j\leq L}b(z_i,z_j)\;\Pf\left[\frac{(5A^2-2z_i^2-z_j^2)}{b(z_i,z_j)}\right]_{1\leq i,j\leq L} &\; L \text{ even},\\
   2^{(L+1)/2}\prod_{1\leq i<j\leq L}b(z_i,z_j)\;\Pf\left[\begin{array}{cc} \left[\frac{(5A^2-2z_i^2-2z_j^2)}{b(z_i,z_j)}\right]_{1\leq i,j\leq L} & \left[1\right]_{1\leq j\leq L}\\[2mm] \left[-1\right]_{1\leq i\leq L} & 0\end{array}\right] &\; L \text{ odd},
  \end{dcases}\\
\nn Z_L^{\mathrm{c}}&=
  \begin{dcases}
  2^{L/2}\prod_{1\leq i<j\leq L} b(z_i,z_j) \Pf{\left[\frac1{b(z_i,z_j)}\right]}_{1\leq i,j\leq L} &\qquad\qquad L \text{ even},\\
   2^{(L-1)/2}\prod_{1\leq i<j\leq L}b(z_i,z_j)\;\Pf\left[\begin{array}{cc} \left[\frac1{b(z_i,z_j)}\right]_{1\leq i,j\leq L} & \left[1\right]_{1\leq j\leq L}\\[2mm] \left[-1\right]_{1\leq i\leq L} & 0\end{array}\right] &\; L \text{ odd},
  \end{dcases}\\
\nn Z_{L}^{\mathrm{o}}&=
  \begin{dcases}
  (2A)^L\prod_{1\leq i<j\leq L}b(z_i,z_j)^2\;\det\left[\frac{(5A^2-2z_i^2-2z_j^2)}{b(z_i,z_j)}\right]_{1\leq i,j\leq L} & L \text{ even},\\
  2(2A)^L\prod_{1\leq i<j\leq L}b(z_i,z_j)^2\;\det\left[\begin{array}{cc} \left[\frac{(5A^2-2z_i^2-2z_j^2)}{b(z_i,z_j)}\right]_{1\leq i,j\leq L} & \left[1\right]_{1\leq j\leq L}\\[2mm] \left[-1\right]_{1\leq i\leq L} & 0\end{array}\right] & L \text{ odd}.
  \end{dcases}
\end{align}

Now we can compare these results to the localization formulae of
\secref{loc}. It is an elementary check that based on them,
the multidegree of $D_N^{\textrm a}$ divided by $A^L$ satisfies the same recurrence relations
\eqref{eq:recsum} and the same initial condition $\mdeg D_0^{\textrm a}=1$.
We therefore conclude

\begin{thm}\label{thm:sumrule}
We have the equality of multidegrees:
\[
\sum_{\pi\in\mathrm{LP}_L^{\mathrm a}} \phi_\pi^{\mathrm a}|_{\epsilon=0}
=\mdeg D^{\mathrm a}_N,
\]
where $N=L,2L,4L$ depending on $\mathrm a=\mathrm p,\mathrm{i/c},\mathrm {o/m}$.
\end{thm}

Recall that this multidegree is also equal to that of the flat limit
of $D^{\mathrm a}_N$, namely $D^{\mathrm a}_{N;0}$
(see \secref{orbclos} for its definition).
It is shown in \appref{mult} that this implies that
$D_{N;0}^{\mathrm a}=E^{\mathrm a}_N$ as sets
and that $m_\pi$ is the multiplicity
of $E_\pi^{\mathrm a}$ in either $D_{N;0}^{\mathrm a}$ or $\tilde E_N^{\mathrm a}$
(which are conjecturally equal as schemes).


\appendix
\section{Small size examples}
\subsection{Identified case}

$L=2$:
\begin{align*}
\psi^{\mathrm i}_{bb}&=4
(2A-s+{z}_{2}-{z}_{1}) (A+{z}_{1}-{z}_{2})
\\
\psi^{\mathrm i}_{21}&=2
(A-s-2 {z}_{1}) (A+2 {z}_{2})
\end{align*}

$L=3$:
\begin{dgroup*}
\begin{dmath*}
\psi^{\mathrm i}_{21b}= 4
(A-s-2 {z}_{1})  (A+{z}_{2}-{z}_{3}) (A+{z}_{2}+{z}_{3})  (7 A^{3}-9 A^{2} s+3 A s^{2}-3 A s {z}_{1}+2 s^{2} {z}_{1}-3 A {z}_{1}^{2}+2 s {z}_{1}^{2}-4 A^{2} {z}_{2}+3 A s {z}_{2}+2 s {z}_{1} {z}_{2}+2 {z}_{1}^{2} {z}_{2}-A {z}_{3}^{2}-2 {z}_{2} {z}_{3}^{2})
\end{dmath*}
\begin{dmath*}
\psi^{\mathrm i}_{bbb}= 8
(2 A-s+{z}_{3}-{z}_{2}) (2 A-s+{z}_{3}-{z}_{1}) (2 A-s+{z}_{2}-{z}_{1})  (A+{z}_{2}-{z}_{3}) (A+{z}_{1}-{z}_{3}) (A+{z}_{1}-{z}_{2})
\end{dmath*}
\begin{dmath*}
\psi^{\mathrm i}_{3b1}= 4
(A-s-{z}_{1}-{z}_{2}) (A-s-2 {z}_{1}) (A+2 {z}_{3}) (A+{z}_{2}-{z}_{3}) (A+{z}_{2}+{z}_{3}) (A+{z}_{1}-{z}_{2})
\end{dmath*}
\begin{dmath*}
\psi^{\mathrm i}_{b32}= 4
(A-s-{z}_{1}-{z}_{2}) (A+2 {z}_{3}) (A+{z}_{1}-{z}_{2})  (7 A^{3}-7 A^{2} s+2 A s^{2}-A s {z}_{1}+s^{2} {z}_{1}-A {z}_{1}^{2}+s {z}_{1}^{2}+4 A^{2} {z}_{2}-3 A s {z}_{2}+s^{2} {z}_{2}+2 s {z}_{1} {z}_{2}+2 {z}_{1}^{2} {z}_{2}-3 A {z}_{3}^{2}+s {z}_{3}^{2}-2 {z}_{2} {z}_{3}^{2})
\end{dmath*}
\end{dgroup*}

\subsection{Closed case}
$L=3$:
\begin{align*}
\psi^{\mathrm c}_{21\bullet}&=4
(A+{z}_{2}-{z}_{3}) (A+{z}_{2}+{z}_{3}) (3 A^{2}-3 A s+s^{2}+s {z}_{1}+{z}_{1}^{2}-2 A {z}_{2}+s {z}_{2}-{z}_{3}^{2})
\\
\psi^{\mathrm c}_{\bullet 32}&=4
(A-s-{z}_{1}-{z}_{2})
(A+{z}_{1}-{z}_{2}) (3 A^{2}-2 A s-s {z}_{1}-{z}_{1}^{2}+2 A {z}_{2}-s {z}_{2}+{z}_{3}^{2})
\\
\psi^{\mathrm c}_{3\bullet 1}&=4
(A-s-{z}_{1}-{z}_{2}) (A+{z}_{2}-{z}_{3}) (A+{z}_{2}+{z}_{3}) (A+{z}_{1}-{z}_{2})
\end{align*}

$L=4$:
\begin{dgroup*}
\footnotesize
\begin{dmath*}
\psi^{\mathrm c}_{3412}=4 (A+z_1-z_2) (A+z_2-z_3) (A+z_3-z_4) (A+z_3+z_4)
   (A-s-z_1-z_2) \left(5 A^3-6 A^2 s+3 A^2 z_2-3 A^2 z_3
   +2
   A s^2-A s z_1-3 A s z_2+2 A s z_3-A
   z_1^2-2 A z_2 z_3-A z_4^2+s^2 z_1+s^2
   z_2+s z_1^2+s z_1 z_2+s z_1
   z_3+s z_2 z_3
   +z_1^2 z_2+z_1^2 z_3-z_2
   z_4^2-z_3 z_4^2\right)
\end{dmath*}
\begin{dmath*}
\psi^{\mathrm c}_{2143}=4 (A+z_2-z_3) \left(23 A^7-59 s A^6-7 z_2 A^6+7 z_3 A^6+60
   s^2 A^5-10 z_1^2 A^5-11 z_2^2 A^5-11 z_3^2 A^5-10 z_4^2 A^5
   -10
   s z_1 A^5+9 s z_2 A^5-19 s z_3 A^5+2 z_2
   z_3 A^5-28 s^3 A^4+3 z_2^3 A^4-3 z_3^3 A^4+28 s z_1^2
   A^4+23 s z_2^2 A^4+19 s z_3^2 A^4
   +z_2 z_3^2 A^4+14
   s z_4^2 A^4-10 z_2 z_4^2 A^4-18 z_3 z_4^2 A^4+28
   s^2 z_1 A^4+4 s^2 z_2 A^4+18 z_1^2 z_2 A^4+18 s
   z_1 z_2 A^4+24 s^2 z_3 A^4
   +10 z_1^2 z_3 A^4-z_2^2
   z_3 A^4+10 s z_1 z_3 A^4+5 s^4 A^3+3 z_1^4 A^3+3
   z_4^4 A^3+6 s z_1^3 A^3-5 s z_2^3 A^3+3 s z_3^3
   A^3-4 z_2 z_3^3 A^3
   -20 s^2 z_1^2 A^3-18 s^2 z_2^2 A^3+4
   z_1^2 z_2^2 A^3+4 s z_1 z_2^2 A^3-11 s^2 z_3^2
   A^3+9 z_1^2 z_3^2 A^3+3 z_2^2 z_3^2 A^3+9 s z_1 z_3^2
   A^3-2 s z_2 z_3^2 A^3
   -9 s^2 z_4^2 A^3-11 z_1^2
   z_4^2 A^3+9 z_2^2 z_4^2 A^3+4 z_3^2 z_4^2 A^3-11 s
   z_1 z_4^2 A^3+21 s z_2 z_4^2 A^3+30 s z_3
   z_4^2 A^3+16 z_2 z_3 z_4^2 A^3
   -23 s^3 z_1 A^3-9
   s^3 z_2 A^3-18 s z_1^2 z_2 A^3-18 s^2 z_1
   z_2 A^3-14 s^3 z_3 A^3-4 z_2^3 z_3 A^3-4 s z_1^2
   z_3 A^3-s z_2^2 z_3 A^3
   -4 s^2 z_1 z_3 A^3+16
   z_1^2 z_2 z_3 A^3+16 s z_1 z_2 z_3 A^3-5 s
   z_1^4 A^2-3 s z_4^4 A^2+z_2 z_4^4 A^2+3 z_3 z_4^4
   A^2-10 s^2 z_1^3 A^2+2 s^2 z_2^3 A^2
   -4 z_1^2 z_2^3 A^2-4
   s z_1 z_2^3 A^2-s^2 z_3^3 A^2-z_1^2 z_3^3
   A^2+z_2^2 z_3^3 A^2-s z_1 z_3^3 A^2+5 s z_2
   z_3^3 A^2+s^3 z_1^2 A^2+5 s^3 z_2^2 A^2
   -11 s
   z_1^2 z_2^2 A^2-11 s^2 z_1 z_2^2 A^2+2 s^3 z_3^2
   A^2-z_2^3 z_3^2 A^2-11 s z_1^2 z_3^2 A^2-3 s z_2^2
   z_3^2 A^2-11 s^2 z_1 z_3^2 A^2+2 s^2 z_2 z_3^2
   A^2
   +z_1^2 z_2 z_3^2 A^2+s z_1 z_2 z_3^2 A^2+4
   s^3 z_4^2 A^2+z_2^3 z_4^2 A^2+4 z_3^3 z_4^2 A^2+14
   s z_1^2 z_4^2 A^2-10 s z_2^2 z_4^2 A^2-3 s
   z_3^2 z_4^2 A^2
   +4 z_2 z_3^2 z_4^2 A^2+14 s^2 z_1
   z_4^2 A^2-15 s^2 z_2 z_4^2 A^2+z_1^2 z_2 z_4^2
   A^2+s z_1 z_2 z_4^2 A^2-19 s^2 z_3 z_4^2
   A^2-z_1^2 z_3 z_4^2 A^2-z_2^2 z_3 z_4^2 A^2
   -s
   z_1 z_3 z_4^2 A^2-19 s z_2 z_3 z_4^2 A^2+6
   s^4 z_1 A^2+3 s^4 z_2 A^2-3 z_1^4 z_2 A^2-6 s
   z_1^3 z_2 A^2-3 s^2 z_1^2 z_2 A^2+3 s^4 z_3
   A^2-z_1^4 z_3 A^2
   -2 s z_1^3 z_3 A^2+4 s z_2^3
   z_3 A^2-7 s^2 z_1^2 z_3 A^2+s^2 z_2^2 z_3 A^2-4
   z_1^2 z_2^2 z_3 A^2-4 s z_1 z_2^2 z_3 A^2-6
   s^3 z_1 z_3 A^2-2 s^3 z_2 z_3 A^2
   -22 s
   z_1^2 z_2 z_3 A^2-22 s^2 z_1 z_2 z_3 A^2+2
   s^2 z_1^4 A+s^2 z_4^4 A+z_1^2 z_4^4 A-2 z_2^2
   z_4^4 A-z_3^2 z_4^4 A+s z_1 z_4^4 A-2 s z_2
   z_4^4 A
   -3 s z_3 z_4^4 A-2 z_2 z_3 z_4^4 A+4
   s^3 z_1^3 A+3 s z_1^2 z_2^3 A+3 s^2 z_1
   z_2^3 A+2 z_2^3 z_3^3 A+s z_2^2 z_3^3 A-2 s^2
   z_2 z_3^3 A-2 z_1^2 z_2 z_3^3 A
   -2 s z_1 z_2
   z_3^3 A+2 s^4 z_1^2 A-z_1^4 z_2^2 A-2 s z_1^3
   z_2^2 A+5 s^2 z_1^2 z_2^2 A+6 s^3 z_1 z_2^2 A-2
   z_1^4 z_3^2 A-4 s z_1^3 z_3^2 A+2 s z_2^3
   z_3^2 A
   +s^2 z_1^2 z_3^2 A+2 s^2 z_2^2 z_3^2
   A+z_1^2 z_2^2 z_3^2 A+s z_1 z_2^2 z_3^2 A+3
   s^3 z_1 z_3^2 A-s^3 z_2 z_3^2 A-2 s z_1^2
   z_2 z_3^2 A-2 s^2 z_1 z_2 z_3^2 A-s^4 z_4^2
   A
   +z_1^4 z_4^2 A+2 s z_1^3 z_4^2 A-s z_2^3
   z_4^2 A-3 s z_3^3 z_4^2 A-4 s^2 z_1^2 z_4^2 A+4
   s^2 z_2^2 z_4^2 A+z_1^2 z_2^2 z_4^2 A+s z_1
   z_2^2 z_4^2 A+s^2 z_3^2 z_4^2 A
   +z_1^2 z_3^2
   z_4^2 A+z_2^2 z_3^2 z_4^2 A+s z_1 z_3^2 z_4^2
   A-2 s z_2 z_3^2 z_4^2 A-5 s^3 z_1 z_4^2 A+5
   s^3 z_2 z_4^2 A+s z_1^2 z_2 z_4^2 A+s^2
   z_1 z_2 z_4^2 A
   +6 s^3 z_3 z_4^2 A-2 z_2^3 z_3
   z_4^2 A+3 s z_1^2 z_3 z_4^2 A+3 s^2 z_1 z_3
   z_4^2 A+9 s^2 z_2 z_3 z_4^2 A+6 z_1^2 z_2 z_3
   z_4^2 A+6 s z_1 z_2 z_3 z_4^2 A+s z_1^4
   z_2 A
   +2 s^2 z_1^3 z_2 A+4 s^3 z_1^2 z_2 A+3
   s^4 z_1 z_2 A-s z_1^4 z_3 A-2 s^2 z_1^3
   z_3 A-s^2 z_2^3 z_3 A+2 s^3 z_1^2 z_3 A+3
   s z_1^2 z_2^2 z_3 A+3 s^2 z_1 z_2^2 z_3 A
   +3
   s^4 z_1 z_3 A+s^4 z_2 z_3 A-2 z_1^4 z_2
   z_3 A-4 s z_1^3 z_2 z_3 A+6 s^2 z_1^2 z_2
   z_3 A+8 s^3 z_1 z_2 z_3 A-z_3^3 z_4^4+s
   z_2^2 z_4^4-z_2 z_3^2 z_4^4
   +s^2 z_2
   z_4^4+z_1^2 z_2 z_4^4+s z_1 z_2
   z_4^4+s^2 z_3 z_4^4+z_1^2 z_3 z_4^4+s
   z_1 z_3 z_4^4+s z_2 z_3 z_4^4+z_1^4
   z_2^3+2 s z_1^3 z_2^3+s^2 z_1^2 z_2^3-s
   z_2^3 z_3^3
   -s^2 z_2^2 z_3^3-z_1^2 z_2^2
   z_3^3-s z_1 z_2^2 z_3^3+2 s z_1^4 z_2^2+4
   s^2 z_1^3 z_2^2+2 s^3 z_1^2 z_2^2+s z_1^4
   z_3^2+2 s^2 z_1^3 z_3^2-s^2 z_2^3
   z_3^2-z_1^2 z_2^3 z_3^2-s z_1 z_2^3
   z_3^2
   +s^3 z_1^2 z_3^2-s^3 z_2^2 z_3^2-2 s
   z_1^2 z_2^2 z_3^2-2 s^2 z_1 z_2^2 z_3^2-s
   z_1^4 z_4^2-2 s^2 z_1^3 z_4^2-z_1^2 z_2^3
   z_4^2-s z_1 z_2^3 z_4^2+s^2 z_3^3
   z_4^2+z_1^2 z_3^3 z_4^2+z_2^2 z_3^3 z_4^2
   +s
   z_1 z_3^3 z_4^2+s z_2 z_3^3 z_4^2-s^3
   z_1^2 z_4^2-s^3 z_2^2 z_4^2-2 s z_1^2 z_2^2
   z_4^2-2 s^2 z_1 z_2^2 z_4^2+z_2^3 z_3^2
   z_4^2+s z_2^2 z_3^2 z_4^2+s^2 z_2 z_3^2
   z_4^2+z_1^2 z_2 z_3^2 z_4^2
   +s z_1 z_2
   z_3^2 z_4^2-s^4 z_2 z_4^2-z_1^4 z_2 z_4^2-2
   s z_1^3 z_2 z_4^2-3 s^2 z_1^2 z_2 z_4^2-2
   s^3 z_1 z_2 z_4^2-s^4 z_3 z_4^2-z_1^4
   z_3 z_4^2-2 s z_1^3 z_3 z_4^2+s z_2^3
   z_3 z_4^2
   -3 s^2 z_1^2 z_3 z_4^2-z_1^2 z_2^2
   z_3 z_4^2-s z_1 z_2^2 z_3 z_4^2-2 s^3
   z_1 z_3 z_4^2-2 s^3 z_2 z_3 z_4^2-4 s
   z_1^2 z_2 z_3 z_4^2-4 s^2 z_1 z_2 z_3
   z_4^2+s^2 z_1^4 z_2+2 s^3 z_1^3 z_2
   +s^4
   z_1^2 z_2+s^2 z_1^4 z_3+2 s^3 z_1^3
   z_3+s^4 z_1^2 z_3+z_1^4 z_2^2 z_3+2 s
   z_1^3 z_2^2 z_3+s^2 z_1^2 z_2^2 z_3+2 s
   z_1^4 z_2 z_3+4 s^2 z_1^3 z_2 z_3+2 s^3
   z_1^2 z_2 z_3\right)
\end{dmath*}
\begin{dmath*}
\psi^{\mathrm c}_{4321}=4 (A+z_1-z_2) (A+z_3-z_4) (A+z_3+z_4)
   (A-s-z_1-z_2) \left(11 A^4-18 A^3 s+8 A^3 z_2-8 A^3
   z_3+10 A^2 s^2
   -3 A^2 s z_1-11 A^2 s z_2+8 A^2
   s z_3-3 A^2 z_1^2+A^2 z_2^2-8 A^2 z_2 z_3+A^2 z_3^2-3
   A^2 z_4^2-2 A s^3+3 A s^2 z_1+5 A s^2 z_2
   -2 A
   s^2 z_3+3 A s z_1^2+2 A s z_1 z_3-A s
   z_2^2+6 A s z_2 z_3+2 A s z_4^2+2 A z_1^2
   z_3-2 A z_2^2 z_3+2 A z_2 z_3^2-2 A z_2
   z_4^2-s^3 z_1
   -s^3 z_2-s^2 z_1^2-s^2
   z_1 z_2-s^2 z_1 z_3-s^2 z_2 z_3-s
   z_1^2 z_2-s z_1^2 z_3-s z_1 z_2^2+s
   z_1 z_4^2+s z_2^2 z_3+s z_2
   z_4^2-z_1^2 z_2^2+z_1^2 z_4^2
   +z_2^2 z_3^2-z_3^2
   z_4^2\right)
\end{dmath*}
\end{dgroup*}

\subsection{Open case}
$L=2$:
\begin{dgroup*}
\begin{dmath*}
\psi^{\mathrm o}_{r\ell}=4
(A-s-2 {z}_{1}) (A+2 {z}_{2}) (A+{z}_{1}-{z}_{2}) (A+{z}_{1}+{z}_{2})  (A-s-{z}_{1}-{z}_{2}) (A-s-{z}_{1}+{z}_{2})
\end{dmath*}
\begin{dmath*}
\psi^{\mathrm o}_{\ell\ell}=4
(A+2 {z}_{2}) (A+{z}_{1}-{z}_{2}) (A+{z}_{1}+{z}_{2}) (A+2 {z}_{1}) (2 A-2 s-{z}_{1}-{z}_{2}) (2 A-2 s-{z}_{1}+{z}_{2})
\end{dmath*}
\begin{dmath*}
\psi^{\mathrm o}_{rr}=4
(A+{z}_{1}-{z}_{2}) (A-s-2 {z}_{2}) (A-s-{z}_{1}-{z}_{2}) (A-s-2 {z}_{1}) (2 A-2 s-{z}_{1}+{z}_{2}) (2 A-s+{z}_{1}+{z}_{2})
\end{dmath*}
\begin{dmath*}
\psi^{\mathrm o}_{\ell r}=4
(A-s-{z}_{1}-{z}_{2}) (A+{z}_{1}-{z}_{2}) (A+{z}_{1}+{z}_{2})  (11 A^{3}-26 A^{2} s+19 A s^{2}-4 s^{3}-3 A^{2} {z}_{1}+A s {z}_{1}+2 s^{2} {z}_{1}-2 A {z}_{1}^{2}+2 s {z}_{1}^{2}+3 A^{2} {z}_{2}-7 A s {z}_{2}+4 s^{2} {z}_{2}-8 A {z}_{1} {z}_{2}+10 s {z}_{1} {z}_{2}+4 {z}_{1}^{2} {z}_{2}-2 A {z}_{2}^{2}-4 {z}_{1} {z}_{2}^{2})
\end{dmath*}
\begin{dmath*}
\psi^{\mathrm o}_{21}=2
(A-s) (A-s-2 {z}_{1}) (A+2 {z}_{2}) (2 A-s)
 (5 A^{2}-7 A s+2 s^{2}-2 s {z}_{1}-2 {z}_{1}^{2}-2 {z}_{2}^{2})
\end{dmath*}
\end{dgroup*}

\subsection{Mixed case}
$L=2$:
\begin{align*}
\psi^{\mathrm m}_{21}&= 2
(A-s) (A+2 {z}_{2}) (4 A^{2}-5 A s+2 s^{2}+2 s {z}_{1}+2 {z}_{1}^{2}-2 {z}_{2}^{2})
\\
\psi^{\mathrm m}_{rr}&= 4
(2 A-s+{z}_{1}+{z}_{2}) (2 A-2 s-{z}_{1}+{z}_{2}) (A-s-{z}_{1}-{z}_{2}) (A+{z}_{1}-{z}_{2})
\end{align*}

\section{Multiplicity computations}\label{app:geomstuff}
\subsection{Multiplicity of \texorpdfstring{$E_{\pi}^{\mathrm a}$}{component} in \texorpdfstring{$\tilde E_{N}^{\mathrm a}$}{scheme}}\label{app:mult}
We consider here the {\em tangent cone}\/ of $\tilde E_N^{\mathrm a}$ at the point $\underline\pi t$ ($t$ generic diagonal) of $E_\pi^{\mathrm a}$, which is defined by taking the leading terms
of the ideal of equations of $\tilde E_N^{\mathrm a}$ expanded around that point. Let us write $M=\underline\pi t+P$.
Among the equations for $P$, we obtain of course all the equations
of the tangent space, which were computed in the proof of \thmref{genred}; furthermore, the diagonal entries now satisfy $M_{ii}^2=0$, or equivalently $P_{ii}^2=0$.
Also, the equations $(P\underline{\pi}t + \underline{\pi}t P)_{ij}=0$ are now nontrivial when $j=\pi(i)$, resulting in (after simplifying with $t_j\ne 0$)
\[
P_{ii}+P_{jj}=0, \qquad i=\pi(j).
\]
In principle there may be more equations in the ideal of leading terms; as we shall see, this is already enough to get
a bound on the multiplicity, which we shall then show is saturated.

According to the above, the other nonlinear equations concern the diagonal entries of $P$; they all satisfy $P_{ii}^2=0$, but those connected by either $\pi$ or by the symplectic symmetry
are equal, so the number of independent variables that square to zero is exactly the number of chords in the link pattern,
plus the number of fixed points (the latter only occur for odd $L$, $\mathrm a \in \{\mathrm p,\mathrm c\}$).
So we find that the tangent cone sits inside a scheme of degree
equal to $m_\pi=2^{\#\{\text{chords}(\pi)\}+\#\{\text{fixed points}(\pi)\}}$, and is of the same dimension (the would-be extra equations cannot change the dimension of the tangent cone,
since the point $\underline{\pi}t$ is smooth in 
$E^{\mathrm a}_N$, as found in the proof of \thmref{genred}).
Therefore the degree of the tangent cone is less than or equal to $m_\pi$.
Since the equations of $\tilde E_N^{\mathrm a}$ are invariant by conjugation
and the union of orbits by conjugation of the $\underline\pi t$ is dense in $E_\pi^{\mathrm a}$ (\thmref{denseorbit}), this is also true of
the multiplicity of $E_\pi^{\mathrm a}$ in $\tilde E_N^{\mathrm a}$.
Therefore, we have
\[
\mdeg \tilde E_N^{\mathrm a} \le \sum_\pi m_\pi \mdeg E^{\mathrm a}_\pi =\sum_\pi \phi^{\mathrm a}_\pi,
\]
where inequality is here in the sense of multidegrees with positive multigrading, see e.g.\ \cite[lemma 12]{artic39} for details. We now specialize the multidegrees by setting $\epsilon=0$ (this corresponds to equivariance w.r.t.\ a codimension 1 subtorus, and does not spoil positivity of the multigrading).

According to \thmref{sumrule} and \eqref{eq:flatlimmdeg},
\[
\mdeg D_{N;0}^{\mathrm a}|_{\epsilon=0}=\sum_\pi \phi^{\mathrm a}_\pi|_{\epsilon=0},
\]
and according to \propref{normalcone}, $D_{N;0}^{\mathrm a}\subset\tilde E_N^{\mathrm A}$ as schemes.
Therefore the inequality, $\mdeg \tilde E_N^{\mathrm a} \le \mdeg D_{N;0}^{\mathrm a}$ is an equality,
and $\tilde E_N^{\mathrm a}$ and $D_{N;0}^{\mathrm a}$ have the same top-dimensional irreducible components, with the same
multiplicities. Finally, the equation above determines them to be:
\[
\text{multiplicity of $E_\pi^{\mathrm a}$ in $\tilde E_N^{\mathrm a}$} = m_\pi=2^{\#\{\text{chords}(\pi)\}+\#\{\text{fixed points}(\pi)\}}.
\]

\subsection{Multiplicity of \texorpdfstring{$X_{\rho,i}^{\mathrm a}$}{X}}
In this section, given a link pattern $\rho$ such that $|\rho|\ge |f_i\rho|$,
 we compute the multiplicity of $X^{\mathrm a}_{\rho,i}$ in
either $E^{\mathrm a}_\rho \cap \{s_i(M)=s_{i+1}(M)\}$ or $E^{\mathrm a}_\pi \cap \{M_{i,i+1}=0\}$ (assuming
$\rho\in\varepsilon(\pi,i)$).

We consider a point of the form $\underline \rho t$ where the $t_j$'s are nonzero
and $t_i t_{\rho(i)}=t_{i+1}t_{\rho(i+1)}$. By direct inspection, $\underline\rho t\in X^{\mathrm a}_{\rho,i}$.
First we compute its Zariski tangent space in $E_N^{\mathrm a}$. This is the exact same calculation that was performed in the proof of \thmref{genred},
so we do not repeat it here. In fact all the cases that we need here are given in the proof.
For $\mathrm a=\mathrm i$:
\begin{enumerate}\setcounter{enumi}{1}
\item If $i$ and $i+1$ are both connected to the boundary in $\rho$, then the counting is the same.

\item If one of the two is connected to the boundary in $\rho$, say $i+1$, then
naively there is one less independent equation because $t_i t_{\rho(i)}=t_{i+1}t_{\rho(i+1)}$. However, the two equations
which become proportional involve the variable $P_{i,i+1}$, which is here equal to $M_{i,i+1}=0$
either from \thmref{defeq} in $E^{\mathrm a}_\rho$ or by definition in $E^{\mathrm a}_\pi \cap \{ M_{i,i+1}=0 \}$,
so that we are back to the original count.

\item Finally, if neither is connected to the boundary, then we lose two equations. Once again, reimposing
$P_{i,i+1}=M_{i,i+1}=0$ gives one more equation. Similarly, consider the rank condition
on the interval $[i+1,\rho(i)]$ for either $E^{\mathrm a}_\pi$ or $E^{\mathrm a}_\rho$:
$\rkm(M)_{i+1,\rho(i)}=
\rkm(\underline\pi)_{i+1,\rho(i)}=\rkm(\underline\rho)_{i+1,\rho(i)}=r$,
where $(j_1,\rho(j_1)),\ldots,(j_r,\rho(j_r))$ are all the chords inside $[i+1,\rho(i)]$.
Now consider the $(r+1)\times(r+1)$ submatrix of $M$ with row indices $i+1,j_1,\ldots,j_r$
and column indices $\rho(j_1),\ldots,\rho(j_r),\rho(i)$. Its determinant must vanish;
expanding at first order $M=\underline\rho t+P$, the only contribution in the expansion of the determinant
is obtained by matching $j_a$ with $\rho(j_a)$, $a=1,\ldots,r$, and therefore $i+1$ with $\rho(i)$.
This implies $P_{i+1,\rho(i)}=0$, which recreates the second missing equation.
\end{enumerate}

For $\mathrm a=\mathrm c$:
\begin{enumerate}
\item In odd size, if either $i$ or $i+1$ is a fixed point of $\rho$, then the counting is the same.
\item Otherwise, we lose two equations involving $P_{i,i+1}$ and $P_{i+1,\rho(i)}$, which just as above,
are recovered by imposing $M_{i,i+1}=0$ and the rank condition in the interval $[i+1,\rho(i)]$.
\end{enumerate}

So in all cases, we find that the number of equations is the same as in the proof of
\thmref{genred}, so at this stage the dimension we would get out of this computation is $\dim E^{\mathrm a}_N$,
which is one more than the target dimension $\dim X^{\mathrm a}_{\rho,i}$.

However, we have not yet used the additional equation: $s_i(M)=s_{i+1}(M)$, which is valid either by definition
$E^{\mathrm a}_\rho \cap \{s_i(M)=s_{i+1}(M)\}$ or from \thmref{defeq} in $E^{\mathrm a}_\pi \cap \{M_{i,i+1}=0\}$.
If we expand at first order we obtain
\[
t_{\pi(i)} P_{\pi(i),i} + P_{i,\pi(i)} t_i =
t_{\pi(i+1)} P_{\pi(i+1),i+1} + P_{i+1,\pi(i+1)} t_{i+1}.
\]
One can check explicitly in all cases that this equation is independent from the ones above, thus showing that
the dimension is equal to that of $X^{\mathrm a}_{\rho,i}$.

Note that the reasoning above {\em fails}\/ in type $\mathrm a=\mathrm o$ in the case that $\rho(i)=r$, $\rho(i+1)=\ell$,
in the sense that the tangent space has dimension one more than that of the space. In this case one needs
to consider the tangent cone itself, which turns out to be of degree $2$.

\subsection{Multiplicity of \texorpdfstring{$Y_{\rho}^{\mathrm a}$}{Y}}
Recall that given a link pattern $\rho$ such that $\rho(L)\ne L+1$, we can associate to it the involution $\rho'$
where the images of $L$ and $L+1$ are swapped. By definition of $Y_\rho$, $\underline{\rho'} t$, $t$ generic diagonal, is in $Y_\rho^{\mathrm a}$.
We wish to calculate the Zariski tangent space of $\underline{\rho'} t$ inside $E^{\mathrm a}_\pi \cap \{M_{L,L+1}=0\}$, where $\pi=e_L(\rho)$.

It is not hard to check that the counting of equations is the same as in the proof of \thmref{genred} for
$\underline{\rho t}$ in type $\mathrm a=\mathrm i$, except the equations involving only $\{L,L+1,\rho(L),\rho(L+1)\}$
which we redo here:
\[\rho'=
\begin{tikzpicture}[baseline=(current  bounding  box.center),scale=0.5,every node/.style={inner sep=0,outer sep=0}]
\draw[thick] (0,0) circle(1);
\draw[smooth] (225:1) to[out=45,in=191.25] (11.25:1);
\draw[smooth] (348.75:1) to[out=168.75,in=315] (135:1);
\node[anchor=south east] at (135:1.2) {${\scriptstyle \rho(L+1)}$};
\node[anchor=west] at (11.25:1.2) {${\scriptstyle L+1}$};
\node[anchor=west] at (348.75:1.2) {${\scriptstyle L}$};
\node[anchor=north east] at (225:1.2) {${\scriptstyle \rho(L)}$};
\end{tikzpicture}\ ,
\qquad
t_L P_{L,L+1} + t_{L+1} P_{\rho(L+1),\rho(L)}=0.
\]
So we have one less equation than for $\underline{\rho}t$, whereas we need one more.

The first extra equation is obviously $M_{L,L+1}=0$ which implies $P_{L,L+1}=0$.
The second equation comes from the rank condition (\thmref{defeq}) for $E^{\mathrm a}_\pi$ in the interval
$[\rho(L),L]$, which implies $P_{\rho(L),L}=0$.

In total, we find that the Zariski tangent space has dimension one less than $E^{\mathrm a}_N$, which is the
dimension of $Y_\rho^{\mathrm a}$.

\section{\texorpdfstring{$SL_2^{(i)}$}{SL(2)}-invariance of certain subvarieties of \texorpdfstring{$E^{\mathrm a}_N$}{scheme}}\label{app:inv}
Consider a subvariety $V\subset E^{\mathrm a}_N$, and its image under conjugation by a subgroup $SL_2^{(i)}$,
as defined in \secref{levy}. As a first remark, to prove $SL_2^{(i)}$-invariance of $V$,
we only need to show that $\dim (SL_2^{(i)}\cdot V)=\dim V$,
since $SL_2^{(i)}\cdot V$ contains $V$ and is irreducible of the same dimension as $V$.

Next, note that if $V$ is to be $SL_2^{(i)}$-invariant, then in particular $SL_2^{(i)}\cdot V \subset E^{\mathrm a}_N$.
It is easy to see that the only equation of $E^{\mathrm a}_N$ that is potentially violated by sweeping with $SL_2^{(i)}$ is
$(M^2)_{i+1,i+N}=0$. Explicitly, if $M=P M' P^{-1}$ with $M'\in V$, then
\begin{equation}\label{eq:inv}
(M^2)_{i+1,i+N} = P_{i+1,i+1} P_{i+1,i} ((M'{}^2)_{ii}-(M'{}^2)_{i+1,i+1}) - P_{i+1,i}^2 (M'{}^2)_{i,i+N+1}.
\end{equation}
Note that $(M^2)_{i,i+N+1}$ is well-defined in the quotient space $\M_N^{\mathrm a}$ only if $M_{i,i+1}=0$ in $V$, which will be the case below.

Also, if, as in all cases below, one has $V \subset \{ s_i(M) = s_{i+1}(M) \}$, then the equation above simplifies to
\begin{equation}\label{eq:invb}
(M^2)_{i+1,i+N} = - P_{i+1,i}^2 (M'{}^2)_{i,i+N+1}.
\end{equation}

\subsection{Bulk case}
Here $i=1,\ldots,L-1$.

\subsubsection{$F_\rho\cap \{s_i=s_{i+1}\}$ for $\rho(i)\ne i+1, f_i\rho=\rho$}
This is case (1) of the proof of \propref{defX} with $f_i\rho=\rho$. Necessarily, $\mathrm a=\mathrm i$.
Call $V=\overline{F_\rho\cap \{s_i(M)=s_{i+1}(M)\}}$.

First we check that $(M^2)_{i+1,i+N}=0$ in $SL_2^{(i)}\cdot V$.
Since all entries of $M^2$ to the left of and below $(i,i+N+1)$ are known to be zero, this equation
is $\tilde B_N$-invariant and so we only need to use \eqref{eq:invb} with $M'=\rho^{\mathrm i}_<+\La$ where $\La\in \g_N^{\mathrm a}/R_N^{\mathrm a}$. Now
\[
(M'{}^2)_{i,i+N+1} = (\rho^{\mathrm i}_<)_{i,N-i+1} \La_{N-i+1,N+i+1}
=\La_{-i+1,i+1}.
\]
By a similar calculation,
\[
0=(M'{}^2)_{i+1,i+N}=\La_{-i,i}.
\]
These two entries are related by the symplectic symmetry $(a,b)\mapsto(1-b,1-a)$ and are therefore equal.
So $(M'{}^2)_{i,i+N+1}=0$.

By density and $\tilde B_N$-invariance, this implies that $SL_2^{(i)}\cdot V \subset
(SL_2^{(i)}\cdot E_\rho^{\mathrm a})\cap E_N^{\mathrm a}=E_\rho^{\mathrm a}$ from \secref{fi}.

We also learnt in the proof of \lmref{basicgeom}
that sweeping can only permute $s_i$ and $s_{i+1}$, so
we conclude that $SL_2^{(i)} \cdot V \subset E_\rho^{\mathrm a} \cap \{ s_i = s_{i+1} \}$. The latter being of the same dimension as the former,
we conclude that $V$ is $SL_2^{(i)}$-invariant.

\subsubsection{$F_\sigma$ where $\sigma$ is a link pattern except $\sigma(i)=N-i$, $\sigma(i+1)=N-i+1$}
This is case (2) of the proof of \propref{eicut}. Here, $\mathrm a=\mathrm c$,
and $V=\overline{F_\sigma}$.

We follow the same reasoning as above, and consider \eqref{eq:inv} with $M'=\sigma^{\mathrm i}_<+\La$.
First we find that $s_i(M')=\La_{-i,i}$ and $s_{i+1}(M')=\La_{-i+1,i+1}$, and once again these entries are related by the symplectic symmetry
so $s_i(M')=s_{i+1}(M')$.

Therefore, we are reduced to \eqref{eq:invb}, and we have:
\[
(M'{}^2)_{i,i+N+1} =
(\sigma^{\mathrm i}_<)_{i,N-i} \La_{N-i,i+N+1}.
\]
But $\La_{N-i,i+N+1}$ is such that the sum of its row and column is equal to $1\bmod N$,
which means it is on the axis of the symplectic symmetry, which means in type $\mathrm a=\mathrm c$ that it is zero.

So $SL_2^{(i)}\cdot V \subset E^{\mathrm a}_N$. Sweeping can at best permute $s_i$ and $s_{i+1}$ and leave the other $s_j$ unchanged, and preserves the equation
$M_{i,i+1}=0$, so
\[
SL_2^{(i)}\cdot V \subset (E^{\mathrm a}_\pi \cap \{M_{i,i+1}=0\}) \cup \bigcup_{\rho\ne\pi} (E^{\mathrm a}_\rho \cap \{ s_j = s_{\pi(j)}, j\ne i,i+1\}).
\]
It is easy to check that the RHS is of dimension $E^{\mathrm a}_N-1$, just like $V$ itself. By the same argument as above,
we conclude that $V=\overline{F_\sigma}$ is $SL_2^{(i)}$-invariant.

\subsection{Boundary case}
We only do the right boundary, i.e., $i=L$. The left boundary can be treated similarly.

\subsubsection{$E^{\mathrm a}_\pi$ where $\pi(L)\ne L+1$}
Here $V=E^{\mathrm a}_\pi$.
If $\pi(L)\ne L+1$, then any $M\in E^{\mathrm a}_\pi$ satisfies $M_{L,L+1}=0$
by \thmref{defeq}. This means that $PMP^{-1}$ is still upper triangular
for $P\in SL_2^{(L)}$. Also, $s_L(M)=s_{L+1}(M)$ by symplectic symmetry, so once again we are reduced to \eqref{eq:invb}.
But $(M'{}^2)_{L,L+N+1}=0$ because $(M'{}^2)^\dagger=M'{}^2$, and the sum of the row and column indices of that entry is equal to $1\bmod N$.

Therefore, $SL_2^{(L)}\cdot E^{\mathrm a}_\pi \subset E^{\mathrm a}_N$.
and by the usual dimension argument, it must be equal to $E^{\mathrm a}_\pi$.

\subsubsection{$E^{\mathrm a}_\pi \cap F_\sigma$, where $\pi(L)=L+1$ and $\sigma$ is a link pattern, $\sigma(L)\ne L+1$}
Let $V$ be an irreducible component of $E^{\mathrm a}_\pi\cap F_\sigma$ of dimension $\dim E^{\mathrm a}_N-1$
(if there exists any, in which case it is necessary top-dimensional).
By \propref{bdryinv}, whose proof is right above, $E^{\mathrm a}_\sigma$ is $SL_2^{(L)}$-invariant.

Now consider the rank equation southwest of the entry $(\sigma(L),L)$.
Being in $F_\sigma$ implies that the rank is {\em equal}\/ to the number of pairings of $\sigma$ inside $\{\sigma(L),\ldots,L\}$, which we call $r$.
For $E^{\mathrm a}_\pi\cap F_\sigma$ to be nonempty, this implies (equations (4) of \thmref{defeq}) that
the number of pairings of $\pi$ inside $\{\sigma(L),\ldots,L\}$ must be {\em at least}\/ $r$. But $\pi(L)=L+1$ and $\sigma(L)\ne L+1$,
so the number of pairings of $\pi$ in $\{\sigma(L),\ldots,L-1\}$ is the same, i.e., at least $r$, whereas that of $\sigma$ is only $r-1$.
This means that $\pi$ possesses at least
one pairing that $\sigma$ does not, say $i\leftrightarrow \pi(i)$, $\sigma(L)\le i<L$.
This means the equation (3) of \thmref{defeq} $s_i(M)=s_{\pi(i)}(M)$ is generically
violated in $E^{\mathrm a}_\sigma$, and since sweeping with $SL_2^{(L)}$ does not affect them, that $\dim(SL_2^{(L)}\cdot V)\le \dim E^{\mathrm a}_N-1$.

\subsubsection{$F_\sigma$, where $\sigma$ is a link pattern except $\sigma(L)=L,\sigma(L+1)=L+1$}
Set $V=\overline{F_\sigma}$, with $\sigma$ as in the third case of the proof of \propref{bdrycut}.
As explained in the proof for $E^{\mathrm a}_\pi$ above, $M_{L,L+1}=0$ in $V$ implies that $SL_2^{(L)}\cdot V \subset E^{\mathrm a}_N$.
Furthermore, as easily checked on a matrix of the form $M=\sigma_<+L$, one has $s_L(M)=s_{L+1}(M)=0$ in $V$.
This equation is preserved by sweeping (easily checked since the whole $2\times 2$ submatrix of entries with row and column equal to $L,L+1\bmod N$ is actually zero). No top-dimensional component of $E^{\mathrm a}_N$ has this equation,
so $E^{\mathrm a}_N \cap \{s_L(M)=s_{L+1}(M)\}$ is of dimension $\dim E^{\mathrm a}_N-1$, which is the dimension of $V$.

\section{Bound on the degree of the polynomials \texorpdfstring{$\phi_{\pi}^{\mathrm a}$}{ }}\label{app:bounddeg}
We wish to bound the degree of $\phi^{\mathrm a}_\pi = m_\pi \mdeg E_\pi^{\mathrm a}$ as a polynomial in one of the variables $z_1,\ldots,z_L$.

Recall that from the definition of multidegrees \cite{MS-book}, any multidegree
in $\M_N^{\mathrm a}$ is a sum of products of (distinct) weights
of $\M_N^{\mathrm a}$. This gives a first ``naive'' bound on the degree
of $\phi^{\mathrm a}_\pi$ in a given variable $z_i$: it is less or equal to
the number of coordinates in $\M^{\mathrm a}_N$ whose weight has a $z_i$-dependence. See \tabref{deg}.

However, this bound is not enough for our purposes. We can refine it as follows.
We focus at first on the multidegree of the whole of $E_N^{\mathrm a}$ rather than $\phi^{\mathrm a}_\pi$.
Suppose we apply the inductive definition of the multidegree by intersecting $E_N^{\mathrm a}$
with hyperplanes given by the vanishing of entries of the form $M_{ij}$ and $M_{ji}$ for fixed $i$. Each time the multidegree is multiplied by
a factor of the weight of $M_{ij}$, the intersection is trivial and the dimension stays constant. Since all variables whose weight have a $z_i$-dependence belong to that row/column, we have a bound on the degree in $z_i$:
\begin{multline*}
\text{degree in $z_i$ of $\mdeg E_N^{\mathrm a}$}\\
\le \text{number of entries of the form $M_{ij}$ or $M_{ji}$} - (\dim E^{\mathrm a}_N - \dim E^{\mathrm a}_{N-(1,2,4)}),
\end{multline*}
since the resulting variety (after all intersections) is simply the Brauer
loop scheme one size below ($L\to L-1$). More precisely, in all cases except $\mathrm a\in\{\mathrm m,\mathrm o\}$,
the entries of the form
$M_{ij}$ or $M_{ji}$ are in fact exactly those whose weight have $z_i$-dependence, and the inequality above
is an equality;
if $\mathrm a\in\{\mathrm m,\mathrm o\}$, $M_{i,i+n}$ does not have such a dependence, so the LHS is equal either
to the RHS, or the RHS minus one.
In other words,
\begin{multline*}
\text{degree in $z_i$ of $\mdeg E_N^{\mathrm a}$}\\
= \text{naive degree bound} - (\dim E^{\mathrm a}_N - \dim E^{\mathrm a}_{N-(1,2,4)}) +(0\text{ or }1) [\mathrm a\in\{\mathrm m,\mathrm o\}].
\end{multline*}

\begin{table}
\begin{tabular}{|c|c|c|c|c|c|c|}
\hline
a & naive degree bound & dim & dim shift & refined degree bound  & $\#$ recurrences
\\
\hline
p & $2(L-1)$ & $L^2/2, (L^2-1)/2$ & $L,L-1$ & $L-2,L-1$ & $2(L-1)$
\\
i & $2(2L-1)$ & $L(L+1)$ & $2L$ & $2(L-1)$ & $2(2L-1)$
\\
c & $4(L-1)$ & $L^2,L^2-1$ & $2L,2(L-1)$ & $2(L-2),2(L-1)$ & $4(L-1)$
\\
o & $4(2L-1)$ & $2L(L+1)$ & $4L$ & $4L-3^\ast$  & $2(2L-1)$
\\
m & $2(4L-3)$ & $L(2L+1)$ & $4L-1$ & $4(L-1)^\ast$ & $2(2L-1)$
\\
\hline
\end{tabular}
\vskip10pt
\caption{Degree bounds for $\phi^{\mathrm a}_\pi$ and number of available recurrences. The refined degree bound (fourth column)
is equal to the first column minus the third, plus one when there is a $^\ast$. If two numbers
are shown they correspond to even/odd cases.}\label{tab:deg}
\end{table}

Finally, we can calculate the number of recurrence relations of the type
of \eqref{eq:recsum} and find that it is always greater than the degree. This allows us to derive the
explicit expression \eqref{eq:norms}, and shows that $1$
is the correct choice for $\mathrm a\in\{\mathrm m,\mathrm o\}$ in the equation above.

Now if we consider individual components $E^{\mathrm a}_\pi$ rather than the whole scheme,
the same argument applies except after intersecting, we only have an upper bound on the resulting scheme
(it is a subscheme of $E^{\mathrm a}_{N-(1,2,4)}$), so that we obtain an upper bound for the degree:
\[
\text{degree in $z_i$ of $\phi_\pi^{\mathrm a}$}
\le
\text{naive degree bound}
-
(\dim E^{\mathrm a}_N - \dim E^{\mathrm a}_{N-(1,2,4)})
+1 [\mathrm a\in\{\mathrm m,\mathrm o\}].
\]

The result is shown in \tabref{deg}.


\gdef\MRshorten#1 #2MRend{#1}%
\gdef\MRfirsttwo#1#2{\if#1M%
MR\else MR#1#2\fi}
\def\MRfix#1{\MRshorten\MRfirsttwo#1 MRend}
\renewcommand\MR[1]{\relax\ifhmode\unskip\spacefactor3000 \space\fi
\MRhref{\MRfix{#1}}{{\scriptsize \MRfix{#1}}}}
\renewcommand{\MRhref}[2]{%
\href{http://www.ams.org/mathscinet-getitem?mr=#1}{#2}}
\bibliographystyle{amsplainhyper}
\bibliography{biblio}

\end{document}